\documentclass[a4paper,USenglish,cleveref,numberwithinsect]{lipics-v2021}

\nolinenumbers
\title{Fair Termination of Multiparty Sessions}

\author{Luca Ciccone}{Universit\`a di Torino,
  Italy}{luca.ciccone@unito.it}{https://orcid.org/0000-0001-9515-5280}{}
\author{Francesco Dagnino}{Universit\`a di Genova,
  Italy}{francesco.dagnino@dibris.unige.it}{https://orcid.org/ 0000-0003-3599-3535}{}
\author{Luca Padovani}{Universit\`a di Torino,
  Italy}{luca.padovani@unito.it}{https://orcid.org/0000-0001-9097-1297}{}

\authorrunning{L. Ciccone, F. Dagnino and L. Padovani}

\Copyright{Luca Ciccone, Francesco Dagnino and Luca Padovani}



\relatedversion{} 

\EventEditors{John Q. Open and Joan R. Access}
\EventNoEds{2}
\EventLongTitle{42nd Conference on Very Important Topics (CVIT 2016)}
\EventShortTitle{CVIT 2016}
\EventAcronym{CVIT}
\EventYear{2016}
\EventDate{December 24--27, 2016}
\EventLocation{Little Whinging, United Kingdom}
\EventLogo{}
\SeriesVolume{42}
\ArticleNo{23}

\usepackage{stmaryrd}
\usepackage{etoolbox}
\usepackage{mathpartir}
\usepackage{mathtools}
\usepackage{sty/prooftree}
\usepackage{amsmath}
\usepackage{sty/mps}
\usepackage{xspace} 
\usepackage{tikz}

\newcommand{\cf}{cf.\xspace} 

\begin{document}

\maketitle

\newcommand{\pcong}{\preccurlyeq}


\newcommand{\eg}{\emph{e.g.}\xspace}
\newcommand{\ie}{\emph{i.e.}\xspace}
\newcommand{\etal}{\emph{et al.}\xspace}


\newif\ifproofs
\proofstrue

\newif\ifcomments
\commentstrue

\newif\ifextension
\extensiontrue


\definecolor{myred}{rgb}{0.5,0,0}
\definecolor{mygreen}{rgb}{0,0.5,0}
\definecolor{myblue}{rgb}{0,0,0.5}


\newcommand{\set}[1]{\{#1\}}
\newcommand{\subst}[2]{\{#1/#2\}}
\newcommand{\eoe}{\hfill$\lrcorner$}
\newcommand{\mkkeyword}[1]{\textsf{\upshape\color{myblue}#1}}
\newcommand{\mktag}[1]{\textsf{\upshape\color{mygreen}#1}}

\newcommand{\RoleSet}{\mathsf{Roles}}

\renewcommand{\DefTirName}[1]{\defrule{#1}}

\newcommand{\infercorule}[3][]{%
  {\mprset{fraction={===}}\inferrule[#1]{#2}{#3}}%
}

\newcommand{\NatSet}{\mathbb{N}}
\newcommand{\LabelSet}{\mathbb{L}}

\newenvironment{lines}[1][t]{
  \begin{array}[#1]{@{}l@{}}
}{
  \end{array}
}

\setlength{\marginparsep}{1cm}
\setlength{\marginparwidth}{2cm}
\newcommand{\marginnote}[2]{%
  \ifcomments%
    $^{\color{magenta}\mathclap\star}$%
    \marginpar[
        \flushright\tiny\sf\textbf{#1}: #2
    ]{
        \flushleft\tiny\sf\textbf{#1}: #2
    }%
  \fi%
}


\newcommand{\Pol}{\pi}
\newcommand{\Out}{{!}}
\newcommand{\In}{{?}}

\newcommand{\action}{\actionA}
\newcommand{\actionA}{\alpha}
\newcommand{\actionB}{\beta}
\newcommand{\actions}{\actionsA}
\newcommand{\actionsA}{\varphi}
\newcommand{\actionsB}{\psi}

\newcommand{\Tag}[1][m]{\mktag{#1}}

\newcommand{\End}[1][\Pol]{#1\mkkeyword{end}}
\newcommand{\Tags}[1][i\in I]{\sum_{#1}}
\newcommand{\JTags}[1][j\in J]{\sum_{#1}}
\newcommand{\Label}[3]{#1#2#3}
\newcommand{\Labels}[3]{#1#2\{#3\}}
\newcommand{\Maps}[1][i\in I]{\prod_{#1}}

\newcommand{\Map}[1]{#1 \triangleright}

\newcommand{\branch}{+}
\newcommand{\choice}{\oplus}

\newcommand{\parop}{\mathrel{|}}


\newcommand{\EmptyCtx}{\emptyset}
\newcommand{\Ctx}{\Gamma}
\newcommand{\CtxD}{\Delta}


\newcommand{\Measure}{\MeasureM}
\newcommand{\MeasureM}{\mu}
\newcommand{\MeasureN}{\nu}


\newcommand{\IInd}{\mathsf{ind}}
\newcommand{\ICoInd}{\mathsf{coind}}
\newcommand{\IGen}{\mathsf{gen}}
\newcommand{\IAlg}{\mathsf{alg}}


\newcommand{\wtpx}[4]{#1 \vdash\ifblank{#2}{}{_{#2}}\ifblank{#4}{}{^{#4}} #3}
\newcommand{\wtp}[3][]{\wtpx{#2}{}{#3}{#1}}
\newcommand{\wtpc}[3][]{\wtpx{#2}\ICoInd{#3}{#1}}
\newcommand{\wtpi}[3][]{\wtpx{#2}\IInd{#3}{#1}}
\newcommand{\wtpn}[4][]{#3 \vDash^{#2} #4}


\newcommand{\rulename}[1]{\textup{\textsc{\small[#1]}}}
\newcommand{\defrule}[1]{\hypertarget{rule:#1}{\rulename{#1}}}
\newcommand{\refrule}[1]{\hyperlink{rule:#1}{\rulename{#1}}}

\newcommand{\proofcase}[1]{\textit{#1}.}
\newcommand{\proofrule}[1]{\proofcase{Case \refrule{#1}}}


\newcommand{\dom}[1]{\mathsf{dom}(#1)}
\newcommand{\targets}[1]{\mathsf{targets}(#1)}
\newcommand{\fn}[1]{\mathsf{fn}(#1)}
\newcommand{\bn}[1]{\mathsf{bn}(#1)}
\newcommand{\roles}[1]{\mathsf{roles}(#1)}
\newcommand{\co}[1]{\overline{#1}}
\newcommand{\paths}[1]{\mathsf{paths}(#1)}
\newcommand{\acombine}{\odot}
\newcommand{\prefixof}[2]{#1|_{#2}}


\newcommand{\Xnf}[1]{#1^{\mathit{nf}}}
\newcommand{\Pnf}{\Xnf{P}}
\newcommand{\Qnf}{\Xnf{Q}}
\newcommand{\Rnf}{\Xnf{R}}
\newcommand{\Xpar}[1]{#1^{\mathit{par}}}
\newcommand{\Ppar}{\Xpar P}
\newcommand{\Qpar}{\Xpar Q}
\newcommand{\Rpar}{\Xpar R}
\newcommand{\Pth}{P^{\mathit{th}}}
\newcommand{\Qth}{Q^{\mathit{th}}}


\newcommand{\PCtxC}{\mathcal{C}}
\newcommand{\PCtxD}{\mathcal{D}}
\newcommand{\Hole}{[~]}


\newcommand{\ft}{~\text{coherent}}
\newcommand{\coherent}{{\#}}

\newcommand{\ssubt}{\sqsubseteq}
\newcommand{\subt}[1][]{\leqslant\ifblank{#1}{}{_{#1}}}
\newcommand{\isubt}{\subt_{\mathsf{Ind}}}
\newcommand{\usubt}{\subt[*]}
\newcommand{\prefix}{\sqsubseteq}
\newcommand{\rsubt}[1][]{\sqsubseteq\ifblank{#1}{}{_{#1}}}

\newcommand{\converge}{\mathrel\downarrow}
\newcommand{\diverge}{\mathrel\uparrow}

\newcommand{\rank}[1]{\|#1\|}

\newcommand{\eqdef}{\stackrel{\text{\tiny\sf def}}=}

\newcommand{\terminated}{\checkmark}
\newcommand{\red}{\rightarrow}
\newcommand{\nred}{\arrownot\red}
\newcommand{\wred}{\Rightarrow}

\newcommand{\lred}[1]{\stackrel{#1}\longrightarrow}
\newcommand{\xred}[2]{\stackrel{#1}{\longrightarrow}_{#2}}

\newcommand{\xlred}[1]{\xrightarrow{#1}}

\newcommand{\nlred}[1]{\longarrownot\lred{#1}}
\newcommand{\nsred}[1]{\longarrownot\sred{#1}}
\newcommand{\nxred}[2]{\longarrownot\xred{#1}{#2}}

\newcommand{\wlred}[1]{\stackrel{#1}\Longrightarrow}
\newcommand{\wxred}[2]{\stackrel{#1}{\Longrightarrow}_{#2}}
\newcommand{\wsred}[1]{\wxred{#1}{\srel}}

\newcommand{\xwlred}[1]{\xRightarrow{#1}}
\newcommand{\xwxred}[2]{\xRightarrow{#1}_{#2}}
\newcommand{\xwsred}[1]{\xwxred{#1}{\srel}}

\newcommand{\nwlred}[1]{\Longarrownot\wlred{#1}}
\newcommand{\nwsred}[1]{\Longarrownot\wsred{#1}}


\newcommand{\Nat}{\mathbb{N}}


\newcommand{\angles}[1]{\langle#1\rangle}
\newcommand{\uset}{\mathcal{U}}
\newcommand{\Rule}[2]{\angles{#1, #2}}
\newcommand{\RuleSet}{\mathcal{I}}
\newcommand{\CoRuleSet}{\mathcal{I}_{\mathsf{co}}}

\begin{abstract}
    There exists a broad family of multiparty sessions in which the progress of
    one session participant is not unconditional, but depends on the choices
    performed by other participants. These sessions fall outside the scope of
    currently available session type systems that guarantee progress.
    In this work we propose the first type system ensuring that well-typed
    multiparty sessions, including those exhibiting the aforementioned
    dependencies, fairly terminate. Fair termination is termination under a
    fairness assumption that disregards those interactions deemed unfair and
    therefore unrealistic.
    Fair termination, combined with the usual safety properties ensured within
    sessions, not only is desirable \emph{per se}, but it entails progress and
    enables a compositional form of static analysis such that the well-typed
    composition of fairly terminating sessions results in a fairly terminating
    program.
\end{abstract}

\newcommand{\buyer}{\role[buyer]}
\newcommand{\seller}{\role[seller]}
\newcommand{\carrier}{\role[carrier]}

\newcommand{\tadd}{\Tag[add]}
\newcommand{\tpay}{\Tag[pay]}
\newcommand{\tship}{\Tag[ship]}

\section{Introduction}
\label{sec:introduction}

Sessions \cite{Honda93,HondaVasconcelosKubo98,HuttelEtAl16} are private
conversations among processes following a protocol specification called session
type. The decomposition of a distributed program into sessions enables its
modular static analysis and the enforcement of useful properties through a type
system. Examples of such properties are \emph{communication safety} (no message
of the wrong type is ever exchanged), \emph{protocol fidelity} (messages are
exchanged in the order prescribed by session types) and \emph{deadlock freedom}
(the program keeps running unless all sessions have terminated). These are all
instances of \emph{safety properties}, implying that ``nothing bad'' happens. In
general, one is also interested in reasoning and possibly enforcing
\emph{liveness properties}, those implying that ``something good'' happens
\cite{OwickiLamport82}. Examples of liveness properties are \emph{junk freedom}
(every message is eventually received), \emph{progress} (every non-terminated
participant of a session eventually performs an action) and \emph{termination}
(every session eventually comes to an end).

An enduring limitation of current type systems for multiparty sessions is that
\emph{they ensure progress for any participant of a session only when such
progress can be established independently of the choices performed by the other
participants}. To illustrate the impact of this limitation, consider a session
made of three participants named $\buyer$, $\seller$ and $\carrier$ in which the
buyer aims at purchasing an unspecified number of items from the seller and the
seller relies on a carrier for delivering the purchased items to the buyer. The
buyer behaves according to the session type $S$ that satisfies the equation
\begin{equation}
  \label{eq:buyer}
  S = \seller\Out\tadd.S + \seller\Out\tpay.\End[\Out]  
\end{equation}
indicating that it either pays the seller or it adds an item to the shopping
cart and then repeats the same behavior. In this session type, $\tadd$ and
$\tpay$ are messages targeted to the participant with role $\seller$. In turn,
the seller accepts $\tadd$ messages from the buyer until a $\tpay$ message is
received, at which point it instructs the carrier to $\tship$ the items. Thus,
its behavior is described by the session type $T$ that satisfies the equation
\begin{equation}
  \label{eq:seller}
  T = \buyer\In\tadd.T + \buyer\In\tpay.\carrier\Out\tship.\End[\Out]
\end{equation}

Finally, the carrier just waits for the $\tship$ message from the seller. So,
its behavior is described by the session type
\begin{equation}
  \label{eq:carrier}
  \seller\In\tship.\End[\In]
\end{equation}

No available type system is able to guarantee progress for every participant of
this multiparty session. What makes this session somewhat difficult to reason
about is that \emph{the progress of the carrier is not unconditional but depends
on the choices performed by the buyer}: the carrier can make progress only if
the buyer eventually pays the seller.

In this work we propose a type system that guarantees the \emph{fair
termination} of sessions, that is termination under a \emph{fairness
assumption}. The assumption we make is an instance of \emph{relative
fairness}~\cite{QueilleSifakis83} and can be roughly spelled out as follows:
\begin{equation}
  \label{eq:fairness_assumption}
  \textit{If termination is always possible, then it is inevitable.}
\end{equation}

The multiparty session sketched above terminates under this fairness assumption:
since it is always possible for the buyer to pay the seller and terminate, in
every fair execution of the session the buyer eventually pays the seller, even
though we do not know (nor do we impose) an upper bound to the number of items
that the buyer may add to the shopping cart. Simply, the non-terminating
execution of the session in which the buyer keeps adding items to the shopping
cart but never pays is assumed unrealistic and so it can be ignored insofar as
termination is concerned.

The reader might wonder why we focus on fair termination instead of considering
some fair version of progress. There are three reasons why we think that fair
termination is overall more appropriate than just progress.
First of all, ensuring that sessions (fairly) terminate is consistent with the
usual interpretation of the word ``session'' as an activity that lasts for a
\emph{finite amount of time}, even when the maximum duration of the activity is
not known \emph{a priori}.
Second, \emph{fair termination implies progress} when it is guaranteed along
with the usual safety properties of sessions. Indeed, if the session eventually
terminates, it must be the case that any non-terminated participant (think of
the carrier waiting for a $\tship$ message) is guaranteed to eventually make
progress, even when such progress \emph{depends} on choices made by other
participants (like the buyer sending $\tpay$ to the seller).
Last but not least, \emph{fair session termination enables compositional
reasoning} in the presence of multiple sessions. This is not true for progress:
if an action on a session $s$ is blocked by actions on a different session $t$,
then knowing that the session $t$ enjoys progress does not necessarily guarantee
that the action on $s$ will eventually be performed (the interaction on $t$
might continue forever). On the contrary, knowing that $t$ fairly terminates
guarantees that the action on $s$ will eventually be scheduled and performed, so
that $s$ may in turn progress towards termination.



Remarkably, the fairness assumption alone does not suffice to turn any
multiparty session type system into one that ensures fair termination. In fact,
there are several sources of potentially non-terminating behaviors that must be
ruled out in well-typed processes:
\begin{enumerate}
  \item\label{problem:new-sessions} Fairly terminating (and even finite)
  sessions may be chained, nested, interleaved in such a way that some pending
  activities are postponed forever. To avoid this problem, our type system makes
  sure that the effort required by a well-typed process in order to terminate
  remains finite. At the same time, it does not (always) prevent the modeling of
  processes that create an unbounded number of sessions.
  \item\label{problem:coherence} The type-level constraints usually imposed to
  well-typed sessions --
  \emph{duality}~\cite{Honda93,HondaVasconcelosKubo98,HuttelEtAl16},
  \emph{liveness}~\cite{ScalasYoshida19},
  \emph{coherence}~\cite{CarboneMontesiSchurmannYoshida17}, just to mention a
  few -- are in general too weak to entail fair session termination. Our type
  system adopts a stronger notion of ``correct multiparty session'' that entails
  fair termination.  Variants of this notion have already appeared in the
  literature \cite{BravettiZavattaro09,Padovani16}, but we use it here for the
  first time to relate types and processes.
  \item\label{problem:subtyping} A certain mismatch is usually allowed between
  the structure of session types and the structure of the processes that adhere
  to those types. This mismatch is formalized by a subtyping relation for
  session types which, in its standard formulation~\cite{GayHole05}, may
  introduce non-terminating behaviors. Our type system adopts \emph{fair
  subtyping} \cite{Padovani16}, a liveness-preserving refinement of the standard
  subtyping relation for session types \cite{GayHole05}.
\end{enumerate}

\subparagraph{Summary of contributions.}
We present the first type system ensuring the fair termination of multiparty
sessions and capable of addressing a number of natural communication patterns
that are out of scope of existing multiparty session type systems
\cite{ScalasYoshida19,GlabbeekHofnerHorne21}.
We exploit the compositional reasoning enabled by fair termination to prove a
strong soundness result whereby a well-typed composition of fairly terminating
sessions is a fairly terminating program (\cref{thm:soundness}).
This result scales smoothly also in presence of session chaining, session
nesting, session interleaving, session delegation and dynamic session creation.
In sharp contrast, the liveness properties ensured by previous multiparty
session type systems are either limited to single-session programs
\cite{ScalasYoshida19,GlabbeekHofnerHorne21} or require a richer type
structure~\cite{PadovaniVasconcelosVieira14,CoppoDezaniYoshidaPadovani16}.
Our contributions extend and generalize previous work on the fair termination of
binary sessions~\cite{CicconePadovani22} and allow for the modeling of
(intra-session) cyclic network topologies and of multiparty sessions that cannot
be decomposed into equivalent (well-typed) binary sessions. Decidability of type
checking is not substantially more difficult than the same problem in the binary
setting~\cite{CicconePadovani22}.
\emph{En passant}, in this paper we also provide a new characterization of fair
subtyping for (multiparty) session types (\cref{tab:subt}) that is substantially
simpler than those appearing in previous works
\cite{Padovani13,Padovani16,CicconePadovani21,CicconePadovani22}.

\subparagraph{Structure of the paper.}
We recall the key notions related to fair termination
(\cref{sec:fair-termination}) before presenting our language of multiparty
sessions (\cref{sect:calculus}). 
Then, we define multiparty session types and fair subtyping (\cref{sec:types})
and present the typing rules and the soundness properties of the type system
(\cref{sec:ts}).
In the latter part of the paper we illustrate a few more advanced examples of
well-typed processes (\cref{sec:ts_ex}), we discuss related work in more detail
(\cref{sec:related-work}) and we provide hints at further developments
(\cref{sec:conclusion}).
Additional technical material and all the proofs of the presented results can be
found in the Appendix, which we provide for completeness but is not necessary
for reviewing the submission.


\newcommand{\States}{\mathcal{S}}
\newcommand{\StateSet}{\mathcal{C}}
\newcommand{\FinalStates}{\mathcal{F}}
\newcommand{\run}{\rho}
\newcommand{\FairA}{\Phi}
\newcommand{\TF}{\mathbb{T}}

\section{Fair Termination}
\label{sec:fair-termination}

Since the notion of fair termination will apply to several different entities
(session types, multiparty sessions, processes) here we define it for a generic
reduction system. Later on we will show various instantiations of this
definition.
A \emph{reduction system} is a pair $(\States, {\red})$ where $\States$ is a set
of \emph{states} and ${\red} \subseteq \States \times \States$ is a
\emph{reduction relation}.
We adopt the following notation:
we let $C$ and $D$ range over states;
we write $C \red$ if there exists $D \in \States$ such that $C \red D$; we write
$C \nred$ if not $C \red$; we write $\wred$ for the reflexive, transitive
closure of $\red$.
We say that $D$ is \emph{reachable} from $C$ if $C \wred D$.

As an example, the reduction system $(\set{A,B}, \set{(A,A),(A,B)})$
models an entity that can be in two states, $A$ or $B$, and such that the entity
may perform a reduction to remain in state $A$ or a reduction to move from state
$A$ to state $B$. To formalize the evolution of an entity from a particular
state we define \emph{runs}.

\begin{definition}[runs and maximal runs]
  \label{def:run}
  A \emph{run} of $C$ is a (finite or infinite) sequence
  $C_0C_1\dots C_i\dots$ of states such that $C_0 = C$ and
  $C_i \red C_{i+1}$ for every valid $i$. A run is \emph{maximal} if
  either it is infinite or if its last state $C_n$ is such that
  $C_n \nred$.
\end{definition}

Hereafter we let $\run$ range over runs. Each run in the previously defined
reduction system is either of the form $A^n$ -- a finite sequence of $A$ -- or
of the form $A^nB$ -- a finite sequence of $A$ followed by one $B$ -- or
$A^\omega$ -- an infinite sequence of $A$. Among these, the runs of the form
$A^nB$ and $A^\omega$ are maximal, whereas no run of the form $A^n$ is maximal.

We now use runs to define different termination properties of states:
we say that $C$ is \emph{weakly terminating} if there exists a maximal run of
$C$ that is finite;
we say that $C$ is \emph{terminating} if every maximal run of $C$ is finite;
we say that $C$ is \emph{diverging} if every maximal run of $C$ is infinite.
\emph{Fair termination}~\cite{Francez86} is a termination property that only
considers a subset of all (maximal) runs of a state, those that are considered
to be ``realistic'' or ``fair'' according to some fairness assumption.
The assumption that we make in this work, and that we stated in words in
\eqref{eq:fairness_assumption}, is formalized thus:

\begin{definition}[fair run]
  \label{def:fair_run}
  A run is \emph{fair} if it contains finitely many weakly terminating states.
  Conversely, a run is \emph{unfair} if it contains infinitely many weakly
  terminating states.
\end{definition}

Continuing with the previous example, the runs of the form $A^n$ and $A^nB$ are
fair, whereas the run $A^\omega$ is unfair. In general, an unfair run is an
execution in which termination is always within reach, but is never reached.

A key requirement of any fairness assumption is that it must be possible to
extend every finite run to a maximal fair one. This property is called
\emph{feasibility}~\cite{AptFrancezKatz87,GlabbeekHofner19} or \emph{machine
closure}~\cite{Lamport00}.
It is easy to see that our fairness assumption is feasible:

\begin{lemma}
  \label{lem:feasibility}
  If $\run$ is a finite run, then there exists $\run'$ such that $\run\run'$ is
  a maximal fair run.
\end{lemma}

Fair termination is finiteness of all maximal fair runs:

\begin{definition}[fair termination]
  \label{def:fair_termination}
  We say that $C$ is \emph{fairly terminating} if every maximal fair run of $C$
  is finite.
\end{definition}

In the reduction system given above, $A$ is fairly terminating. Indeed, all the
maximal runs of the form $A^nB$ are finite whereas $A^\omega$, which is the only
infinite fair run of $A$, is unfair.




For the particular fairness assumption that we make, it is possible to provide a
sound and complete characterization of fair termination that does not mention
fair runs. This characterization will be useful to relate fair termination with
the notion of correct multiparty session (\cref{def:coherence}) and the
soundness property of the type system (\cref{thm:soundness}).

\begin{theorem}
  \label{thm:fair_termination}
  Let $(\States, {\red})$ be a reduction system and $C\in\States$. Then $C$ is
  fairly terminating if and only if every state reachable from $C$ is weakly
  terminating.
\end{theorem}

\begin{remark}[fair reachability of predicates~\cite{QueilleSifakis83}]
  Most fairness assumptions have the form ``if \emph{something} is infinitely
  often possible then \emph{something} happens infinitely often'' and, in this
  respect, our formulation of fair run (\cref{def:fair_run}) looks slightly
  unconventional. However, it is not difficult to realize that
  \cref{def:fair_run} is an instance of the notion of fair reachability of
  predicates as defined by Queille and Sifakis~\cite[Definition
  3]{QueilleSifakis83}. According to Queille and Sifakis, a run $\run$ is fair
  with respect to some predicate $\StateSet \subseteq \States$ if, whenever in
  $\run$ there are infinitely many states from which a state in $\StateSet$ is
  reachable, then in $\run$ there are infinitely many occurrences of states in
  $\StateSet$. When we take $\StateSet$ to be $\nred$, that is the set of
  terminated states that do not reduce, pretending that irreducible states
  should occur infinitely often in the run is nonsensical. So, the fairness
  assumption boils down to assuming that such states should \emph{not} be
  reachable infinitely often, which is precisely the formulation of
  \cref{def:fair_run}.
  \eoe
\end{remark}


\section{A Calculus of Multiparty Sessions} 
\label{sect:calculus} 

In this section we define the calculus for multiparty sessions on which we apply
our static analysis technique. The calculus is an extension of the one presented
by Ciccone and Padovani~\cite{CicconePadovani22} to multiparty sessions in the
style of Scalas and Yoshida~\cite{ScalasYoshida19}. 

We use an infinite set of \emph{variables} ranged over by $x$, $y$, $z$, an
infinite set of \emph{session names} ranged over by $s$ and $t$, a set of
\emph{roles} ranged over by $\rolep$, $\roleq$, $\roler$, a set of \emph{message
tags} ranged over by $\Tag$, and a set of \emph{process names} ranged over by
$A$, $B$, $C$. In the literature of sessions tags are usually called labels. We
adopt a different terminology to avoid confusion with another notion of label
that we introduce in \cref{sec:types}.
We use roles to distinguish the participants of a session. In particular, an
\emph{endpoint} $\ep\sn\role$ consists of a session name $\sn$ and a role
$\role$ and is used by the participant with role $\role$ to interact with the
other participants of the session $s$.
We use $u$ and $v$ to range over \emph{channels}, which are either variables or
session endpoints.
We write $\seqof x$ and $\seqof u$ to denote possibly empty sequences of
variables and channels, extending this notation to other entities.
We use $\Pol$ to range over the elements of the set $\set{\iact,\oact}$ of
\emph{polarities}, distinguishing input actions ($\iact$) from output actions
($\oact$).

\begin{table}
  \caption{\label{tab:proc-syntax}Syntax of processes.}
  \centering
  \begin{math}
    \displaystyle
    \begin{array}[t]{@{}rcll@{}}
      P, Q, R & ::= & & \textbf{Process} \\
      &   & \pdone & \text{termination} \\
      & | & \pwait\chvar{P} & \text{signal input} \\
      & | & \pich\chvar\role{x}{P} & \text{channel input} \\
      & | & \pbranch[i\in I]\chvar\role\Pol{\Tag_i}{P_i} & \text{tag input/output} \\
      & | & \pres\sn{P_1\ppar\cdots\ppar P_n} & \text{session} \\
    \end{array}
    ~
    \begin{array}[t]{@{}rcll@{}}
      \\
      & | & \pinvk\pdn{\seqof\chvar} & \text{invocation} \\
      & | & \pclose\chvar & \text{signal output} \\
      & | & \poch\chvar\role\achvar{P} & \text{channel output} \\
      & | & P \pchoice Q &\text{choice} \\
      & | & \pcast\chvar P & \text{cast} \\
    \end{array}
  \end{math}
\end{table}

A \emph{program} is a finite set of \emph{definitions} of the form
$\pdef\pdn{\seqof\var}{P}$, at most one for each process name, where $P$ is a
term generated by the syntax shown in \cref{tab:proc-syntax}.
The term $\pdone$ denotes the terminated process that performs no action.
The term $\pinvk\pdn{\seqof\chvar}$ denotes the invocation of the process with
name $\pdn$ passing the channels $\seqof\chvar$ as arguments. When
$\seqof\chvar$ is empty we just write $A$ instead of $\pinvk{A}{}$.
The term $\pclose\chvar$ denotes the process that sends a termination signal on
the channel $\chvar$, whereas $\pwait\chvar P$ denotes the process that waits
for a termination signal from channel $\chvar$ and then continues as $P$.
The term $\poch\chvar\role\achvar{P}$ denotes the process that sends the channel
$\achvar$ on the channel $\chvar$ to the role $\rolep$ and then continues as
$P$. Dually, $\pich\chvar\role\var{P}$ denotes the process that receives a
channel from the role $\rolep$ on the channel $\chvar$ and then continues as $P$
where $\var$ is replaced with the received channel.
The term $\pbranch[i\in I]\chvar\role\Pol{\Tag_i}{P_i}$ denotes a process that
exchanges one of the tags $\Tag_i$ on the channel $\chvar$ with the role $\role$
and then continues as $P_i$. Whether the tag is sent or received depends on the
polarity $\Pol$ and, as it will be clear from the operational semantics, the
polarity $\Pol$ also determines whether the process behaves as an internal
choice (when $\Pol$ is $\oact$) or an external choice (when $\Pol$ is $\iact$).
In the first case the process chooses \emph{actively} the tag being sent,
whereas in the second case the process reacts \emph{passively} to the tag being
received.
We assume that $I$ is finite and non-empty and also that the tags $\Tag_i$ are
pairwise distinct. For brevity, we write $\pbranch\chvar\role\Pol{\Tag_k}{P_k}$
instead of $\pbranch[i\in I]\chvar\role\Pol{\Tag_i}{P_i}$ when $I$ is the
singleton set $\set{k}$.
The term $P \pchoice Q$ denotes a process that non-deterministically behaves
either as $P$ or as $Q$.

A term $\pres{s}{P_1\ppar\cdots\ppar P_n}$ with $n\geq 1$ denotes the parallel
composition of $n$ processes, each of them being a participant of the session
$s$. Each process is associated with a distinct a role $\role_i$ and
communicates in $s$ through the endpoint $\ep{s}{\role_i}$. Combining session
creation and parallel composition in a single form is common in session type
systems based on linear
logic~\cite{CairesPfenningToninho16,Wadler14,LindleyMorris16} and helps
guaranteeing deadlock freedom. 
Finally, a \emph{cast} $\pcast\chvar P$ denotes a process that behaves exactly
as $P$. This form is only relevant for the type system (\cref{sec:ts}) and
denotes the fact that the type of $\chvar$ is subject to an application of
subtyping.

The free and bound names of a process are defined as usual, the latter ones
being easily recognizable as they occur within round parenteses. We write
$\fn{P}$ for the set of free names of $P$ and we identify processes modulo
renaming of bound names. Note that $\fn{P}$ may contain variables and session
names, but not endpoints.
Occasionally we write $\pdef{A}{\seqof{x}}{P}$ as a predicate or side condition,
meaning that $P$ is the process associated with the process name $A$. For each
of such definitions we assume that $\fn{P} \subseteq \set{\seqof{x}}$.

\begin{table}
  \caption{Structural precongruence of processes.}
  \label{tab:pcong}
  \centering
  \begin{math}
    \displaystyle
    \begin{array}{@{}lr@{~}c@{~}ll@{}}
      \defrule{s-par-comm} & \pres\sn{\procs{P} \ppar P \ppar Q \ppar \procs{Q}} 
      & \pcong & 
      \pres\sn{\procs{P} \ppar Q \ppar P \ppar \procs{Q}}
      \\
      \defrule{s-par-assoc} & \pres\sn{\procs{P} \ppar \pres\asn{R \ppar \procs{Q}}} 
      & \pcong & 
      \pres\asn{\pres\sn{\procs{P} \ppar R} \ppar \procs{Q}}
      & \text{if $s \in \fn{R}$}
      \\
      \defrule{s-cast-comm} & \pcast{u}{\pcast{v}{P}} & \pcong & \pcast{v}{\pcast{u}{P}}
      \\
      \defrule{s-cast-new} & \pres{s}{\pcast{\ep{s}\role} P \ppar \procs{Q}}
      & \pcong & 
      \pres{s}{P \ppar \procs{Q}}
      \\
      \defrule{s-cast-swap} & \pres{s}{\pcast{\ep{t}\role}{P} \ppar \procs{Q}} 
      & \pcong & 
      \pcast{\ep{t}\role}{\pres{s}{P \ppar \procs{Q}}}
      & \text{if $s \ne t$} 
      \\
      \defrule{s-call} & \pinvk{A}{\seqof{u}} & \pcong & P\subst{\seqof{u}}{\seqof{x}}
      & \text{if $\pdef{A}{\seqof{x}}{P}$}
    \end{array}
  \end{math}
\end{table}

\begin{table}
  \caption{Reduction of processes.}
  \label{tab:red}
  \begin{mathpar}
    \inferrule[r-choice]{ }{
      P_1 \choice P_2 \red P_k
    }
    ~ k\in\set{1,2}
    \and
    \inferrule[r-signal]{ }{
      \pres{s}{\pwait{\ep{s}\role}{P} \parop \pclose{\ep{s}{\roleq_1}} \parop \cdots \parop \pclose{\ep{s}{\roleq_n}}} 
      \red
      P
    }
    \and
    \inferrule[r-channel]{ }{
      \pres{s}{\poch{\ep{s}{\rolep}}{\roleq}{v}{P} \parop \pich{\ep{s}{\roleq}}{\rolep}{x}{Q} \parop \procs{R}}
      \red
      \pres{s}{P \parop Q\subst{v}{x}	\parop \procs{R}}
    }
    \and
    \inferrule[r-pick]{ }{
      \pres{s}{\pobranch[i\in I]{\ep{s}\role}\roleq{\Tag_i}{P_i} \parop \procs{Q}} 
      \red
      \pres{s}{\pobranch{\ep{s}\role}\roleq{\Tag_k}{P_k}\parop \procs{Q}}
    }
    ~ k\in I
    \and
    \inferrule[r-tag]{ }{
      \pres{s}{\pobranch{\ep{s}\role}{\roleq}{\Tag_k}{P} \parop \pibranch[i\in I]{\ep{s}\roleq}\rolep{\Tag_i}{Q_i} \parop \procs{R}} 
      \red
      \pres{s}{P \parop Q_k \parop \procs{R}}
    }
    ~ k\in I
    \and
    \inferrule[r-par]{
      P \red Q
    }{
      \pres{s}{P \parop \procs{R}} \red \pres{s}{Q \parop \procs{R}}
    }
    \and
    \inferrule[r-cast]{
      P \red Q
    }{
      \pcast{u}{P} \red \pcast{u}{Q}
    }
    \and
    \inferrule[r-struct]{
      P \pcong P'
      \\
      P' \red Q'
      \\
      Q' \pcong Q
    }{
      P \red Q
    }
  \end{mathpar}
\end{table}

The operational semantics of processes is given by the structural precongruence
relation $\pcong$ defined in \cref{tab:pcong} and the reduction relation $\red$
defined in \cref{tab:red}. As usual, structural precongruence allows us to
rearrange the structure of processes without altering their meaning, whereas
reduction expresses an actual computation or interaction step.
The adoption of a structural \emph{pre}congruence (as opposed to a more common
congruence relation) is not strictly necessary, but it simplifies the technical
development by reducing the number of cases we have to consider in proofs
without affecting the properties of the calculus in any way.

Rules \refrule{s-par-comm} and \refrule{s-par-assoc} state commutativity and
associativity of parallel composition of processes (we write $\seqof{P}$ to
denote possibly empty parallel compositions of processes). In
\refrule{s-par-assoc}, the side condition $s \in \fn{R}$ makes sure that $R$ is
indeed a participant of the session $s$. Note that this rule only states
right-to-left associativity. Left-to-right associativity is derivable from this
rule and repeated uses of \refrule{s-par-comm}.
Rule \refrule{s-cast-comm} allows us to swap two consecutive casts.
Rule \refrule{s-cast-new} removes an unguarded cast on an endpoint of the
restricted session (we refer to this operation as ``performing the cast'').
Rule \refrule{s-cast-swap} swaps a cast and a restricted session as long as the
endpoint in the cast refers to a different session.
Finally, rule \refrule{s-call} unfolds a process invocation to its definition.
Hereafter, we write $\subst{u}{x}$ for the capture-avoiding substitution of each
free occurrence of $x$ with $u$ and $\subst{\seqof{u}}{\seqof{x}}$ for its
natural extension to equal-length tuples of variables and names.
The rules \refrule{s-cast-new}, \refrule{s-cast-swap} and \refrule{s-call} are
not invertible: by \refrule{s-cast-new} casts can only be removed but never
added; by \refrule{s-cast-swap} casts can only be moved closer to their
restriction, so that they can be eventually performed by \refrule{s-cast-new};
by \refrule{s-call} process invocations can only be unfolded.
%

The reduction relation is quite standard.
Rule \refrule{r-choice} reduces $P_1\pchoice P_2$ to either $P_1$ or $P_2$, non
deterministically.
Rule \refrule{r-signal} terminates a session in which all participants
($\roleq_1,\ldots,\roleq_n$) but one ($\rolep$) are sending a termination signal
and $\rolep$ is waiting for it; the resulting process is the continuation of the
participant $\rolep$.
Rule \refrule{r-channel} models the exchange of a channel among two participants
of a session.
Rule \refrule{r-pick} models an internal choice whereby a process picks one
particular tag $\Tag_k$ to send on a session.
Rule \refrule{r-tag} synchronizes two participants $\rolep$ and $\roleq$ on
the tag chosen by $\rolep$.
Finally, rules \refrule{r-par}, \refrule{r-cast} and \refrule{r-struct} close
reductions under parallel compositions and casts and by structural precongruence. 

%

\newcommand{\Main}{\textit{Main}}
\newcommand{\Buyer}{\textit{Buyer}}
\newcommand{\Seller}{\textit{Seller}}
\newcommand{\Carrier}{\textit{Carrier}}

In the rest of this section we illustrate the main features of the calculus with
some examples. For none of them the existing multiparty session type systems are
able to guarantee progress.

\begin{example}[purchase]
  \label{ex:bsc}
  We model a particular instance of the buyer-seller-carrier interaction that we
  have informally discussed in \cref{sec:introduction} with the following
  definitions:
  \begin{align*}
    \Main & \peq \pres\sn{ \pinvk\Buyer{\ep\sn\buyer} \ppar \pinvk\Seller{\ep\sn\seller} \ppar \pinvk\Carrier{\ep\sn\carrier}} \\ 
    \Buyer(x) & \peq \act{x}\seller\oact\set{
    			  \tadd.\act{x}\seller\oact\tadd.\pinvk\Buyer{x},
                  \tpay.\pclose{x}
                } \\ 
    \Seller(x) & \peq \act{x}\buyer\iact\set{
                  \tadd.\pinvk\Seller{x},
                  \tpay.\act{x}\carrier\oact\tship.\pclose{x} 
                } \\ 
    \Carrier(x) & \peq \act{x}\seller\iact\tship.\pwait{x}\pdone
  \end{align*}

  Note that the buyer either sends $\tpay$ or it sends two $\tadd$ messages in a
  row before repeating this behavior. That is, this particular buyer always adds
  an even number of items to the shopping cart.
  Nonetheless, the buyer periodically has a chance to send a $\tpay$ message and
  terminate. Therefore, the execution of the program in which the buyer only
  sends $\tadd$ is unfair according to \cref{def:fair_run} hence this program is
  fairly terminating. 
  \eoe
\end{example}


\newcommand{\rbuyer}{\role[b]}
\newcommand{\rseller}{\role[s]}
\newcommand{\rcarrier}{\role[c]}

\newcommand{\tquery}{\Tag[query]}
\newcommand{\tprice}{\Tag[price]}
\newcommand{\tok}{\Tag[ok]}
\newcommand{\tcancel}{\Tag[cancel]}
\newcommand{\tbox}{\Tag[box]}
\newcommand{\tsplit}{\Tag[split]}
\newcommand{\tgiveup}{\Tag[giveup]}
\newcommand{\tyes}{\Tag[yes]}
\newcommand{\tno}{\Tag[no]}

\begin{example}[purchase with negotiation]
  \label{ex:2bsc}
  Consider a variation of \cref{ex:bsc} in which the buyer, before making the
  payment, negotiates with a secondary buyer for an arbitrarily long time. The
  interaction happens in two nested sessions, an outer one involving the primary
  buyer, the seller and the carrier, and an inner one involving only the two
  buyers. We model the interaction as the program below, in which we collapse
  role names to their initials.
  \begin{align*}
    \Main & \peq \pres\sn{ \pinvk\Buyer{\ep\sn\rbuyer} \ppar \pinvk\Seller{\ep\sn\rseller} \ppar \pinvk\Carrier{\ep\sn\rcarrier} }
    \\ 
    \Buyer(x) & \peq \act{x}\rseller\oact\tquery.
                \act{x}\rseller\iact\tprice.
                \pres{t}{ \pinvk{\Buyer_1}{x,\ep\asn{\rbuyer_1}} \ppar \pinvk{\Buyer_2}{\ep\asn{\rbuyer_2}}} 
    \\ 
    \Seller(x) & \peq \act{x}\rbuyer\iact\tquery.
                \act{x}\rbuyer\oact\tprice.
                \act{x}\rbuyer\iact\set{
                  \tpay.\act{x}\rcarrier\oact\tship.\pclose{x},
                  \tcancel.\act{x}\rcarrier\oact\tcancel.\pclose{x}
                }
    \\     
    \Carrier(x) & \peq \act{x}\rseller\iact\set{
                  \tship.\act{x}\rbuyer\oact\tbox.\pclose{x},
                  \tcancel.\pclose{x} 
                }
    \\
    \Buyer_1(x,y) & \peq \act{y}{\rbuyer_2}\oact\{
                      \begin{lines}
                        \tsplit.\act{y}{\rbuyer_2}\iact\{
                          \begin{lines}
                            \tyes.\pcast{x}
                                  \act{x}\rseller\oact\tok.
                                  \act{x}\rcarrier\iact\tbox.
                                  \pwait{x}
                                  \pwait{y}
                                  \pdone,
                                  \\
                            \tno.\pinvk{\Buyer_1}{x,y} \},
                          \end{lines}
                        \\
                        \tgiveup.
                          \pwait{y}
                          \pcast{x}
                          \act{x}\rseller\oact\tcancel.
                          \pwait{x}
                          \pdone \}
                      \end{lines}
    \\
    \Buyer_2(y) & \peq \act{y}{\rbuyer_1}\iact\set{
                    \tsplit.\act{y}{\rbuyer_1}\oact\set{
                      \tyes.\pclose{y},
                      \tno.\pinvk{\Buyer_2}{y} 
                    },
                    \tgiveup.\pclose{y} 
                  }
  \end{align*} 

  The buyer queries the seller which replies with a price. At this point,
  $\Buyer$ creates a new session $t$ and forks as a primary buyer $\Buyer_1$ and
  a secondary buyer $\Buyer_2$. The interaction between the two sub-buyers goes
  on until either $\Buyer_1$ gives up or $\Buyer_2$ accepts its share of the
  price. In the former case, the primary buyer waits for the internal session to
  terminate and $\tcancel$s the order with the seller which, in turn, aborts the
  transaction with the carrier. In the latter case, the buyer confirms the order
  to the seller, which then instructs the carrier to $\tship$ a $\tbox$ to the
  buyer.
  
  Note that the outermost session $s$, taken in isolation, terminates in a
  bounded number of interactions, but its progress cannot be established without
  assuming that the innermost session $t$ terminates. In particular, if the two
  buyers keep negotiating forever, the seller and the carrier starve.
  However, the innermost session can terminate if $\Buyer_1$ sends $\tgiveup$ to
  $\Buyer_2$ or if $\Buyer_2$ sends $\tyes$ to $\Buyer_1$. Thus, the run in
  which the two buyers negotiate forever is unfair, the session $t$ fairly
  terminates and the session $s$ terminates as well.
  
  On the technical side, note that the definition of $\Buyer_1$ contains two
  casts on the variable $x$. As we will see in \cref{ex:2bsc-ts}, these casts
  are necessary for the typeability of $\Buyer_1$ to account for the fact that
  $x$ is used \emph{differently} in two distinct branches of the process.
  \eoe
\end{example}


\newcommand{\Sort}{\textit{Sort}}
\newcommand{\Merge}{\textit{Merge}}

\newcommand{\rworker}{\role[w]}
\newcommand{\rmaster}{\role[m]}

\newcommand{\treq}{\Tag[req]}
\newcommand{\tres}{\Tag[res]}

\begin{example}[parallel merge sort]
  \label{ex:pms}
  To illustrate an example of program that creates an unbounded number of
  sessions we model a parallel version of the merge sort algorithm.
  \begin{align*}
    \Main & \peq \pres{s}{
                    \act{\ep{s}\rmaster}\rworker\oact\treq.
                    \act{\ep{s}\rmaster}\rworker\iact\tres.
                    \pwait{s}
                    \pdone \parop
                    \pinvk\Sort{\ep{s}\rworker}
                  } \\ 
    \Sort(x) & \peq \act{x}\rmaster\iact\treq.(
                  \pres{t}{
                    \pinvk\Merge{x,\ep{t}\rmaster} \parop
                    \pinvk\Sort{\ep{t}{\rworker_1}} \parop
                    \pinvk\Sort{\ep{t}{\rworker_2}}
                  }
                  \pchoice
                  \act{x}\rmaster\oact\tres.\pclose{x}
                )
    \\ 
    \Merge(x,y) & \peq \act{y}{\rworker_1}\oact\treq.
                  \act{y}{\rworker_2}\oact\treq.
                  \act{y}{\rworker_1}\iact\tres.
                  \act{y}{\rworker_2}\iact\tres.
                  \pwait{y}
                  \act{x}\rmaster\oact\tres.
                  \pclose{x}
  \end{align*}

  The program starts as a single session $s$ in which a master $\rmaster$ sends
  the initial collection of data to the worker $\rworker$ as a $\treq$ message
  and waits for the $\tres$ult. The worker is modeled as a process $\Sort$ that
  decides whether to sort the data by itself (right branch of the choice in
  $\Sort$), in which case it sends the $\tres$ult directly to the master, or to
  partition the collection (left branch of the choice in $\Sort$). In the latter
  case, it creates a new session $t$ in which it sends $\treq$uests to two
  sub-workers $\rworker_1$ and $\rworker_2$, it gathers the partial $\tres$ults
  from them and gets back to the master with the complete $\tres$ult.

  Since a worker may always choose to start two sub-workers in a new session,
  the number of sessions that may be created by this program is unbounded. At
  the same time, each worker may also choose to complete its task without
  creating new sessions. So, while in principle there exists a run of this
  program that keeps creating new sessions forever, this run is unfair according
  to \cref{def:fair_run}.
  \eoe
\end{example}

\newcommand{\srel}{\mathcal{S}}

\section{Multiparty Session Types and Fair Subtyping}
\label{sec:types}

In this section we define syntax and semantics of multiparty session types
(\cref{sec:types_syntax_semantics}) as well as an inference system for fair
subtyping (\cref{sec:types_inference_system}).

\subsection{Syntax and Semantics}
\label{sec:types_syntax_semantics}

A \emph{session type} is a regular tree~\cite{Courcelle83} coinductively
generated by the productions below:
\[
  \textstyle
  \textbf{Session type}
  \qquad
  S, T, U, V ::= \End \mid \Tags\rolep\Pol\Tag_i.S_i \mid \rolep\Pol{S}.T
\]

The session type $\End$ describes the behavior of a process that sends/receives
a termination signal.
The session type $\Tags\rolep\Pol\Tag_i.S_i$ describes the behavior of a process
that sends to or receives from the participant $\rolep$ one of the tags $\Tag_i$
and then behaves according to $S_i$. Note that the source or destination role
$\rolep$ and the polarity $\Pol$ are the same in every branch. We require that
$I$ is not empty and $i, j \in I$ with $i \ne j$ implies $\Tag_i \ne \Tag_j$.
Occasionally we write $\rolep\Pol\Tag_1.S_1 + \cdots + \rolep\Pol\Tag_n.S_n$
instead of $\Tags[i=1]^n \rolep\Pol\Tag_i.S_i$.
Finally, a session type $\rolep\Pol{S}.T$ describes the behavior of a process
that sends to or receives from the participant $\rolep$ an endpoint of type $S$
and then behaves according to $T$.
We often specify infinite session types as solutions of equations of the form $S
= \cdots$ where the metavariable $S$ may occur on the right hand side of $=$
guarded by at least one prefix. A regular tree satisfying such equation is
guaranteed to exist and to be unique~\cite{Courcelle83}.

In order to describe a whole multiparty session at the level of types we
introduce the notion of \emph{session map}.

\begin{definition}[session map]
  \label{def:session_map}
  A \emph{session map} is a finite, partial map from roles to session types
  written $\set{\Map{\role_i} S_i}_{i\in I}$.  We let $M$ and $N$ range over
  session maps, we write $\dom{M}$ for the domain of $M$, we write $M \parop N$
  for the union of $M$ and $N$ when $\dom{M} \cap \dom{N} = \emptyset$, and we
  abbreviate the singleton map $\set{\Map\rolep{S}}$ as $\Map\rolep{S}$.
\end{definition}

We describe the evolution of a session at the level of types by means of a
\emph{labeled transition system} for session maps. Labels are generated by the
grammar below:
\[
  \textbf{Label}
  \qquad
  \ell ::= \tau \mid \action
  \qquad\qquad
  \textbf{Action}
  \qquad
  \actionA, \actionB ::= \Pol\terminated \mid \Map\rolep{\roleq\Pol\Tag} \mid \Map\rolep{\roleq\Pol S}
\]

The label $\tau$ represents either an internal action performed by a participant
independently of the others or a synchronization between two participants.
The labels of the form $\Pol\terminated$ describe the input/output of
termination signals, whereas the labels of the form $\Map\rolep\roleq\Pol\Tag$
and $\Map\rolep\roleq\Pol S$ represent the input/output of a tag $\Tag$ or of an
endpoint of type $S$. 

\begin{table}
  \caption{\label{tab:lts} Labeled transition system for session maps.}
  \begin{mathpar}
    \inferrule[l-end]{ }{
      \Map\role\End \xlred{\Pol\terminated} \Map\role\End
    }
    \and
    \inferrule[l-channel]{ }{
      \Map\rolep\roleq\Pol U.S \xlred{\Map\rolep\roleq\Pol U} \Map\rolep{S}
    }
    \and
    \inferrule[l-pick]{ }{
      \textstyle
      \Map\rolep{\Tags\roleq\Out\Tag_i.S_i}
      \lred\tau
      \Map\rolep{\roleq\Out\Tag_k.S_k}
    }
    ~k\in I
    \and
    \inferrule[l-tag]{ }{
      \textstyle
      \Map\rolep{\Tags\roleq\Pol\Tag_i.S_i}
      \xlred{\Map\rolep{\roleq\Pol\Tag_k}}
      \Map\rolep{S_k}
    }
    ~ k \in I
    \and
    \inferrule[l-tau]{
      M \lred\tau M'
    }{
      M \parop N \lred\tau M' \parop N
    }
    \and
    \inferrule[l-terminate]{
      M \lred{\In\terminated} M'
      \\
      N \lred{\Out\terminated} N'
    }{
      M \parop N \lred{\In\terminated} M' \parop N'
    }
    \and
    \inferrule[l-sync]{
      M \lred{\co\action} M'
      \\
      N \lred{\action} N'
    }{
      M \parop N \lred\tau M' \parop N'
    }
  \end{mathpar}
\end{table}

The labeled transition system is defined by the rules in \cref{tab:lts}, most of
which are straightforward. Rule \refrule{l-pick} models the fact that the
participant $\rolep$ may internally choose one particular tag $\Tag_k$ before
sending it to $\roleq$. The chosen tag is not negotiable with the receiver.
Rule \refrule{l-terminate} models termination of a session. A session terminates
when there is exactly one participant waiting for the termination signal and all
the others are sending it. This property follows from a straightforward
induction on the derivation of $M \lred{\In\terminated} N$ using
\refrule{l-terminate} and \refrule{l-end}.
%
%
The existence of a single participant waiting for the termination signal ensures
that there is a uniquely determined continuation process after the session has
been closed.
Finally, rule \refrule{l-sync} models the synchronization between two
participants performing complementary actions. The complement of an action
$\action$, denoted by $\co\action$, is the partial operation defined by the
equations
\[
  \co{\Map\rolep\roleq\Pol\Tag} \eqdef \Map\roleq\rolep{\co\Pol}\Tag
  \qquad
  \co{\Map\rolep\roleq\Pol S} \eqdef \Map\roleq\rolep{\co\Pol} S
\]
where $\co\Pol$ denotes the complement of the polarity $\Pol$. The complement of
actions of the form $\Pol\terminated$ is undefined, so rule \refrule{l-sync}
cannot be applied to terminated sessions.
Hereafter we write $\wred$ for the reflexive, transitive closure of $\lred\tau$
and $\wlred\action$ for the composition ${\wred}{\lred\action}$.

We call \emph{coherence} the property of multiparty sessions that we wish to
enforce with our type system, namely the fact that a session can always
terminate no matter how it evolves.
We formulate coherence directly on the transition system of session maps, in
line with the approach of Scalas and Yoshida~\cite{ScalasYoshida19} and without
introducing global types.

\begin{definition}
  \label{def:coherence}
  We say that $M$ is \emph{coherent}, notation $\coherent M$, if $M \wred N$
  implies $N \wlred{\In\terminated}$.
\end{definition}

The term ``coherence'' is borrowed from Carbone
\etal~\cite{CarboneLMSW16,CarboneMontesiSchurmannYoshida17}, although the
property is actually stronger than the one of Carbone \etal as it entails fair
termination of multiparty sessions through \cref{thm:fair_termination}.
In particular, if we consider the reduction system whose states are session maps
and whose reduction relation is $\lred\tau$, then $\coherent M$ implies $M$
fairly terminating.

\begin{example}[buyer-seller-carrier session map]
  \label{ex:bsc_coherent}
  Consider the session types
  \[
  \begin{array}{l@{~}c@{~}l}
    S_b & = & \seller\Out\tadd.\seller\Out\tadd.S_b + \seller\Out\tpay.\End[\Out]\\
    S_s & = & \buyer\In\tadd.S_s + \buyer\In\tpay.\carrier\Out\tship.\End[\Out]\\
    S_c & = & \seller\In\tship.\End[\In]
  \end{array} 
  \]
  which describe the behavior of the processes $\Buyer$, $\Seller$ and
  $\Carrier$ in \cref{ex:bsc}. The session map $\Map{\buyer}{S_b} \parop
  \Map{\seller}{S_s} \parop  \Map{\carrier}{S_c}$ is coherent. To see that,
  consider any interaction between the buyer and the seller. One of two cases
  applies: either the buyer has sent an even number of $\tadd$ messages to the
  seller, in which case it can send $\tpay$ and the session eventually
  terminates, or the buyer has sent an odd number of $\tadd$ messages to the
  seller, in which case it can send one more $\tadd$ message followed by a
  $\tpay$ message and once again the session eventually terminates.
  \eoe
\end{example} 

Coherence allows us to provide a semantic definition of \emph{fair subtyping},
the relation that defines the safe substitution principle for session endpoints
in our type system.

\begin{definition}[fair subtyping]
  \label{def:ssubt}
  We say that $S$ is a \emph{fair subtype} of\/ $T$, notation
  $S \ssubt T$, if\/ $M \parop \Map\role{S}$ coherent implies
  $M \parop \Map\role{T}$ coherent for every $M$ and $\role$.
\end{definition}

\Cref{def:ssubt} does not say much about the properties of fair subtyping except
for the fact that it is a coherence-preserving preorder. For this reason, we
devote \cref{sec:types_inference_system} to defining an alternative
characterization of fair subtyping that highlights its relationship with the
standard subtyping relation for session types~\cite{GayHole05}.


\subsection{Inference System for Fair Subtyping}
\label{sec:types_inference_system}

\begin{table}
  \caption{\label{tab:subt} Inference system for fair subtyping.}
  \begin{mathpar}
    \inferrule[f-end]{\mathstrut}{
      \End \subt[n] \End
    }
    \and
    \inferrule[f-channel]{
      S \subt[n] T
    }{
      \role\Pol U.S \subt[n] \role\Pol U.T
    }
    \and
    \inferrule[f-tag-in]{
      \forall i\in I: S_i \subt[n_i] T_i
      \\
      \forall i\in I: n_i \leq n
    }{
      \textstyle
      \Tags\rolep\In\Tag_i.S_i \subt[n] \Tags[i\in I \cup J]\rolep\In\Tag_i.T_i
    }
    \and
    \inferrule[f-tag-out-1]{
      \forall i\in I: S_i \subt[n_i] T_i
      \\
      \forall i\in I: n_i \leq n
    }{
      \textstyle
      \Tags[i\in I]\role\Out\Tag_i.S_i \subt[n] \Tags\role\Out\Tag_i.T_i
    }
    \and
    \inferrule[f-tag-out-2]{
      \forall i\in I: S_i \subt[n_i] T_i
      \\
      \exists i\in I: n_i < n
    }{
      \textstyle
      \Tags[i \in I \cup J]\role\Out\Tag_i.S_i \subt[n] \Tags\role\Out\Tag_i.T_i
    }
  \end{mathpar}
\end{table}

Consider the relation $\subt[n]$ coinductively defined by the inference system
in \cref{tab:subt}, where $n$ ranges over natural numbers. The characterization
of fair subtyping that we consider is the relation ${\subt} \eqdef
\bigcup_{n\in\NatSet} {\subt[n]}$.
The rules for deriving $S \subt[n] T$ are quite similar to those of the standard
subtyping relation for session types~\cite{GayHole05}: \refrule{f-end} states
reflexivity of subtyping on terminated session types; \refrule{f-channel}
relates higher-order session types with the same polarity and payload type;
\refrule{f-tag-in} is the usual covariant rule for the input of tags (the set
of tags in the larger session type includes those in the smaller one);
\refrule{f-tag-out-2} is the usual contravariant rule for the output of tags
(the set of tags in the smaller session type includes those in the larger one).
Overall, these rules entail a ``simulation'' between the behaviors described by
$S$ and $T$ whereby all inputs offered by $S$ are also offered by $T$ and all
outputs performed by $T$ are also performed by $S$.
The main differences between $\subt$ and the subtyping relation of Gay and
Hole~\cite{GayHole05} are the presence of an invariant rule for outputs
\refrule{f-tag-out-1} and the natural number $n$ annotating each subtyping
judgment $S \subt[n] T$. Intuitively, this number estimates how much $S$ and $T$
differ in terms of performed outputs. In all rules but \refrule{f-tag-out-2},
the annotation in the conclusion of the rule is just an upper bound of the
annotations found in the premises. In \refrule{f-tag-out-2}, where the sets of
output tags in related session types may differ, the annotation $n$ is required
to be a \emph{strict} upper bound for at least one of the premises. That is,
there must be at least one premise in which the annotation strictly decreases,
while no restriction is imposed on the others. Intuitively, this ensures the
existence of a tag shared by the two related session types whose corresponding
continuations are slightly less different. So, the annotation $n$ provides an
upper bound to the number of applications of \refrule{f-tag-out-2} along any
path (\ie any sequence of actions) shared by $S$ and $T$ that leads to
termination.  In the particular case when $n=0$, the rule \refrule{f-tag-out-2}
cannot be applied, so that $T$ may perform all the outputs also performed by
$S$.

\begin{example}
\label{ex:buyer_types}
  Consider the session type $S = \seller\Out\tadd.S +
  \seller\Out\tpay.\End[\Out]$, which describes the behavior of the buyer in
  \cref{eq:buyer} purchasing an arbitrary number of items, $T =
  \seller\Out\tadd.\seller\Out\tadd.T + \seller\Out\tpay.\End[\Out]$, which
  describes the behavior of the buyer in \cref{ex:bsc} always purchasing an even
  number of items, and $U = \seller\Out\tadd.U$, which describes the behavior of
  a buyer attempting to purchase an infinite number of items without ever paying
  the seller.
  We have $S \subt T$ and $S \not\subt U$. Indeed, we can derive
  \[
    \begin{prooftree}
      \[
        \[
          \mathstrut\smash\vdots
          \justifies
          S \subt[1] T
        \]
        \justifies
        S \subt[2] \seller\Out\tadd.T
        \using\refrule{f-tag-out-2}
      \]
      \[
        \justifies
        \End[\Out] \subt[0] \End[\Out]
        \using\refrule{f-end}
      \]
      \justifies
      S \subt[1] T
      \using\refrule{f-tag-out-2}
    \end{prooftree}
  \]
  but there is no derivation for $S \subt[n] U$ no matter how large $n$ is
  chosen.
  Note that there are infinitely many sequences of actions of $S$ that cannot be
  performed by both $T$ and $U$. In particular, $T$ cannot perform any sequence
  of actions consisting of an odd number of $\Tag[add]$ outputs followed by a
  $\Tag[pay]$ output, whereas $U$ cannot perform any sequence of $\Tag[add]$
  outputs followed by a $\Tag[pay]$ output. Nonetheless, there is a path shared
  by $S$ and $T$ that leads into a region of $S$ and $T$ in which no more
  differences are detectable. The annotations in the derivation tree measures
  the distance of each judgment from such region. In the case of $S$ and $U$,
  there is no shared path that leads to a region where no differences are
  detectable.
  \eoe
\end{example}

\newcommand{\player}{\role[player]}
\newcommand{\tplay}{\Tag[play]}
\newcommand{\tquit}{\Tag[quit]}
\newcommand{\twin}{\Tag[win]}
\newcommand{\tlose}{\Tag[lose]}

\begin{example}
  \label{ex:slot-machine}
  Consider the session types $S = \player\In\tplay.(\player\Out\twin.S +
  \player\Out\tlose.S) + \player\In\tquit.\End[\Out]$ and $T =
  \player\In\tplay.\player\Out\tlose.T + \player\In\tquit.\End[\Out]$ describing
  the behavior of two slot machines, an unbiased one in which the player may win
  at every play and a biased one in which the player never wins. If we try to
  build a derivation for $S \subt[n] T$ we obtain
  \[
    \begin{prooftree}
      \[
        \[
          \mathstrut\smash\vdots
          \justifies
          S \subt[n-1] T
        \]
        \justifies
        \player\Out\twin.S + \player\Out\tlose.S \subt[n] \player\Out\tlose.T
        \using\refrule{f-tag-out-1}
      \]
      \[
        \justifies
        \End[\Out] \subt[n] \End[\Out]
        \using\refrule{f-end}
      \]
      \justifies
      S \subt[n] T
      \using\refrule{f-tag-in}
    \end{prooftree}
  \]
  which would contain an infinite branch with strictly decreasing annotations.
  Therefore, we have $S \not\subt T$.
  In this case there exists a shared path leading into a region of $S$ and $T$
  in which no more differences are detectable between the two protocols, but
  this path starts from an input. The fact that $S$ is \emph{not} a fair subtype
  of $T$ has a semantic justification. Think of a $\player$ that deliberately
  insists on playing until it wins. This is possible when $\player$ interacts
  with the unbiased slot machine $S$ but not with the biased one $T$.
  \eoe
\end{example}

In the rest of this section we study the fundamental properties of $\subt$,
starting from the non-obvious fact that it is a preorder.

\begin{theorem}
	\label{thm:subt-preorder}
  	$\subt$ is a preorder.
\end{theorem}

While reflexivity of $\subt$ is trivial to prove (since \refrule{f-tag-out-2} is
never necessary, it suffices to only consider judgments with a $0$ annotation),
transitivity is surprisingly complex. The challenging part of proving that from
$S \subt[m] U$ and $U \subt[n] T$ we can derive $S \subt[k] T$ is to come up
with a feasible annotation $k$. As it turns out, such $k$ depends not only on
$m$ and $n$, but also on annotations found in different regions of the
derivation trees that prove $S \subt[m] U$ and $U \subt[n] T$. In particular,
the ``difference'' of $S$ and $T$ is not simply the ``maximum difference'' or
``the sum of the differences'' of $S$ and $U$ and of $U$ and $T$.
More in detail, we first show that we can always find a derivation of $S\subt[m]
U$ where the rank annotations of all judgements occurring in it are below some
$h \geq m$; then, the judgement $S\subt[k] T$ is provable for $k = m + (1+h)n$. 
For previous characterizations of fair
subtyping~\cite{Padovani13,Padovani16,CicconePadovani21,CicconePadovani22},
transitivity has been established indirectly by relating the inference system of
fair subtyping (\cref{tab:subt}) with its semantic definition
(\cref{def:ssubt}). For \cref{thm:subt-preorder} we are able to provide a direct
proof (\cf \cref{sec:subt-preorder}). 


Now we establish the connection between $\subt$ and $\ssubt$ (\cref{def:ssubt}).
First of all, we prove that $\subt$ is coherence-preserving just like $\ssubt$
is.

\begin{theorem}[soundness]
	\label{thm:subt-soundness}
 	If $S \subt T$ then $S \ssubt T$.
\end{theorem}

The proof of this result relies on a key property of $\subt$ not enjoyed by the
usual subtyping relation on session types~\cite{GayHole05}: when $S\subt T$ and
$M \parop \Map\rolep S$ is coherent, the session map $M \parop \Map\rolep T$ can
successfully terminate. The rank annotation on subtyping judgements is used to
set up an appropriate inductive argument for proving this property.


\cref{thm:subt-soundness} alone suffices to justify the adoption of $\subt$ as
fair subtyping relation, but we are interested in understanding to which extent
$\subt$ covers $\ssubt$. In this respect, it is quite easy to see that there
exist session types that are related by $\ssubt$ but not by $\subt$. For
example, consider $S = \role\Out\Tag[a].S$ and $T = \role\In\Tag[b].T$ and
observe that these two session types describe completely different protocols
(the output of infinitely many $\Tag[a]$'s in the case of $S$ and the input of
infinitely many $\Tag[b]$'s in the case of $T$). In particular, we have $S
\not\subt T$ and $T \not\subt S$ but also $S \ssubt T$ and $T \ssubt S$. That
is, $S$ and $T$ are \emph{unrelated} according to $\subt$ but they are
\emph{equivalent} according to $\ssubt$. This equivalence is justified by the
fact that there exists no coherent session map in which $S$ and $T$ could play
any role, because none of them can ever terminate.

This discussion hints at the possibility that, if we restrict the attention to
those session types that \emph{can} terminate, which are the interesting ones as
far as this work is concerned, then we can establish a tighter correspondence
between $\subt$ and $\ssubt$. We call such session types \emph{bounded}, because
they describe protocols for which termination is always within reach.

\begin{definition}[bounded session type]
	\label{def:bounded_type}
	We say that a session type is \emph{bounded} if all of its subtrees contain a
	$\End$ leaf.
\end{definition}

Note that a \emph{finite} session type is always bounded but not every bounded
session type is finite. If we consider the reduction system in which states are
session types and we have $S \red T$ if $T$ is an immediate subtree of $S$, then
$S$ is bounded if and only if $S$ is fairly terminating.
Now, for the family of bounded session types we can prove a \emph{relative
completeness} result for $\subt$ with respect to $\ssubt$.

\begin{theorem}[relative completeness]
	\label{thm:subt_completeness}
  	If $S$ is bounded and $S \ssubt T$ then $S \subt T$.
\end{theorem}

The proof of \cref{thm:subt_completeness} is done by contradiction. We show
that, for any bounded $S$, if $S\subt T$ does not hold then we can build a
session map $M$ called \emph{discriminator} such that $M\parop \Map\rolep S$ is
coherent and $M\parop \Map\rolep T$ is not, which contraddicts the hypothesis $S
\ssubt T$.
The boundedness of $S$ is necessary to make sure that it is always possible to
find a session map $N$ such that $N\parop \Map\rolep S$ is coherent.




\section{Type System} 
\label{sec:ts}

In this section we describe the type system for the calculus of multiparty
sessions of \cref{sect:calculus}. 
The typing judgments have the form $\wtp[n]\Ctx{P}$, meaning that the process
$P$ is well typed in the typing context $\Ctx$ and has rank $n$. As usual, the
\emph{typing context} is a map associating channels with session types and is
meant to contain an association for each name in $\fn{P}$. We write $u_1 : S_1,
\dots, u_n : S_n$ for the map with domain $\set{u_1,\dots,u_n}$ that associates
$u_i$ with $S_i$. Occationally we write $\seqof{u : S}$ for the same context,
when the number and the specific associations are unimportant. We also assume
that endpoints occurring in a typing context have different session names. That
is, $\ep{s}\rolep, \ep{s}\roleq \in \dom\Ctx$ implies $\rolep = \roleq$. 
This constraint makes sure that each well-typed process plays exactly one role
in each of the sessions in which it participates. It is also a common assumption
made in all multiparty session calculi.
We use $\Ctx$ and $\CtxD$ to range over typing contexts, we write $\EmptyCtx$
for the empty context and $\Ctx,\CtxD$ for the union of $\Ctx$ and $\CtxD$ when
they have disjoint domains and disjoint sets of session names.
The \emph{rank} $n$ in a typing judgment estimates the number of sessions that
$P$ has to create and the number of casts that $P$ has to perform in order to
terminate. The fact that the rank is finite suggests that so is the effort
required by $P$ to terminate.

\begin{table}
  \caption{Typing rules.}
  \label{tab:ts}
  \begin{mathpar}
    \inferrule[t-done]{\mathstrut}{
          \wtp[n]\EmptyCtx\pdone
        }
        \and
        \inferrule[t-call]{
          \wtp[n]{\seqof{u : S}}{P\subst{\seqof{u}}{\seqof{x}}}
        }{
          \wtp[n+m]{\seqof{u : S}}{\pinvk{A}{\seqof u}}
        }~ \tass{A}{\seqof{S}}{n}, \Definition{A}{\seqof x}{P}
        \and
        \inferrule[t-wait]
        {
          \wtp[n]{\Ctx}{P}
        }{
          \wtp[n]{\Ctx, u : \End[\In]}{\pwait{u}{P}}
        }
        \and
        \inferrule[t-close]{\mathstrut}
        {
          \wtp[n]{u : \End[\Out]}{\pclose{u}}
        }
        \and
        \inferrule[t-channel-in]{
        \wtp[n]{\Ctx, u : T, x : S}{P}
        }{
        \wtp[n]{\Ctx, u : \rolep\In{S}.T}{\pich{u}{\rolep}{x}{P}}
        }
        \and
        \inferrule[t-channel-out]{
        \wtp[n]{\Ctx, u : T}{P}
        }{
        \wtp[n]{\Ctx, u : \role\Out{S}.T , v : S}{\poch{u}{\rolep}{v}{P}}
        }
        \and
        \inferrule[t-tag]
        {
          \forall i\in I:
          \wtp[n]{\Ctx, u : S_i}{P_i}
        }{
          \textstyle
          \wtp[n]{
            \Ctx, u : \Tags\rolep\Pol \Tag_i.S_i
          }{
            \pbranch[i \in I]{u}{\role}{\Pol}{\Tag_i}{P_i}
          }
        }
        \and
        \inferrule[t-choice]{
          \wtp[n_1]\Ctx{P_1}
          \\
          \wtp[n_2]\Ctx{P_2}
        }{
          \wtp[n_k]\Ctx{P_1 \choice P_2}
        } ~ k \in \set{1,2}
        \and
        \inferrule[t-cast]{
          \wtp[n]{\Ctx, u : T}{P}
        }{
          \wtp[m+n]{\Ctx, u : S}{\pcast{u}{P}}
        }
        ~ S \subt[m] T
        \and
        \inferrule[t-par]{
          \forall i\in\set{1,\dots,h}:
          \wtp[n_i]{\Ctx_i, \ep{s}{\role_i} : S_i}{P_i}
        }{
          \wtp[1+n_1+\cdots+n_h]{
            \Ctx_1, \dots, \Ctx_h
          }{
            \pres{s}{P_1 \parop \cdots \parop P_h}
          }
        }
        ~ \coherent\set{\Map{\role_i} S_i}_{i=1..h}
        \and
        \infercorule[co-tag]{
          \wtp[n]{\Ctx, u : S_k}{P_k}
        }{
          \textstyle
          \wtp[n]{\Ctx, u : \Tags\rolep\Pol \Tag_i.S_i}{\pbranch[i \in I]{u}{\role}{\Pol}{\Tag_i}{P_i}}
        }~ k \in I
        \and
        \infercorule[co-choice]{
          \wtp[n]{\Ctx}{P_k}
        }{
          \wtp[n]{\Ctx}{P_1 \choice P_2}
        } ~ k \in \set{1,2} 
  \end{mathpar}
\end{table}

The typing rules are shown in \cref{tab:ts} as a \emph{generalized inference
system}~\cite{AnconaDagninoZucca17,Dagnino19,CicconeDagninoZucca21,Dagnino21th}
in which, roughly speaking, the singly-lined rules are interpreted coinductively
and the doubly-lined rules -- called \emph{corules} -- are interpreted
inductively. We will come back with a more detailed intuition later on
(\cref{def:well-typed-process}), although we will not provide a formal
definition of the interpretation of a generalized inference system in this
paper. The interested reader may refer to the cited literature for details.
We type check a program $\set{\Definition{A_i}{\seqof{x_i}}{P_i}}_{i\in I}$
under a global set of assignments $\set{\tass{A_i}{\seqof{S_i}}{n_i}}_{i\in I}$
associating each process name $A_i$ with a tuple of session types $\seqof{S_i}$,
one for each of the variables in $\seqof{x_i}$, and a rank $n_i$. The program is
well typed if $\wtp[n_i]{\seqof{x_i : S_i}}{P_i}$ is derivable for every $i\in
I$, establishing that the tuple $\seqof{S_i}$ corresponds to the way the
variables $\seqof{x_i}$ are used by $P_i$ and that $n_i$ is a feasible rank
annotation for $P_i$. We now describe the typing rules in detail.


The rule \refrule{t-done} states that the terminated process is well typed in
the empty context, to make sure that no unused channels are left behind. Note
that $\pdone$ can be given any rank, since it performs no casts and it creates
no new sessions. 
The rule \refrule{t-call} checks that a process invocation
$\pinvk{A}{\seqof{u}}$ is well typed by unfolding $A$ into the process
associated with $A$. The types associated with $\seqof{u}$ must match those of
the global assignment $\tass{A}{\seqof{S}}{n}$ and the rank of the process must
be no greater than that of the invocation. The potential mismatch between the
two ranks improves typeability in some corner cases.
The rules \refrule{t-wait} and \refrule{t-close} concern processes that exchange
termination signals. The channel being closed is consumed and, in the case of
\refrule{t-wait}, no longer available in the continuation $P$. Again,
$\pclose{u}$ can be typed with any rank whereas the rank of $\pwait{u}P$
coincides with that of $P$.
The rules \refrule{t-channel-in} and \refrule{t-channel-out} deal with the
exchange of channels in a quite standard way. Note that the actual type of the
exchanged channel is required to coincide with the expected one. In particular,
no covariance or contravariance of input and output respectively is allowed.
Relaxing the typing rule in this way would introduce implicit applications of
subtyping that may compromise fair termination~\cite{CicconePadovani22}. In our
type system, each application of subtyping must be explicitly accounted for as
we will see when discussing \refrule{t-cast}.
Rule \refrule{t-tag} deals with the exchange of tags. Channels that are not used for such
communication must be used in the same way in all branches, whereas the type of
the channel on which the message is exchanged changes accordingly. All branches
are required to have the same rank, which also corresponds to the rank of the
process. Unlike other presentations of this typing rule~\cite{GayHole05}, we
require the branches in the process to be matched exactly by those in the type.
Again, this is to avoid implicit application of subtyping, which might
jeopardize fair termination. 
The rule \refrule{t-choice} deals with non-deterministic choices and requires
both continuations to be well typed in the same typing context. The judgment in
the conclusion inherits the rank of one of the processes, typically the one with
minimum rank. As we will see in \cref{ex:pms_ts}, this makes it possible to
model finite-rank processes that may create an unbounded number of sessions or
that perform an unbounded number of casts.

The rule \refrule{t-cast} models the substitution principle induced by fair
subtyping: when $S \subt[m] T$, a channel of type $S$ can be used where a
channel of type $T$ is expected or, in dual fashion~\cite{Gay16}, a process
using $u$ according to $T$ can be used in place of a process using $u$ according
to $S$. To keep track of this cast, the rank in the conclusion is augmented by
the weight $m$ of the subtyping relation between $S$ and $T$.
Note that the typing rule guesses the target type of the cast.

Finally, the rule \refrule{t-par} deals with session creation and parallel
composition. This rule is inspired to the \emph{multiparty cut} rule found in
linear logic interpretations of multiparty session
types~\cite{CarboneLMSW16,CarboneMontesiSchurmannYoshida17} and provides a
straightforward way for enforcing deadlock freedom. Each process in the
composition must be well typed in a slice of the typing context augmented with
the endpoint corresponding to its role. The session map of the new session must
be coherent, implying that it fairly terminates. The rank of the composition is
one plus the aggregated rank of the composed processes, to account for the fact
that one more session has been created. Recall that coherence is a property
expressed on the LTS of session maps (\cref{def:coherence}) in line with the
approach of Scalas and Yoshida~\cite{ScalasYoshida19}.

The typing rules described so far are interpreted \emph{coinductively}. That is,
in order for a rank $n$ process $P$ to be well typed in $\Ctx$ there must be a
\emph{possibly infinite} derivation tree built with these rules and whose
conclusion is the judgment $\wtp[n]\Ctx{P}$. But in a generalized inference
system like the one we are defining, this is not enough to establish that $P$ is
well typed. In addition, it must be possible to find \emph{finite} derivation
trees for all of the judgments occurring in this possibly infinite derivation
tree using the discussed rules \emph{and possibly} the corules, which we are
about to describe. 
Since the additional derivation trees must be finite, all of their branches must
end up with an application of \refrule{t-done} or \refrule{t-close}, which are
the only axioms in \cref{tab:ts} corresponding to the only terminated processes
in \cref{tab:proc-syntax}. So, the purpose of these finite typing derivations is
to make sure that in every well-typed (sub-)process there exists a path that
leads to termination. On the one hand, this is a sensible condition to require
as our type system is meant to enforce fair process termination. On the other
hand, insisting that these finite derivations can be built using only the typing
rules discusses thus far is overly restrictive, for a process might have
\emph{one} path that leads to termination, but also alternative paths that lead
to (recursive) process invocations. In fact, all of the processes we have
discussed in \cref{ex:bsc,ex:2bsc,ex:pms} are structured like this. The two
corules \refrule{co-choice} and \refrule{co-tag} in \cref{tab:ts} establish
that, whenever a multi-branch process is dealt with, it suffices for \emph{one}
of the branches to lead to termination. A key detail to note in the case of
\refrule{co-choice} is that the rank of the non-deterministic choice coincides
with that of the branch that leads to termination. This makes sense recalling
that the rank associated with a process represents the overall effort required
for that process to terminate.

Let us recap the notion of well-typed process resulting from the typing rules of
\cref{tab:ts}.

\begin{definition}[well-typed process]
  \label{def:well-typed-process}
  We say that $P$ is \emph{well typed} in the context $\Ctx$ and has rank $n$ if
  (1) there exists an arbitrary (possibly infinite) derivation tree obtained
  using the (singly-lined) rules in \cref{tab:ts} and whose conclusion is
  $\wtp[n]\Ctx{P}$ and (2) for each judgment in such tree there is a finite
  derivation obtained using the rules and the (doubly-lined) corules.
\end{definition}

\begin{remark}
The term ``\emph{co}rule'' seems to suggest that the rule should be
\emph{co}inductively interpreted. As we have seen above
(\cref{def:well-typed-process}), corules are interpreted inductively. We have
chosen to stick with the terminology used in the works that introduced
generalized inference systems \cite{AnconaDagninoZucca17,Dagnino19}.
\eoe
\end{remark}


\begin{example}
  \label{ex:bsc-ts} 
  Let us show some typing derivations for fragments of \cref{ex:bsc} using the
  types $S_b$, $S_s$ and $S_c$ from \cref{ex:bsc_coherent}. Concerning $\Buyer$,
  we obtain the infinite derivation
  \[
    \begin{prooftree}
      \[
        \[
          \mathstrut\smash\vdots
          \justifies
          \wtp[0]{
            x : S_b
          }{
            \pinvk\Buyer{x}
          }
          \using\refrule{t-call}
        \]
        \justifies
        \wtp[0]{
          x : \seller\Out\tadd.S_b
        }{
          \act{x}\seller\oact\tadd.\pinvk\Buyer{x}
        }
        \using\refrule{t-tag}
      \]
      \[
        \justifies
        \wtp[0]{
          x : \End[\Out]
        }{
          \pclose{x}
        }
        \using\refrule{t-close}
      \]
      \justifies
      \wtp[0]{
        x : S_b
      }{
        \act{x}\seller\oact\set{
          \tadd.\act{x}\seller\oact\tadd.\pinvk\Buyer{x},
          \tpay.\pclose{x}
        }
      }
      \using\refrule{t-tag}
    \end{prooftree}
  \]
  and, for each judgment in it, it is easy to find a finite derivation possibly
  using \refrule{co-tag}. Concerning $\Main$ we obtain
  \[
    \begin{prooftree}
      \[
        \mathstrut\smash\vdots
        \justifies
        \wtp[0]{
          \ep{s}\buyer : S_b
        }{
          \pinvk\Buyer{\ep{s}\buyer}
        }
        \using\refrule{t-call}
      \]
      \[
        \smash\vdots
        \justifies
        \wtp[0]{
          \ep{s}\seller : S_s
        }{
          \pinvk\Seller{\ep{s}\seller}
        }
        \using\refrule{t-call}
      \]
      \vdots
      \justifies
      \wtp[1]{
        \EmptyCtx
      }{
        \pres\sn{
          \pinvk\Buyer{\ep{s}\buyer} \ppar \pinvk\Seller{\ep{s}\seller} \ppar \pinvk\Carrier{\ep{s}\carrier}
        }
      }
      \using\refrule{t-par}
    \end{prooftree}
  \]
  where the application of \refrule{t-par} is justified by the fact that
  $\Map\buyer{S_b} \parop \Map\seller{S_s} \parop \Map\carrier{S_c}$ is coherent
  (\cref{ex:bsc_coherent}).
  No participant creates new sessions or performs casts, so they all have zero
  rank. The rank of $\Main$ is 1 since it creates the session $s$.
  \eoe
\end{example}

We can prove a strong soundness result for our type system, stating that 
well-typed, closed processes can always successfully terminate no matter how they reduce.

\begin{theorem}[soundness]
	\label{thm:soundness}
	If $\wtp[n]\EmptyCtx P$ and $P \wred Q$, then $Q \wred\pcong \pdone$.
\end{theorem}

There are several valuable implications of \cref{thm:soundness} on a well-typed,
closed process $P$:
\begin{description}
  \item[Deadlock freedom.] If $Q$ cannot reduce any further, then it must be
  (structurally precongruent to) $\pdone$, namely there are no residual
  input/output actions.
  \item[Fair termination.] Under the fairness assumption,
  \cref{thm:fair_termination} assures that $P$ eventually reduces to $\pdone$.
  This also implies that every session created by $P$ eventually terminates.
  \item[Progress.] If $Q$ contains a sub-process with pending input/output
  actions, the fact that $Q$ may reduce to $\pdone$ means that these actions are
  eventually performed.
\end{description}

The proof of \cref{thm:soundness} is essentially composed of a standard subject
reduction result showing that typing is preserved by reductions and a proof that
every well-typed process other than $\pdone$ may always reduce in such a way
that a suitably defined \emph{well-founded measure} strictly decreases. The
measure is a lexicographically ordered pair of natural numbers with the
following meaning: the first component measures the number of sessions that must
be created and the total weight of casts that must be performed in order for the
process to terminate (this information is essentially the rank we associate with
typing judgments); the second component measures the overall effort required to
terminate every session that has already been created (these sessions are
identified by the fact that their restriction occurs unguarded in the process).
We account for this effort by measuring the shortest reduction that terminates a
coherent session map (\cref{def:coherence}). The reason why we need two
quantities in the measure is that in general every application of fair subtyping
may \emph{increase} the length of the shortest reduction that terminates a
coherent session map. So, when casts are performed the second component of the
measure may increase, but the first component reduces.
As a final remark, it should be noted that the overall measure associated with a
well-typed process \emph{may also increase}, for example if new sessions are
created (\cref{ex:pms}). However, one particular reduction that decreases the
measure is always guaranteed to exist.

We conclude this section discussing a few more examples that motivate the
features of the type system that are key for ensuring fair program termination.


\begin{example}
  \label{ex:corules}
  To see simple examples of processes whose ill/well typing crucially depends on
  the fact that we use a generalized inference system consider the definitions
  \[
    \Definition{A}{}{A}
    \qquad
    \Definition{B}{}{B \pchoice B}
    \qquad
    \Definition{C}{}{C \pchoice \pdone}
  \]
  which define a stuck process $A$, a diverging process $B$ and a fairly
  terminating process $C$ that admits an infinite reduction. For them we can
  find the infinite typing derivations below: 
  \[
    \begin{prooftree}
      \[
        \mathstrut\smash\vdots
        \justifies
        \wtp[0]{
          \EmptyCtx
        }{
          A
        }
        \using\refrule{t-call}
      \]
      \justifies
      \wtp[0]{
        \EmptyCtx
      }{
        A
      }
      \using\refrule{t-call}
    \end{prooftree}
    \qquad
    \begin{prooftree}
      \[
        \mathstrut\smash\vdots
        \qquad
        \[
          \mathstrut\smash\vdots
          \justifies
          \wtp[0]{
            \EmptyCtx
          }{
            B
          }
          \using\refrule{t-call}
        \]
        \justifies
        \wtp[0]{
          \EmptyCtx
        }{
          B \pchoice B
        }
        \using\refrule{t-choice}
      \]
      \justifies
      \wtp[0]{
        \EmptyCtx
      }{
        B
      }
      \using\refrule{t-call}
    \end{prooftree}
    \qquad
    \begin{prooftree}
      \[
        \[
          \mathstrut\smash\vdots
          \justifies
          \wtp[0]{
            \EmptyCtx
          }{
            C
          }
          \using\refrule{t-call}
        \]
        \[
          \justifies
          \wtp[0]{
            \EmptyCtx
          }{
            \pdone
          }
          \using\refrule{t-done}
        \]
        \justifies
        \wtp[0]{
          \EmptyCtx
        }{
          C \pchoice \pdone
        }
        \using\refrule{t-choice}
      \]
      \justifies
      \wtp[0]{
        \EmptyCtx
      }{
        C
      }
    \end{prooftree}
  \]

  However, only for $C$ it is possible to find a finite typing derivation using
  the corule~\refrule{co-choice}. So, $A$ and $B$ are ill typed, whereas $C$ is
  well typed. This is consistent with the fact that only $C$ can always reduce
  to the successfully terminated process $\pdone$.
  \eoe 
\end{example}

\begin{example}[infinitely ranked processes]
  \label{ex:rank_inf}
  The mere existence of a path that leads to termination ensured by the
  generalized interpretation of the typing rules in \cref{tab:ts} does not
  always guarantee that the process is actually able to terminate. An example
  where this is the case is shown by the process $A$ defined as
  \[
    \Definition{A}{}
      \pres{s}{
        \act{\ep{s}{\rolep}}{\roleq}\oact\set{
          \Tag[a].\pclose{\ep{s}{\rolep}},
          \Tag[b].\pwait{\ep{s}{\rolep}}{A}
        }
        \parop
        \act{\ep{s}{\roleq}}{\rolep}\iact\set{
          \Tag[a].\pwait{\ep{s}{\roleq}}{A},
          \Tag[b].\pclose{\ep{s}{\roleq}}
        }
      }
  \]
  which creates a session $s$ and splits as two parallel sub-processes connected
  by $s$. Each sub-process has a path that leads to termination but, because of
  the way they synchronize, when one sub-process terminates the other one
  restarts $A$. For $A$ it would be possible to build a finite typing derivation
  with the help of \refrule{co-tag}, but $A$ is ill typed because it cannot be
  assigned a finite rank, since it creates a new session at each recursive
  invocation.
  
  Further examples of infinitely ranked processes, including ones where the rank
  is affected by the presence of casts, are discussed by Ciccone and
  Padovani~\cite{CicconePadovani22} for binary sessions and can be easily
  reframed in our multiparty setting.
  %
  %
  \eoe
\end{example}


\section{Advanced Examples}
\label{sec:ts_ex}


\begin{example}
\label{ex:2bsc-ts}
In this example we show that the process $\Buyer_1$ playing the role $\rbuyer_1$
in the inner session of \cref{ex:2bsc} is well typed. For clarity, we recall its
definition here: 
\[
Buyer_1(x,y) \peq \act{y}{\rbuyer_2}\oact\{
	\begin{lines}
	  \tsplit.\act{y}{\rbuyer_2}\iact\{
		\begin{lines}
			\tyes.\pcast{x}
				\act{x}\rseller\oact\tok.
				\act{x}\rcarrier\iact\tbox.
				\pwait{x}
				\pwait{y}
				\pdone,
				\\
			\tno.\pinvk{Buyer_1}{x,y} \},
		\end{lines}
	  \\
	  \tgiveup.
		\pwait{y}
		\pcast{x}
		\act{x}\rseller\oact\tcancel.
		\pwait{x}
		\pdone \}
	\end{lines}
\]

We wish to build a typing derivation showing that $Buyer_1$ has rank $1$ and
uses $x$ and $y$ respectively according to $S$ and $T$, where $S =
\rseller\Out\tok.\rcarrier\In\tbox.\End[\In] + \rseller\Out\tcancel.\End[\In]$
and $T = \rbuyer_2\Out\tsplit.(\rbuyer_2\In\tyes.\End[\In] + \rbuyer_2\In\tno.T)
+ \rbuyer_2\Out\tgiveup.\End[\In]$.
As it has been noted previously, what makes this process interesting is that it
uses the endpoint $x$ differently depending on the messages it exchanges with
$\rbuyer_2$ on $y$. Since rule \refrule{t-tag} requires any endpoint other
than the one on which messages are exchanged to have the same type, the only way
$\Buyer_2$ can be declared well typed is by means of the casts that occur in its
body.
For the branch in which $\Buyer_1$ proposes to $\tsplit$ the payment we obtain
the following derivation tree:
\[
	\begin{prooftree}
		\[
			\[
				\[
					\[
						\[
							\[
								\justifies
								\wtp[0]\EmptyCtx\pdone
								\using\refrule{t-done}
							\]
							\justifies
							\wtp[0]{
								y : \End[\In]
							}{
								\pwait{y}\pdone
							}
							\using\refrule{t-wait}
						\]
						\justifies
						\wtp[0]{
							x : \End[\In],
							y : \End[\In]
						}{
							\pwait[\dots]{x}
						}
						\using\refrule{t-wait}
					\]
					\justifies
					\wtp[0]{
						x : \rcarrier\In\tbox.\End[\In],
						y : \End[\In]
					}{
						\act{x}\rcarrier\iact\tbox\dots
					}
					\using\refrule{t-tag}
				\]
				\justifies
				\wtp[0]{
					x : \rseller\Out\tok.\rcarrier\In\tbox.\End[\In],
					y : \End[\In]
				}{
					\act{x}\rseller\oact\tok\dots
				}
				\using\refrule{t-tag}
			\]
			\justifies
			\wtp[1]{
				x : S,
				y : \End[\In]
			}{
				\pcast{x}\cdots
			}
			\using\refrule{t-cast}
		\]
		\[
			\vdots
			\justifies
			\wtp[1]{
				x : S,
				y : T
			}{
				\pinvk{\Buyer_1}{x,y}
			}
			\using\refrule{t-call}
		\]
		\justifies
		\wtp[1]{
			x : S,
			y : \rbuyer_2\In\tyes.\End[\In] + \rbuyer_2\In\tno.T
		}{
			\act{y}{\rbuyer_2}\iact\set{\tyes\dots, \tno\dots}
		}
		\using\refrule{t-tag}
	\end{prooftree}
\]

Note how the application of \refrule{t-cast} is key to change the type of $x$ in
the branch where the proposed split is accepted by $\rbuyer_2$. In that branch,
$x$ is deterministically used to send an $\tok$ message and we leverage on the
fair subtyping relation $S \subt[1] \rseller\Out\tok.\rcarrier\In\tbox.\End[\In]$.

For the branch in which $\Buyer_1$ sends $\tgiveup$ we obtain the following
derivation tree:
\[
	\begin{prooftree}
		\[
			\[
				\[
					\[
						\justifies
						\wtp[0]\EmptyCtx{
							\pdone
						}
						\using\refrule{t-done}
					\]
					\justifies
					\wtp[0]{
						x : \End[\In]
					}{
						\pwait{x}\pdone
					}
					\using\refrule{t-wait}
				\]
				\justifies
				\wtp[0]{
					x : \rseller\Out\tcancel.\End[\In]
				}{
					\act{x}\rseller\oact\tcancel.
					\pwait{x}
					\pdone
				}
			\]
			\justifies
			\wtp[1]{
				x : S
			}{
				\pcast{x}
				\act{x}\rseller\oact\tcancel.
				\pwait{x}
				\pdone
			}
			\using\refrule{t-cast}
		\]
		\justifies
		\wtp[1]{
			x : S,
			y : \End[\In]
		}{
			\pwait{y}
			\pcast{x}
			\act{x}\rseller\oact\tcancel.
			\pwait{x}
			\pdone
		}
		\using\refrule{t-wait}
	\end{prooftree}
\]

Once again the cast is necessary to change the type of $x$, but this time
leveraging on the fair subtyping relation $S \subt[1]
\rseller\Out\tcancel.\End[\In]$.
These two derivations can then be combined to complete the proof that the body
of $\Buyer_1$ is well typed:
\[
	\begin{prooftree}
		\qquad
		\mathstrut\smash\vdots
		\qquad
		\qquad
		\qquad
		\smash\vdots
		\qquad
		\justifies
		\wtp[1]{
			x : S,
			y : T
		}{
			\act{y}{\rbuyer_2}\oact\set{\tsplit\dots, \tgiveup\dots}
		}
		\using\refrule{t-tag}
	\end{prooftree}
\]

Clearly, it is also necessary to find finite derivation trees for all of the
judgments shown above. This can be easily achieved using the corule
\refrule{co-tag}.
\eoe
\end{example}

\begin{example}
	\label{ex:non-det}
	\newcommand{\SlotMachine}{\mathit{Slot}}
	\renewcommand{\seller}{\rseller}
	Casts can be useful to reconcile the types of a channel that is used
	differently in different branches of a non-deterministic choice. For
	example, below is an alternative modeling of $\Buyer$ from \cref{ex:bsc}
	where we abbreviate $\role[seller]$ to $\seller$ for convenience:
	\[
		\Definition{B}{x}{
			\pcast{x}
			\act{x}\seller\oact\tadd.
			\act{x}\seller\oact\tadd.\pinvk{B}{x}
			\pchoice
			\pcast{x}
			\act{x}\seller\oact\tpay.
			\pclose{x}
		}
	\]

	Note that $x$ is used for sending two $\tadd$ messages in the left branch of
	the non-deterministic choice and for sending a single $\tpay$ message in the
	right branch. Given the session type $S = \seller\Out\tadd.S +
	\seller\Out\tpay.\End[\Out]$ and using the fair subtyping relations $S
	\subt[2] \seller\Out\tadd.\seller\Out\tadd.S$ and $S \subt[1]
	\seller\Out\tpay.\End[\Out]$ we can obtain the following typing derivation
	for the body of $B$:
	\[
		\begin{prooftree}
			\[
				\[
					\[
						\[
							\mathstrut\smash\vdots
							\justifies
							\wtp[1]{
								x : S
							}{
								\pinvk{B}{x}
							}
							\using\refrule{t-call}
						\]
						\justifies
						\wtp[1]{
							x : \seller\Out\tadd.S
						}{
							\act{x}\seller\oact\tadd.\pinvk{B}{x}
						}
						\using\refrule{t-tag}
					\]
					\justifies
					\wtp[1]{
						x : \seller\Out\tadd.\seller\Out\tadd.S
					}{
						\act{x}\seller\oact\tadd.
						\act{x}\seller\oact\tadd.\pinvk{B}{x}
					}
					\using\refrule{t-tag}
				\]
				\justifies
				\wtp[3]{
					x : S
				}{
					\pcast{x}
					\act{x}\seller\oact\tadd.
					\act{x}\seller\oact\tadd.\pinvk{B}{x}	
				}
				\using\refrule{t-cast}
			\]
			\!
			\[
				\[
					\[
						\justifies
						\wtp[0]{
							x : \End[\Out]
						}{
							\pclose{x}
						}
						\using\refrule{t-close}
					\]
					\justifies
					\wtp[0]{
						x : \seller\Out\tpay.\End[\Out]
					}{
						\act{x}\seller\oact\tpay.
						\pclose{x}
					}
					\using\refrule{t-tag}
				\]
				\justifies
				\wtp[1]{
					x : S
				}{
					\pcast{x}
					\act{x}\seller\oact\tpay.
					\pclose{x}
				}
				\using\refrule{t-cast}
			\]
			\justifies
			\wtp[1]{
				x : S
			}{
				\pcast{x}
				\act{x}\seller\oact\tadd.
				\act{x}\seller\oact\tadd.\pinvk{B}{x}
				\pchoice
				\pcast{x}
				\act{x}\seller\oact\tpay.
				\pclose{x}	
			}
			\using\refrule{t-choice}
		\end{prooftree}
	\]
	In general, the transformation $\pbranch[i=1..n]{u}{\role}\oact{\Tag_i}{P_i}
	\leadsto \pcast{u}\act{u}{\role}\oact{\Tag_1}.P_1 \pchoice \cdots \pchoice
	\pcast{u}\act{u}\role\oact{\Tag_n}.P_n$ does not always preserve typing, so
	it is not always possible to encode the output of tags using casts and
	non-deterministic choices. As an example, the definition
	\[
		\Definition\SlotMachine{x}{
			\act{x}\player\iact\set{
				\tplay.\act{x}\player\oact\set{
					\twin.\pinvk\SlotMachine{x},
					\tlose.\pinvk\SlotMachine{x}
				},
				\tquit.\pclose{x}
			}
		}
	\]
	implements the unbiased slot machine of \cref{ex:slot-machine}.
	It is easy to see that $\SlotMachine$ is well typed under the global type
	assignment $\tass\SlotMachine{T}{0}$ where $T =
	\player\In\tplay.(\player\Out\twin.T + \player\Out\tlose.T) +
	\player\In\tquit.\End[\Out]$. In particular, $\SlotMachine$ has rank $0$
	since it performs no casts and it creates no sessions. If we encode the tag
	output in $\SlotMachine$ using casts and non-deterministic choices we end up
	with the following process definition, which is ill typed because it cannot
	be given a finite rank:
	\[
		\Definition{\SlotMachine}{x}{
			\act{x}\player\iact\set{
				\tplay.(
					\pcast{x}
					\act{x}\player\oact\twin.
					\pinvk{\SlotMachine}{x}
					\pchoice
					\pcast{x}
					\act{x}\player\oact\tlose.
					\pinvk{\SlotMachine}{x}
				),
				\tquit.\pclose{x}
			}
		}
	\]

	The difference between this version of $\SlotMachine$ and the above
	definition of $B$ is that $\SlotMachine$ always recurs after a cast, so it
	is not obvious that finitely many casts suffice in order for $\SlotMachine$
	to terminate. 
	\eoe
\end{example}

\begin{example}
	\label{ex:pms_ts}
	Here we provide evidence that the process definitions in \cref{ex:pms} are
	well typed, even if they model processes that can open arbitrarily many
	sessions. In that example, the most interesting process definition is that
	of the worker $\Sort$, which is recursive and may create a new session. In
	contrast, $\Merge$ is finite and $\Main$ only refers to $\Sort$. We claim
	that these process definitions are well typed under the global type
	assignments
	\[
		\tass\Main{}{1}
		\qquad
		\tass\Sort{U}{0}
		\qquad
		\tass\Merge{T, V}{0}
	\]
	where $T = \rmaster\Out\tres.\End[\Out]$, $U = \rmaster\In\treq.T$ and $V =
	\rworker_1\Out\treq.\rworker_2\Out\treq.\rworker_1\In\tres.\rworker_2\In\tres.\End[\In]$.

	For the branch of $\Sort$ that creates a new session we obtain the
	derivation tree
	\[
		\begin{prooftree}
			\[
				\mathstrut\smash\vdots
				\justifies
				\wtp[0]{
					x : T,
					\ep{t}\rmaster : V
				}{
					\pinvk\Merge{x,\ep{t}\rmaster}
				}
				\using\refrule{t-call}
			\]
			\[
				\smash\vdots
				\justifies
				\wtp[0]{
					\ep{t}{\rworker_i} : U
				}{
					\pinvk\Sort{\ep{t}{\rworker_i}}
				}
				\using\refrule{t-call}, i=1,2
			\]
			\justifies
			\wtp[1]{
				x : T
			}{
				\pres{t}{
					\pinvk\Merge{x,\ep{t}\rmaster} \parop
					\pinvk\Sort{\ep{t}{\rworker_1}} \parop
					\pinvk\Sort{\ep{t}{\rworker_2}}
				}
			}
			\using\refrule{t-par}
		\end{prooftree}
	\]
	where the rank $1$ derives from the fact that the created session involves
	three zero-ranked participants.
	For the body of $\Sort$ we obtain the following derivation tree:
	\[
		\begin{prooftree}
			\[
				\[
					\smash\vdots
					\justifies
					\wtp[1]{
						x : T
					}{
						\pres{t}{
							\pinvk\Merge{x,\ep{t}\rmaster} \parop
							\pinvk\Sort{\ep{t}{\rworker_1}} \parop
							\cdots
						}
					}
					\using\refrule{t-par}
				\]
				\[
					\[
						\justifies
						\wtp[0]{
							x : \End[\Out]
						}{
							\pclose{x}
						}
						\using\refrule{t-close}
					\]
					\justifies
					\wtp[0]{
						x : T
					}{
						\act{x}\rmaster\oact\tres.\pclose{x}
					}
					\using\refrule{t-tag}
				\]
				\justifies
				\wtp[0]{
					x : T
				}{
					\pres{t}{
						\pinvk\Merge{x,\ep{t}\rmaster} \parop
						\pinvk\Sort{\ep{t}{\rworker_1}} \parop
						\cdots
					}
					\pchoice
					\act{x}\rmaster\oact\tres.\pclose{x}
				}
				\using\refrule{t-choice}
			\]
			\justifies
			\wtp[0]{
				x : U
			}{
				\act{x}\rmaster\iact\treq.(
					\pres{t}{
						\pinvk\Merge{x,\ep{t}\rmaster} \parop
						\pinvk\Sort{\ep{t}{\rworker_1}} \parop
						\cdots
					}
					\pchoice
					\act{x}\rmaster\oact\tres.\pclose{x}
				)
			}
			\using\refrule{t-tag}
		\end{prooftree}
	\]
	
	In the application of the rule \refrule{t-choice}, the rank of the whole
	choice coincides with that of the branch in which no new sessions are
	created. This way we account for the fact that, even though $\Sort$
	\emph{may} create a new session, it does not \emph{have to} do so in order
	to terminate.
	\eoe
\end{example}

\section{Related Work}
\label{sec:related-work}

\subparagraph{Fair termination of binary sessions.}
Our type system is both a refinement and an extension of the one presented by
Ciccone and Padovani~\cite{CicconePadovani22}, which ensures the fair
termination of \emph{binary} sessions. The main elements of the two type systems
are closely related, but there are some key differences.
In that work, the fairness assumption being made is \emph{strong fairness}
\cite{Francez86,AptFrancezKatz87,Kwiatkowska89,GlabbeekHofner19} which
guarantees fair termination of binary sessions \emph{at the level of types} but
not necessarily \emph{at the level of processes}.
The key difference between types and processes is that types generate
\emph{finite-state} reduction systems (because of their regularity) whereas
processes may generate \emph{infinite-state} reduction systems. While strong
fairness is known to be the strongest possible fairness assumption for
finite-state systems~\cite{GlabbeekHofnerHorne21}, it is not strong enough to
make the right-to-left direction of \cref{thm:fair_termination} hold for
infinite-state systems.
In fact, it can be shown that strong fairness and the fairness assumption we
make in this work (\cref{def:fair_run}) are unrelated for infinite-state
reduction systems, in the sense that there exist fair runs that are not strongly
fair and there exist strongly fair runs that are not fair runs. The fairness
assumption we make in this work is general enough so that it can be related to
both types (\cref{def:coherence}) and processes (\cref{thm:soundness}) through
\cref{thm:fair_termination}.
The main advantage of working with native multiparty sessions is that they
enable the natural modeling of interactions involving multiple participants in
possibly cyclic network topologies, like those in \cref{ex:2bsc,ex:pms}.
Another difference and contribution of our work compared to the one of Ciccone
and Padovani~\cite{CicconePadovani22} is that the definition of the fair
subtyping relation is simpler. In particular, the inference system we provide
(\cref{tab:subt}) does not make use of
corules~\cite{CicconePadovani21,CicconePadovani22} nor does it require auxiliary
predicates~\cite{Padovani13,Padovani16}.

\subparagraph{Liveness properties of multiparty sessions.}
The enforcement of liveness properties has always been a key aspect of session
type systems, although previous works have almost exclusively focused on
progress rather than on (fair) termination.
Scalas and Yoshida~\cite{ScalasYoshida19} define a general framework for
ensuring safety and liveness properties of multiparty sessions. In particular,
they define a hierarchy of three liveness predicates to characterize ``live''
sessions that enjoy progress. They also point out that the coarsest liveness
property in this hierarchy, which is the one more closely related to fair
termination, cannot be enforced by their type system. In part, this is due to
the fact that their type system relies on a standard subtyping relation for
session types \cite{GayHole05} instead of fair subtyping
\cite{Padovani13,Padovani16}. As we have seen in \cref{sec:ts}, even for
single-session programs the mere adoption of fair subtyping is not enough and it
is necessary to meet additional requirements (\cref{ex:corules,ex:rank_inf}).
The work of van Glabbeek \etal~\cite{GlabbeekHofnerHorne21} presents a type
system for multiparty sessions that ensures progress and is not only sound but
also complete. The fairness assumption they make -- called \emph{justness} -- is
substantially weaker than our own (\cref{def:fair_run}) and such that the unfair
runs are those in which some interactions between participants are
systematically discriminated in favor of other interactions involving a disjoint
set of independent participants. For this reason, their progress property is in
between the two more restrictive liveness predicates of Scalas and
Yoshida~\cite{ScalasYoshida19} and can only be guaranteed when it is independent
of the behavior of the other participants of the same session.
In the end, simple sessions like those described in \cref{ex:bsc,ex:2bsc,ex:pms}
fall outside the scope of these works as far as liveness properties are
concerned.

Another major difference between our work and the ones cited
above~\cite{ScalasYoshida19,GlabbeekHofnerHorne21} is that fair termination,
unlike progress, enables compositional reasoning and so we are able to enforce a
global liveness property (\cref{thm:soundness}) \emph{even in the presence of
multiple sessions} (see \cref{ex:2bsc,ex:pms}). Notable examples of multiparty
session type systems ensuring progress also in the presence of multiple
(possibly interleaved) sessions are provided by Padovani
\etal~\cite{PadovaniVasconcelosVieira14} and by Coppo
\etal~\cite{CoppoDezaniYoshidaPadovani16}. This is achieved by a rich type
structure that prevents mutual dependencies between different sessions. In any
case, these works do not address sessions in which progress may depend on
choices made by session participants.

\subparagraph{Termination of binary sessions.}
Termination is a liveness property that can be guaranteed when finite session
types are considered~\cite{PerezCairesPfenningToninho12}. As soon as infinite
session types are considered, many session type systems weaken the guaranteed
property to deadlock freedom.
Lindley and Morris~\cite{LindleyMorris16} define a type system for a functional
language with session primitives and recursive session types that is strongly
normalizing. That is, a well-typed program along with all the sessions it
creates is guaranteed to terminate. This strong result is due to the fact that
the type language is equipped with least and greatest fixed point operators that
are required to match each other by duality.
Termination is strictly stronger than fair termination. In particular, there
exist fairly terminating programs that are not terminating because they allow
reductions of unbounded length (see \cref{ex:bsc,ex:2bsc,ex:pms}).

\subparagraph{Liveness properties in the $\pi$-calculus.}
Kobayashi~\cite{Kobayashi02} defines a behavioral type system that guarantees
lock freedom in the $\pi$-calculus. Lock freedom is a liveness property akin to
progress for sessions, except that it applies to \emph{any} communication
channel (shared or private). 
Padovani~\cite{Padovani14} adapts and extends the type system of
Kobayashi~\cite{Kobayashi02} to enforce lock freedom in the \emph{linear
$\pi$-calculus}~\cite{KobayashiPierceTurner99}, into which binary sessions can
be encoded~\cite{DardhaGiachinoSangiorgi17}.
All of these works annotate types with numbers representing finite upper bounds
to the number of interactions needed to unblock a particular input/output
action. For this reason, none of our key examples (\cref{ex:bsc,ex:2bsc,ex:pms})
is in the scope of these analysis techniques.
Kobayashi and Sangiorgi~\cite{KobayashiSangiorgi10} show how to enforce lock
freedom by combining deadlock freedom and termination. Our work can be seen as a
generalization of this approach whereby we enforce lock freedom by combining
deadlock freedom (through a mostly conventional session type system) and
\emph{fair} termination. Since fair termination is coarser than termination, the
family of programs for which lock freedom can be proved is larger as well.

\subparagraph{Deadlock freedom.}
Our type system enforces deadlock freedom essentially thanks to the shape of the
rule \refrule{t-par} which is inspired to the cut rule of linear logic. This
rule has been applied to session type systems for binary
sessions~\cite{Wadler14,CairesPfenningToninho16,LindleyMorris16} and
subsequently extended to multiparty
sessions~\cite{CarboneLMSW16,CarboneMontesiSchurmannYoshida17}. In the latter
case, the rule -- dubbed \emph{multiparty cut} -- requires a coherence condition
among cut types establishing that the session types followed by the single
participants adhere to a so-called global type describing the multiparty session
as a whole. The rule \refrule{t-par} adopts with schema, except that the
coherence condition is stronger to entail fair session termination. 
The key principle of these formulations of the cut rule as a typing rule for
parallel processes is to impose a tree-like network topology, whereby two
parallel processes can share at most one channel. In the multiparty case, cyclic
network topologies can be modeled within each session (\cref{ex:pms}) since
coherence implies deadlock freedom.

Having a single construct that merges session restriction and parallel
composition allows for a simple formulation of the typing rules so that dealock
freedom is easily guaranteed. However, many session calculi separate these two
forms in line with the original presentation of the $\pi$-calculus. We think
that our type system can be easily reformulated to support distinct session
restriction and parallel composition by means of
hypersequents~\cite{KokkeMontesiPeressotti18,KokkeMontesiPeressotti19}.

A more liberal version of the cut rule, named multi-cut and inspired to
Gentzen's ``mix'' rule, is considered by Abramsky \etal~\cite{AbramskyGN96}
enabling processes to share more than one channel. In this setting,
deadlock freedom is lost but can be recovered by means of a richer type structure
that keeps track of the dependencies between different channels. This approach
has been pioneered by Kobayashi~\cite{Kobayashi02, Kobayashi06} for the
$\pi$-calculus and later on refined by Padovani~\cite{Padovani14}.
Other approaches to ensure deadlock freedom based on
\emph{dependency/connectivity graphs} that capture the network topology
implemented by processes have been studied by Carbone and
Debois~\cite{CarboneDebois10}, Kobayashi and Laneve~\cite{KobayashiLaneve17},
de'Liguoro and Padovani~\cite{deLiguoroP18}, and Jacobs
\etal~\cite{JacobsBalzerKrebbers22}.

\section{Concluding Remarks}
\label{sec:conclusion}

Sessions ought to terminate. Until recently this property has been granted only
for sessions whose duration is bounded. In this work we have presented the first
type system ensuring the \emph{fair termination} of multiparty sessions, that is
a termination property under the assumption that, if termination is always
reachable, then it is eventually achieved.
Fair termination is stronger than weak termination but substantially weaker than
strong normalization. In particular, fair termination does not rule out infinite
runs of well-typed processes as long as they purposefully eschew termination.
When fair termination is combined with the usual safety properties of sessions,
it entails a strong progress property whereby \emph{any} pending action is
eventually performed. Our type system is the first ensuring such strong progress
property for multiparty (and possibly multiple) sessions.


A cornerstone element of the type system is \emph{fair subtyping}, a
coherence-preserving refinement of the standard subtyping relation for session
types \cite{GayHole05}. 
In this work, we have also contributed a new characterization of fair subtyping
(\cref{tab:subt,thm:subt-soundness,thm:subt_completeness}) that is substantially
simpler than previous ones
\cite{Padovani13,Padovani16,CicconePadovani21,CicconePadovani22} since it does
not require auxiliary predicates nor the use of a generalized inference
system~\cite{AnconaDagninoZucca17,Dagnino19,CicconePadovani21,CicconePadovani22}.
Thanks to this new characterization we have been able to prove the transitivity
of fair subtyping (\cref{thm:subt-preorder}) without relying on its (relative)
completeness with respect to its semantic counterpart (\cref{def:ssubt}).
%


The decidability of fair subtyping and of type checking follow from analogous
results for binary sessions~\cite{CicconePadovani22}. The rank of processes can
be inferred using the same algorithm that works for the binary
case~\cite[auxiliary material]{CicconePadovani22}. Considering that fair
subtyping for multiparty session types coincides with fair subtyping for binary
session types except for the presence of roles, it would be easy to adapt the
type checking tool \textsf{FairCheck}~\cite{FairCheck} to the process language
we consider in this paper. The most relevant difference would be the algorithm
for deciding the coherence of a session map, which is somewhat more complex than
that for the compatibility between two session types.
As for the binary setting, to which extent the type system is amenable to full
type reconstruction is yet to be established. In particular, a hypothetical type
inference algorithm would have to be able to solve fair subtyping inequations
and this problem has not been investigated yet. 
Another open question that may have a relevant practical impact is whether the
type system remains \emph{sound} in a setting where communications are
\emph{asynchronous}. We expect the answer to be positive, as is the case for
other synchronous multiparty session types systems~\cite{ScalasYoshida19}, but
we have not worked out the details yet.

In this paper we have focused on the theoretical aspects of fairly terminating
multiparty sessions. A natural development of this work is its application to a
real programming environment. We envision two approaches that can be followed to
this aim. A bottom-up approach may apply our static analysis technique to a
program (in our process calculus) that is extracted from actual code and that
captures the code's communication semantics. We expect that suitable annotations
may be necessary to identify those branching parts of the code that represent
non-deterministic choices in the program. Most typically, these branches will
correspond to finite loops or to queries made to the human user of the program
that have several different continuations. A top-down approach may provide
programmers with a \emph{generative tool} that, starting from a global
specification in the form of a \emph{global type}~\cite{HondaYoshidaCarbone16},
produces template code that is ``well-typed by design'' and that the programmer
subsequently instantiates to a specific application.
Scribble~\cite{YoshidaHuNeykovaNg13,AnconaEtAl16} is an example of such a tool.
Interestingly, the usual notion of global type projectability is not sufficient
to entail that the session map resulting from a projection is coherent. However,
coherence would be guaranteed by requiring that the projected global type is
fairly terminating.

Finally, we plan to investigate the adaptation of the type system for ensuring
the fair termination in the popular actor-based model. This is a drastically
different setting in which the order of messages is not as controllable as in
the case of sessions. As a consequence, type based analyses require radically
different formalisms such as mailbox types~\cite{deLiguoroP18}, for which the
study of fair subtyping and of type systems enforcing fair termination is
unexplored.


\bibliography{main}

\clearpage
\appendix


\section{Supplement to Section \ref{sec:fair-termination}}

\begin{proof}[Proof of \cref{lem:feasibility}]
    Let $D$ be the last state of $\run$. We distinguish two possibilities: if $D$
    is weakly terminating, then there exists a finite maximal run $D\run'$ of $D$;
    if $D$ is diverging, then there exists an infinite run $D\run'$ of $D$ such
    that no state in $\run'$ is weakly terminating. In both cases we conclude by
    noting that $\run\run'$ is a maximal fair run. 
\end{proof}

\begin{proof}[Proof of \cref{thm:fair_termination}]
  ($\Rightarrow$) Let $D$ be a state reachable from $C$. That is, there exists a
  finite run $\run$ of $C$ ending with $D$. By \cref{lem:feasibility} we deduce
  that this run can be extended to a maximal fair one $\run\run'$. From the
  hypothesis that $C$ is fairly terminating we deduce that $\run\run'$ is
  finite. Hence, $D$ is weakly terminating.

  ($\Leftarrow$) Let $C_0C_1\dots$ be an infinite fair run of $C$.  Using the
  hypothesis we deduce that each $C_i$ is weakly terminating, which is absurd.
  Hence, either there are no maximal fair runs or 
  every maximal fair run of $C$ is finite, but the first case is not possible by \cref{lem:feasibility}, 
  thus $C$ is fairly terminating. 
\end{proof}


\section{Supplement to Section \ref{sec:types_inference_system}}
\label{sec:fs-proof}

In this section we prove that $\subt$ (\cref{tab:subt}) is a preorder and also
that it is sound and (relatively) complete with respect to $\ssubt$
(\cref{def:ssubt}). To do so it is convenient to introduce some more notation
concerning the lebeled transition system of sessions (\cref{tab:lts}). We write
$\xwlred{\action_1\cdots\action_n}$ for the composition
$\wlred{\action_1}\cdots\wlred{\action_n}$;
we let $\actionsA$ and $\actionsB$ range over strings of actions; we write
$\varepsilon$ for the empty string and $|\actions|$ for the length of
$\actions$;
we write $M \lred\ell$ if there exists $N$ such that $M \lred\ell N$ and $M
\nlred\ell$ if not $M \lred\ell$; similarly for $M \wlred\actions$ and $M
\nwlred\actions$.


\subsection{Proof of Theorem \ref{thm:subt-preorder}}
\label{sec:subt-preorder}

\begin{lemma}
  \label{lem:bounded_derivation}
  Let $S \subt[n]^m T$ if and only if $S \subt[n] T$ is the
  conclusion of a derivation in which every rank annotation is at
  most $m$. Then $S \subt[n] T$ if and only $S \subt[n]^m T$ for
  some $m$.
\end{lemma}
\begin{proof}
  \newcommand{\msubt}[1]{\subt[#1{\leq}]}
  The ``if'' part is obvious. Concerning the ``only if'' part, it
  suffices to show that each judgment in the set
  \[
    \srel \eqdef \set{U \subt[m] V \mid U \subt[m] V \wedge \nexists
      n < m: U \subt[n] V}
  \]
  is derivable from premises that are also in $\srel$. This is
  enough to prove $S \subt[n]^m T$ from $S \subt[n] T$, because in
  $\srel$ there is at most one judgment $U \subt[n] V$ for each pair
  of session types $U$ and $V$ and, by regularity of $U$ and $V$,
  the derivation of $U \subt[n] V$ obtained using judgments in
  $\srel$ contains finitely many annotations, which must have a
  maximum.

  Suppose $U \subt[m] V \in \srel$. Then $U \subt[m] V$ is
  derivable. We reason by cases on the last rule applied to derive
  this judgment.

  \proofrule{f-end}
  Then $U = V = \End$ and there is nothing left to prove since
  \refrule{f-end} has no premises.

  \proofrule{f-tag-in}
  Then $U = \Tags\role\In\Tag_i.U_i$ and
  $V = \Tags[i\in J]\role\In\Tag_i.V_i$ and $I \subseteq J$ and
  $U_i \subt[n_i] V_i$ and $n_i \leq m$ for every $i\in I$.
  By definition of $\srel$ we have that, for every $i\in I$, there
  exists $m_i \leq n_i$ such that $U \subt[m_i] V_i \in \srel$.
  Then $U \subt[m] V$ is derivable by \refrule{f-tag-in} using
  premises in $\srel$.

  \proofrule{f-tag-out-1}
  Analogous to the previous case.

  \proofrule{f-tag-out-2}
  Then $U = \Tags\role\Out\Tag_i.U_i$ and
  $V = \Tags[i\in J]\role\Out\Tag_i.V_i$ and $J \subseteq I$ and
  $U_i \subt[n_i] V_i$ for every $i\in J$ and $n_k < m$ for some
  $k\in I$.
  By definition of $\srel$ we have that, for every $i\in J$, there
  exists $m_i \leq n_i$ such that $U \subt[m_i] V_i \in \srel$.
  In particular, $m_k \leq n_k < m$.
  Then $U \subt[m] V$ is derivable by \refrule{f-tag-out-2} using
  premises in $\srel$.
\end{proof}

\begin{proof}[Proof of \cref{thm:subt-preorder}]
  The proof that $\subt$ is reflexive is trivial, since
  $S \subt[n] S$ is derivable for every $n$. Concerning
  transitivity, by \cref{lem:bounded_derivation} it suffices to show
  that each judgment in the set
  \[
    \srel \eqdef \set{
      S \subt[n_1 + (1 + m)n_2] T \mid S \subt[n_1]^m U \wedge U \subt[n_2] T
    }
  \]
  is derivable using the rules in \cref{tab:subt} from premises that
  are also in $\srel$.
  Suppose $S \subt[n] T \in \srel$. Then there exist $U$, $n_1$, $m$
  and $n_2$ such that $S \subt[n_1]^m U$ and $U \subt[n_2] T$ and
  $n = n_1 + (1 + m)n_2$.
  We reason by cases on the last rules applied to derive
  $S \subt[n_1] U$ and $U \subt[n_2] T$.

  \proofrule{f-end}
  Then $S = U = T = \End$ hence $S \subt[n] T$ is derivable by \refrule{f-end}.

  \proofrule{f-tag-in}
  Then $S = \Tags\role\In\Tag_i.S_i$ and $U = \Tags[i\in J]\role\In\Tag_i.U_i$
  and $T = \Tags[i\in K]\role\In\Tag_i.T_i$ and $I \subseteq J \subseteq K$ and
  $S_i \subt[n_{1i}]^m U_i$ and $n_{1i} \leq n_1$ for every $i\in I$ and $U_i
  \subt[n_{2i}] T_i$ and $n_{2i} \leq n_2$ for every $i\in J$.
  By definition of $\srel$ we have that $S_i \subt[n_{1i} + (1 + m)n_{2i}] T_i
  \in \srel$ for every $i\in I$.
  Observe that $n_{1i} + (1 + m)n_{2i} \leq n_1 + (1 + m)n_2 = n$ for every
  $i\in I$ hence $S \subt[n] T$ is derivable by \refrule{f-tag-in}.

  \proofrule{f-tag-out-1}
  Then $S = \Tags\role\Out\Tag_i.S_i$ and
  $U = \Tags\role\Out\Tag_i.U_i$ and $T = \Tags\role\Out\Tag_i.T_i$
  and $S_i \subt[n_{1i}]^m U_i$ and $n_{1i} \leq n_1$ and
  $U_i \subt[n_{2i}] T_i$ and $n_{2i} \leq n_2$ for every $i\in I$.
  By definition of $\srel$ we have that
  $S_i \subt[n_{1i} + (1 + m)n_{2i}] T_i \in \srel$ for every
  $i\in I$.
  Observe that $n_{1i} + (1 + m)n_{2i} \leq n_1 + (1 + m)n_2 = n$
  for every $i\in I$ hence $S \subt[n] T$ is derivable by
  \refrule{f-tag-out-1}.

  \proofcase{Case \refrule{f-tag-out-1} and \refrule{f-tag-out-2}}
  Then $S = \Tags\role\Out\Tag_i.S_i$ and
  $U = \Tags\role\Out\Tag_i.U_i$ and
  $T = \Tags[i\in J]\role\Out\Tag_i.T_i$ and $J \subseteq I$ and
  $S_i \subt[n_{1i}]^m U_i$ and $n_{1i} \leq n_1$ for every $i\in I$
  and $U_i \subt[n_{2i}] T_i$ for every $i\in J$ and $n_{2k} < n_2$
  for some $k\in J$.
  By definition of $\srel$ we have that
  $S_i \subt[n_{1i} + (1 + m)n_{2i}] T_i \in \srel$ for every
  $i\in J$.
  Observe that $n_{1k} + (1 + m)n_{2k} < n_1 + (1 + m)n_2 = n$ hence
  $S \subt[n] T$ is derivable by \refrule{f-tag-out-2}.

  \proofcase{Case \refrule{f-tag-out-2} and \refrule{f-tag-out-1}}
  Then $S = \Tags\role\Out\Tag_i.S_i$ and
  $U = \Tags[i\in J]\role\Out\Tag_i.U_i$ and
  $T = \Tags[i\in J]\role\Out\Tag_i.T_i$ and $J \subseteq I$ and
  $S_i \subt[n_{1i}]^m U_i$ for every $i\in J$ and $n_{1k} < n_1$
  for some $k\in J$ and $U_i \subt[n_{2i}] T_i$ for every $i\in J$
  and $n_{2i} \leq n_2$ for every $i\in J$.
  By definition of $\srel$ we have that
  $S_i \subt[n_{1i} + (1 + m)n_{2i}] T_i \in \srel$ for every
  $i\in J$.
  Observe that $n_{1k} + (1 + m)n_{2k} < n_1 + (1 + m)n_2 = n$ hence
  $S \subt[n] T$ is derivable by \refrule{f-tag-out-2}.

  \proofrule{f-tag-out-2}
  Then $S = \Tags\role\Out\Tag_i.S_i$,
  $U = \Tags[i\in J]\role\Out\Tag_i.U_i$ and
  $T = \Tags[i\in K]\role\Out\Tag_i.T_i$ and
  $K \subseteq J \subseteq I$ and $S_i \subt[n_{1i}]^m U_i$ for
  every $i\in J$ and $n_{1j} < n_1$ for some $j\in J$ and
  $U_i \subt[n_{2i}] T_i$ for every $i\in K$ and $n_{2k} < n_2$ for
  some $k\in K$.
  By definition of $\srel$ we have that
  $S_i \subt[n_{1i} + (1 + m)n_{2i}] T_i \in \srel$ for every
  $i\in K$.
  Observe that
  \[
    \begin{array}{rcll}
      n_{1k} + (1 + m)n_{2k} & \leq & m + (1 + m)n_{2k} & \text{since $n_{1k} \leq m$}
      \\
      & < & 1 + m + (1 + m)n_{2k}
      \\
      & = & (1 + m)(1 + n_{2k})
      \\
      & \leq & (1 + m)n_2 & \text{since $n_{2k} < n_2$}
      \\
      & < & n_1 + (1 + m)n_2 & \text{since $n_{1j} < n_1$}
    \end{array}
  \]
  hence $S \subt[n] T$ is derivable by \refrule{f-tag-out-2}.
\end{proof}

\subsection{Proof of Theorem \ref{thm:subt-soundness}}

We start with an auxiliary result formalizing the simulation entailed by the
relation $S \subt T$.

\begin{lemma}
	\label{lem:subt_sim}
  	If\/ $S \subt T$ and $M \parop \Map\rolep{S}$ is coherent and $M \parop
  	\Map\rolep{T} \wred N \parop \Map\rolep{T'}$, then $M \parop \Map\rolep{S}
  	\wred N \parop \Map\rolep{S'}$ for some $S' \subt T'$.
\end{lemma}
\begin{proof}
  We prove the result for a single reduction
  $M \parop \Map\rolep{T} \lred\tau N \parop \Map\rolep{T'}$. The
  general statement then follows by a straightforward induction on
  the length of the reduction
  $M \parop \Map\rolep{T} \wred N \parop \Map\rolep{T'}$ using the
  fact that coherence is preserved by reductions.

  \proofcase{Case $M \lred\tau N$}
  Then $T' = T$ and we conclude by taking $S' \eqdef S$.

  \proofcase{Case $\Map\rolep{T} \lred\tau \Map\rolep{T''}$}
  Then
  $\Map\rolep{T} = \Map\rolep{\Tags\roleq\Out\Tag_i.T_i} \lred\tau
  \Map\rolep{\Label\roleq\Out{\Tag_k}.T_k} = \Map\rolep{T'}$ for
  some $k\in I$.
  From the hypothesis $S \subt T$ we deduce
  $S = \Tags[i\in J]\roleq\Out\Tag_i.S_i$ where $I \subseteq J$ and
  $S_i \subt T_i$ for every $i\in I$.
  Now we have
  $M \parop \Map\rolep{S} \lred\tau M \parop
  \Map\rolep{\roleq\Out\Tag_k.S_k}$ and also
  $\roleq\Out\Tag_k.S_k \subt T'$.
  We conclude by taking $S' \eqdef \roleq\Out\Tag_k.S_k$.

  \proofcase{Case $M \xlred{\Map\roleq{\rolep\Out\Tag}} N$ and
    $\Map\rolep{T} \xlred{\Map\rolep{\roleq\In\Tag}}
    \Map\rolep{T'}$}
  Then $T = \Tags\roleq\In\Tag_i.T_i$ and $\Tag = \Tag_k$ and
  $T' = T_k$ for some $k\in I$.
  From the hypothesis $S \subt T$ we deduce
  $S = \Tags[i\in J]\roleq\In\Tag_i.S_i$ and $J \subseteq I$ and
  $S_i \subt T_i$ for every $i\in J$.
  From the hypothesis $M \parop \Map\rolep{S}$ coherent we deduce
  $k\in J$ or else the participant $\rolep$ would not be able to
  receive the $\Tag$ tag.
  We conclude by taking $S' \eqdef S_k$.

  \proofcase{Case $M \xlred{\Map\roleq{\rolep\In\Tag}} N$ and
    $\Map\rolep{T} \xlred{\Map\rolep{\roleq\Out\Tag}}
    \Map\rolep{T'}$}
  Then $T = \Tags\roleq\Out\Tag_i.T_i$ and $\Tag = \Tag_k$ and
  $T' = T_k$ for some $k\in I$.
  From the hypothesis $S \subt T$ we deduce
  $S = \Tags[i\in J]\roleq\Out\Tag_i.S_i$ and $I \subseteq J$ and
  $S_i \subt T_i$ for every $i\in I$.
  We conclude by taking $S' \eqdef S_k$.

  \proofcase{Case $M \xlred{\Map\roleq\rolep\Out U} N$ and
    $\Map\rolep{T} \xlred{\Map\rolep\roleq\In U} \Map\rolep{T'}$}
  Then $T = \roleq\In U.T'$.
  From the hypothesis $S \subt T$ we deduce $S = \roleq\In U.S'$
  and $S' \subt T'$.
  We conclude by observing that
  $M \parop \Map\rolep{S} \lred\tau N \parop \Map\rolep{S'}$.

  \proofcase{Case $M \xlred{\Map\roleq\rolep\In U} N$ and
    $\Map\rolep{T} \xlred{\Map\rolep\roleq\Out U} \Map\rolep{T'}$}
  Then $T = \roleq\Out U.T'$.
  From the hypothesis $S \subt T$ we deduce $S = \roleq\Out U.S'$
  and $S' \subt T'$.
  We conclude by observing that
  $M \parop \Map\rolep{S} \lred\tau N \parop \Map\rolep{S'}$.
\end{proof}

Next we show that $S \subt T$ preserves the termination of any session map that
completes $S$ into a coherent one.

\begin{lemma}
	\label{lem:subt_term}
  	If\/ $S \subt[n] T$ and $M \parop \Map\rolep{S}$ is coherent, then $M \parop
  	\Map\rolep{T} \wlred{\In\terminated}$.
\end{lemma}
\begin{proof}
  By induction on the lexicographically ordered tuple
  $(n, |\actions|)$ where $\actions$ is any string of actions such
  that $M \xwlred{\co\actions\co\Pol\terminated}$ and
  $\Map\role{S} \xwlred{\actions\Pol\terminated}$.  We know that at
  least one such $\actions$ least does exist from the hypothesis
  $M \parop \Map\rolep{S}$ is coherent.
  We now reason by cases on the shape of $\actions$.

  \proofcase{Case $\actions = \varepsilon$}
  Then $S = \End$.
  From the hypothesis $S \subt[n] T$ and \refrule{f-end} we deduce
  $T = \End$ and we conclude
  $M \parop \Map\rolep{T} \wlred{\In\terminated}$.

  \proofcase{Case $\actions = \Map\rolep\roleq\In\Tag\actionsB$}
  Then $S = \Tags\roleq\In\Tag_i.S_i$ and $\Tag = \Tag_k$ for some
  $k\in I$.
  From the hypothesis $S \subt[n] T$ and \refrule{f-tag-in} we
  deduce $T = \Tags[i\in J]\roleq\In\Tag_i.T_i$ and $I \subseteq J$
  and $S_i \subt[n_i] T_i$ and $n_i \leq n$ for every $i\in I$.
  We conclude using the induction hypothesis.

  \proofcase{Case $\actions = \Map\rolep\roleq\Out\Tag\actionsB$}
  Then $S = \Tags\roleq\Out\Tag_i.S_i$ and $\Tag = \Tag_k$ for some
  $k\in I$. We distinguish two sub-cases, according to the last rule
  used in the derivation of $S \subt[n] T$.
  If the last rule was \refrule{f-tag-out-1}, then
  $T = \Tags[i\in I]\roleq\Out\Tag_i.T_i$ and $S_i \subt[n_i] T_i$
  and $n_i \leq n$ for every $i\in I$.
  In particular, $S_k \subt[n_k] T_k$ and $n_k \leq n$ and we
  conclude using the induction hypothesis.
  If the last rule was \refrule{f-tag-out-2}, then
  $T = \Tags[i\in J]\roleq\Out\Tag_i.T_i$ with $J \subseteq I$ and
  $S_i \subt[n_i] T_i$ for every $i\in J$ and $n_j < n$ for some
  $j\in J$.
  In particular, we have $S_j \subt[n_j] T_j$ and we conclude using
  the induction hypothesis.
\end{proof}

\begin{proof}[Proof of \cref{thm:subt-soundness}]
  Consider a run
  $M \parop \Map\rolep{T} \wred N \parop \Map\rolep{T'}$.
  From \cref{lem:subt_sim} we deduce that there exists $S' \subt T'$
  such that $M \parop \Map\rolep{S} \wred N \parop \Map\rolep{S'}$.
  From the hypothesis that $M \parop \Map\rolep{S}$ is coherent we
  deduce $N \parop \Map\rolep{S'}$ is also coherent.
  From \cref{lem:subt_term} we conclude
  $N \parop \Map\rolep{T'} \wlred{\In\terminated}$.
\end{proof}


\subsection{Proof of Theorem~\ref{thm:subt_completeness}}

\begin{table}
  \caption{\label{tab:usubt} Inference system for unfair subtyping.}
  \begin{mathpar}
    \inferrule[u-end]{\mathstrut}{
      \End \usubt \End
    }
    \and
    \inferrule[u-channel]{
      S \usubt T
    }{
      \role\Pol U.S \usubt \role\Pol U.T
    }
    \and
    \inferrule[u-tag-in]{
      \forall i\in I: S_i \usubt T_i
    }{
      \textstyle
      \Tags\rolep\In\Tag_i.S_i \usubt \Tags[i\in I \cup J]\rolep\In\Tag_i.T_i
    }
    \and
    \inferrule[u-tag-out]{
      \forall i\in I: S_i \usubt T_i
    }{
      \textstyle
      \Tags[i \in I \cup J]\role\Out\Tag_i.S_i \usubt \Tags\role\Out\Tag_i.T_i
    }
  \end{mathpar}
\end{table}

The proof of \cref{thm:subt_completeness} is by contradiction, showing that from
the hypothesis $S \not\subt T$ we are able to find a session map $M$ that is
coherent when completed by $S$ but not when it is completed by $T$.
For this proof we need some auxiliary notions and notation.
First of all, we consider \emph{unfair subtyping} as the subtyping relation
$\usubt$ coinductively defined by the rules in \cref{tab:usubt}. It is
straightforward to see that ${\subt} \subseteq {\usubt}$.
Then, we introduce some convenient notation for building session maps. To do
this, we assume the existence of an arbitrary total order $<$ on the set of
roles.
Now, given a finite set of roles $\set{\role_1,\dots,\role_n}$ where $\role_1 <
\cdots < \role_n$, we write $\set{\role_1,\dots,\role_n}\Out\Tag.S$ for the
session type $\role_1\Out\Tag\cdots\role_n\Out\Tag.S$.
Given a finite family $\set{M_i}_{i\in I}$ of session maps all having the same
domain $\set\roleq \subseteq D \subseteq \RoleSet\setminus\set\rolep$, we write
$\Map\roleq{\Tags{\rolep\Pol\Tag_i.M_i}}$ for the session map $M$ having domain
$D$ and such that
\[
  M(\roler) \eqdef
  \begin{cases}
    \Tags\rolep\Pol\Tag_i.  D\setminus\set{\roleq}\Out\Tag_i.
    M_i(\roleq) & \text{if $\roler = \roleq$}
  \\
  \Tags\roleq\In\Tag_i.M_i(\roler) & \text{if
    $\roler \neq \roleq$}
  \end{cases}
\]
for every $\roler \in D$.
As suggested by the notation, this session map realizes a conversation in which
$\roleq$ first interacts with $\rolep$ by exchanging a tag $\Tag_i$ and then it
informs all the other participants about the tag that has been exchanged. This
session map has the property
\[
  M \xlred{\roleq:\rolep\Pol\Tag_k}\wred M_i
\]
for every $k\in I$.

Similarly, given a session map $N$ with domain $D \supseteq\set\roleq$, we write
$\Map{\roleq}{\rolep\Pol{U}.N}$ for the session map $M$ with domain $D$ such
that
\[
  M(\roler) \eqdef
  \begin{cases}
    \rolep\Pol{U}.N(\roleq) & \text{if $\roler = \roleq$}
    \\
    N(\roler) & \text{if $\roler \in D\setminus\set{\rolep,\roleq}$}
  \end{cases}
\]
for every $\roler \in D$. Note that $M$ has the property
\[
  M \xlred{\roleq:\rolep\Pol{U}} N
\]

The first key step is showing that $S \ssubt T$ implies $S \usubt T$ when $S$ is
a bounded session type. That is, unfair subtyping is a \emph{necessary
condition} for fair subtyping to hold.

\begin{lemma}
  \label{lem:usubt_completeness}
  If\/ $S$ is bounded and $S \ssubt T$ then $S \usubt T$.
\end{lemma}
\begin{proof}
  Using the coinduction principle it suffices to show that each
  judgment in the set
  \[
    \srel \eqdef \set{ S \usubt T \mid \text{$S$ is bounded and $S \ssubt T$} }
  \]
  is derivable by the rules in \cref{tab:subt} from premises that
  satisfy the same property. Let $S \usubt T \in \srel$. Then $S$ is
  bounded and $S \ssubt T$.
  We reason by cases on the shape of $S$.

  \proofcase{Case $S = \End$}
  Consider $M \eqdef \Map{\roleq}{\End[\co\Pol]}$ and observe that
  $M \parop \Map{\rolep} S$ is coherent. Then $M \parop \Map{\rolep} T$ is
  coherent as well, which implies $T = \End$.
  We conclude by observing that $\End \usubt \End$ is derivable with
  \refrule{u-end}.

  \proofcase{Case $S = \Tags\roleq\Out\Tag_i.S_i$}
  Let $\set{M_i}_{i\in I}$ be a family of session maps such that
  $M_i \parop \Map{\rolep} S_i$ is coherent for every $i\in I$. Such
  family is guaranteed to exist from the hypothesis that $S$ is
  bounded.
  Without loss of generality we may assume that the $M_i$ all have
  the same domain $D \supseteq \set\roleq$.
  Let $M \eqdef \Map{\roleq}{\Tags\rolep\In\Tag_i.M_i}$ and observe that
  $M \parop \Map{\rolep}{S}$ is coherent by definition of $M$.  Then
  $M \parop \Map{\rolep}{T}$ is coherent as well.
  We deduce that $T = \Tags[i\in J]\roleq\Out\Tag_i.T_i$ and
  $J \subseteq I$ and also that $M_i \parop \Map{\rolep}{T_i}$ is coherent
  for every $i\in J$. Hence $S_i \ssubt T_i$ for every $i\in J$,
  namely $S_i \usubt T_i \in \srel$ for every $i\in J$ by definition
  of $\srel$.
  We conclude by observing that $S \usubt T$ is derivable by
  \refrule{u-tag-out}.

  \proofcase{Case $S = \Tags\roleq\In\Tag_i.S_i$}
  Let $\set{M_i}_{i\in I}$ be a family of session maps such that
  $M_i \parop \Map{\rolep}{S_i}$ is coherent for every $i\in I$. Such
  family is guaranteed to exist from the hypothesis that $S$ is
  bounded. Without loss of generality we may assume that the $M_i$
  all have the same domain $D \supseteq \set\roleq$.
  Let $M \eqdef \Map{\roleq}{\Tags\rolep\Out\Tag_i.M_i}$ and observe that
  $M \parop \Map{\rolep}{S}$ is coherent by definition of $M$.
  We deduce that $T = \Tags[i\in J]\roleq\In\Tag_i.T_i$ and
  $I \subseteq J$ and also that $M_i \parop \Map{\rolep}{T_i}$ is coherent
  for every $i\in I$. Hence $S_i \ssubt T_i$ for every $i\in I$,
  namely $S_i \usubt T_i \in \srel$ for every $i\in I$ by definition
  of $\srel$.
  We conclude by observing that $S \usubt T$ is derivable by
  \refrule{u-tag-in}.

  \proofcase{Case $S = \roleq\Pol{U}.S'$}
  Let $N$ be a session map such that $N \parop \Map{\rolep}{S'}$ is
  coherent. Such $N$ is guaranteed to exist from the hypothesis that
  $S$ is bounded.
  Let $M \eqdef \Map{\roleq}{\rolep\co\Pol{U}.N}$ and observe that
  $M \parop \Map{\rolep}{S}$ is coherent by definition of $M$.
  We deduce that $T = \roleq\Pol{U}.T'$ and also that
  $N \parop \Map{\rolep}{T'}$ is coherent. Hence $S' \ssubt T'$, namely
  $S' \usubt T' \in \srel$ by definition of $\srel$.
  We conclude by observing that $S \usubt T$ is derivable by
  \refrule{u-channel}.
\end{proof}

\newcommand{\dualof}[3][D]{\mathsf{dual}_{#1}(\Map{#2}#3)}

Next we show that every bounded session type may be part of a coherent session
map. This result is somewhat related to the notion of \emph{duality} in binary
session type theories~\cite{Honda93,HondaVasconcelosKubo98}, showing that every
behavior can be completed by a matching -- dual -- one.

\begin{definition}[duality]
  Let $\targets{\cdot}$ be the function that yields the set of roles occurring
  in a session type, let $S$ be a bounded session type and $D$ be a non-empty
  set of roles that includes $\targets{S}$ but not $\rolep$.
  Let $\dualof\rolep{S}$ be the session map corecursively defined by
  the following equations:
  \[
    \begin{array}{r@{~}c@{~}ll}
      \dualof\rolep{\End[\In]} & = & \set{\Map\roleq \End[\Out]}_{\roleq\in D}
      \\
      \dualof\rolep{\End[\Out]} & = & \Map{\min D} \End[\In] \parop \set{\Map\roleq \End[\Out]}_{\roleq\in D \setminus \set{\min D}}
      \\
      \dualof\rolep{\Tags\roleq\Pol\Tag_i.S_i} & = &
      \Map\roleq \Tags\rolep\co\Pol\Tag_i.\dualof\rolep{S_i}
      \\
      \dualof\rolep{\roleq\Pol{U}.S} & = &
      \Map\roleq \rolep\co\Pol{U}.\dualof\rolep{S}
    \end{array}
  \]
\end{definition}

\begin{lemma}[duality]
	\label{lem:duality}
	$\dualof\rolep{S} \parop \Map\rolep S$ is coherent.
\end{lemma}
\begin{proof}
  Follows from the definition of $\dualof\rolep{S}$.
\end{proof}

Now we provide an algorithmic way of computing the ``difference'' between two
session types related by unfair subtyping.

\begin{definition}[subtyping weight]
  Under the hypothesis $S\usubt T$, let $\rk(S,T)\in\N\cup\set\infty$ be the
  least solution of the system of equations below:
\[
  \begin{array}{@{}r@{~}c@{~}ll@{}}
    \rk(\End, \End) & = & 0
    \\
    \rk(\Tags\role\In\Tag_i.S_i, \Tags[i\in J]\role\In\Tag_i.T_i)
    & = & \max_{i\in I} \rk(S_i, T_i)
    & I \subseteq J
    \\
    \rk(\Tags\role\Out\Tag_i.S_i, \Tags[i\in J]\role\Out\Tag_i.T_i)
    & = & 1 + \min_{i\in J} \rk(S_i, T_i)
    & J \subsetneq I
    \\
    \rk(\Tags\role\Out\Tag_i.S_i, \Tags\role\Out\Tag_i.T_i)
    & = & \min\set{ 1 + \min_{i\in I} \rk(S_i,T_i), \max_{i\in I} \rk(S_i,T_i)}
    \\
    \rk(\role\Pol{U}.S', \role\Pol{U}.T') & = & \rk(S', T')
  \end{array}
\]
\end{definition}

To see that $\rk(S,T)$ is well defined, observe that the system of equations
defining $\rk(S,T)$ under the hypothesis $S \usubt T$ contains finitely many
equations, say $n$, by regularity of $S$ and $T$. The system is representable as
a monotone endofunction $F$ on the complete lattice $(\Nat\cup\set\infty)^n$.
Thus, $F$ has a least solution of which $\rk(S,T)$ is a component.

We call two session types $S$ and $T$ divergent if they are related by unfair
subtyping and have infinite rank.

\begin{definition}[divergence]
	\label{def:diverge}
 	 We write $S \diverge T$ if $S \usubt T$ and $\rk(S, T) = \infty$.
\end{definition}

\begin{lemma}
	\label{lem:diverge}
  	If $S \diverge T$ then the derivation of $S \usubt T$ contains at least one application of \refrule{u-tag-out} with $J \subsetneq I$ and one of the following holds:
  	\begin{enumerate}
  	\item $S = \Tags \role\In\Tag_i.S_i$ and
    	$T = \Tags[i \in J] \role\In\Tag_i.T_i$ with $I \subseteq J$ and
    	$S_k \diverge T_k$ for some $k\in I$, or
  	\item $S = \Tags \role\Out\Tag_i.S_i$ and
    	$T = \Tags[i\in J] \role\Out\Tag_j.T_j$ with $J \subseteq I$ and
    	$S_i \diverge T_i$ for every $i\in J$, or
  	\item $S = \role\Pol{U}.S'$ and $T = \role\Pol{U}.T'$ and
    	$S' \diverge T'$.
  	\end{enumerate}
\end{lemma}
\begin{proof}
  If the derivation of $S \usubt T$ contained no application of
  \refrule{u-tag-out} with $J \subsetneq I$ we would have
  $\rk(S, T) = 0$. Now we reason by cases on the last rule used to
  derive $S \usubt T$.

  \proofrule{u-end}
  Then $S = T = \End$.  This case is impossible because
  $\rk(\End, \End) = 0$ by definition.

  \proofrule{u-tag-in}
  Then $S = \Tags\role\In\Tag_i.S_i$ and
  $T = \Tags[i\in J] \role\In\Tag_i.T_i$ with $I \subseteq J$ and
  $S_i \usubt T_i$ for every $i\in I$ and
  $\infty = \rk(S, T) = \max_{i\in I} \rk(S_i, T_i)$.
  That is, $\rk(S_k,T_k) = \infty$ for some $k \in I$, hence we
  conclude $S_k \diverge T_k$.

  \proofrule{u-tag-out}
  Then $S = \Tags\role\Out\Tag_i.S_i$ and
  $T = \Tags[i\in J]\role\Out\Tag_i.T_i$ with $J \subseteq I$ and
  $S_i \usubt T_i$ for every $i\in J$.
  We distinguish two sub-cases.
  If $J \subsetneq I$ then
  $\infty = \rk(S, T) = 1 + \min_{i\in J} \rk(S_i, T_i)$, that is
  $\rk(S_i, T_i) = \infty$ for every $i\in J$.
  If $J = I$ then
  $\infty = \rk(S, T) = \min\set{1 + \min_{i\in I} \rk(S_i, T_i),
    \max_{i\in I} \rk(S_i, T_i)} \leq 1 + \min_{i\in I} \rk(S_i,
  T_i)$ and we have $\rk(S_i, T_i) = \infty$ for every $i\in I$.
  Therefore, in both cases, we conclude $S_i \diverge T_i$ for every
  $i \in J$.

  \proofrule{u-channel}
  Then $S = \role\Pol{U}.S'$ and $T = \role\Pol{U}.T'$ and
  $S' \usubt T'$ and $\infty = \rk(S, T) = \rk(S', T')$ hence we
  conclude $S' \diverge T'$.
\end{proof}

\newcommand{\discriminator}[3]{\mathsf{disc}(#1,#2,#3)}

Finally, the key aspect of the proof of \cref{lem:divergence} is how we build the session map $M$ such that $M \parop \Map{\rolep}{S}$ is coherent while $M \parop \Map{\rolep}{T}$ is not. 

\begin{definition}[discriminator]
	\label{def:discriminator}
	Let $\discriminator\rolep{S}{T}$ be the session map
  	corecursively defined by the following equations:
  	\[
    	\begin{array}{r@{~}c@{~}ll}
      	\discriminator\rolep{
        	\Tags\roleq\In\Tag_i.S_i
      	}{
        	\Tags[i\in J]\roleq\In\Tag_i.T_i
      	}
      	& = &
      	\Map\roleq \Tags[i\in I, S_i \diverge T_i]\rolep\Out\Tag_i.\discriminator\rolep{S_i}{T_i}
      	& \text{if $I \subseteq J$}
      	\\
      	\discriminator\rolep{
        	\Tags\roleq\Out\Tag_i.S_i
      	}{
        	\Tags[i\in J]\roleq\Out\Tag_i.T_i
      	}
      	& = &
      	\Map\roleq \Tags[i\in J]\roleq\In\Tag_i.\discriminator\rolep{S_i}{T_i} 
        \\
        & + & \Tags[i\in I\setminus J] \roleq\In\Tag_i.\dualof\rolep{S_i}
      	& \text{if $J \subseteq I$}
      	\\
      	\discriminator\rolep{
        	\roleq\Pol U.S 
      	}{
        	\roleq\Pol U.T
      	}
      	& = &
      	\Map\roleq \rolep\co\Pol.\discriminator\rolep{S}{T}
    	\end{array}
  	\]
\end{definition}

\begin{lemma}
	\label{lem:divergence}
  	If $S$ is bounded and $S \diverge T$ then $S \not\ssubt T$.
\end{lemma}
\begin{proof}
  From the hypothesis that
  $S \diverge T$ and \cref{lem:diverge} we deduce that the
  derivation of $S \usubt T$ contains at least one application of
  \refrule{u-tag-out} with $J \subsetneq I$.
  Consider $\discriminator\rolep{S}{T}$ from \cref{def:discriminator}.
  Note that $\discriminator\rolep{S}{T}$ always sends a subset of
  the labels accepted by $S$, it is willing to receive any label
  sent by $S$, and it can always terminate successfully when
  interacting with $S$. Note also that it terminates successfully
  only after receiving a label from $S$ that $T$ cannot sent.
  Therefore, we have
  $\discriminator\rolep{S}{T} \parop \Map\rolep{S}$ coherent and
  $\discriminator\rolep{S}{T} \parop \Map\rolep{T}$ incoherent,
  which proves $S \not\ssubt T$.
\end{proof}

\begin{proof}[Proof of \cref{thm:subt_completeness}]
  \newcommand{\foosubt}{\subt[@]}
  Let $S \foosubt T$ if $S \usubt T$ and $\rk(U, T) < \infty$ for
  every judgment $U \usubt V$ in the derivation of $S \usubt T$.
  From \cref{lem:usubt_completeness,lem:divergence} we have that
  $S \ssubt T$ implies $S \foosubt T$. Indeed, if there is a
  judgment $U \usubt V$ in the derivation of $S \usubt T$ such that
  $\rk(U,V) = \infty$, then it is possible to build a session $M$
  such that $M \parop \Map\role{S}$ is coherent and
  $M \parop \Map\role{T}$ is not by induction on the minimum depth
  of the judgment $U \usubt V$ in the derivation using the
  hypothesis that $S$ is bounded and \cref{lem:divergence}.

  Now, using the principle of coinduction, it suffices to show that
  each judgment in the set
  \[
    \srel \eqdef \set{ S\subt[\rk(S,T)] T \mid S \foosubt T}
  \]
  is derivable using one of the rules in \cref{tab:subt} whose
  premises all belong to $\srel$.
\end{proof}


\section{Type System Soundness}
\label{sec:soundness-proof}

In this section we provide lemmas and proofs required to prove
\cref{thm:soundness}. We write $\prod_{i=1}^h \Map{\role_i}{S_i}$ for
$\Map{\role_1}{S_1} \parop \dots \parop \Map{\role_1}{S_1} = \set{\Map{\role_h}
S_h}_{i=1,\dots,h}$. Concerning the notation about the transitions of session
maps, we refer to \cref{sec:fs-proof}.
Finally, we write $\wtpi{\Ctx}{P}$ for the inductive type derivation.  

\subsection{Subject Reduction}

\begin{lemma}
\label{lem:substitution}
	If $\wtp[n]{\Ctx, x : S}{P}$ and $\Ctx, u : S$ is defined, then $\wtp[n]{\Ctx, u : S}{P \subst{u}{x}}$.
	A typing context is \emph{defined} if the endpoints occurring in it all have different session names.
\end{lemma}
\begin{proof}
By bounded coinduction.
\end{proof}

\begin{lemma}[Subject Congruence]
\label{lem:subj_cong}
	If\/ $\wtp[n] \Ctx {P}$ and $P \pcong Q$, then $\wtp[m] \Ctx {Q}$ for some $m \leq n$.
\end{lemma}
\begin{proof}
By induction on the derivation of $P \pcong Q$ and by cases on the last rule applied.

\proofrule{s-par-comm} 
Then $P = \pres{s}{\procs{P} \ppar P' \ppar Q' \ppar \procs{Q}} \pcong \pres{s}{\procs{P} \ppar Q' \ppar P' \ppar \procs{Q}} = Q$.
From rule \refrule{t-par} we deduce that there exist $\Ctx_i, \role_i, S_i, n_i$ for $i = 1,\dots,h$ such that
\begin{itemize}
\item $\Ctx = \Ctx_1,\dots,\Ctx_h$
\item $n = 1 + \sum_{i=1}^h n_i$
\item $\prod_{i=1}^h \Map{\role_i}{S_i} \ft$
\item $\wtp[n_i]{\Ctx_i, \ep{s}{\role_i} : S_i}{P_i}$ for $i = 1,\dots,k$
\item $\wtp[n_{k+1}]{\Ctx_{k+1}, \ep{s}{\role_{k+1}} : S_{k+1}}{P'}$
\item $\wtp[n_{k+2}]{\Ctx_{k+2}, \ep{s}{\role_{k+2}} : S_{k+2}}{Q'}$
\item $\wtp[n_i]{\Ctx_i, \ep{s}{\role_i} : S_i}{Q_i}$ for $i = k+3,\dots,h$
\end{itemize}

We conclude $\wtp[m]\Ctx{Q}$ with one application of \refrule{t-par} by taking $m \eqdef n$.

\proofrule{s-par-assoc}
Then $P = \pres{s}{\procs{P} \ppar \pres{t}{R \ppar \procs{Q}}} \pcong \pres{t}{\pres{s}{\procs{P} \ppar R} \ppar \procs{Q}} = Q$ and $s \in \fn{R}$.
From rule \refrule{t-par} we deduce that there exist $\Ctx_i, \role_i, S_i, n_i$ for $i = 1,\dots,h$ such that
\begin{itemize}
\item $\Ctx = \Ctx_1,\dots,\Ctx_h$
\item $n = 1 + \sum_{i=1}^h n_i$
\item $\prod_{i=1}^h \Map{\role_i}{S_i} \ft$
\item $\wtp[n_i]{\Ctx_i, \ep{s}{\role_i} : S_i}{P_i}$ for $i = 1,\dots,h - 1$
\item $\wtp[n_h]{\Ctx_h, \ep{s}{\role_h} : S_h}{\pres{t}{R \ppar \procs{Q}}}$
\end{itemize}
From rule \refrule{t-par} and the hypothesis that $s \in \fn{R}$ we deduce that there exist $\CtxD_i, \roleq_i, T_i, m_i$ for $i = 1,\dots,k$ such that
\begin{itemize}
\item $\Ctx_h = \CtxD_1,\dots,\CtxD_k$
\item $n_h = 1 + \sum_1^k m_i$
\item $\prod_1^k \Map{\roleq_i}{T_i} \ft$
\item $\wtp[m_1]{\CtxD_1, \ep{s}{\role_h} : S_h, \ep{t}{\roleq_1} : T_1}{R}$
\item $\wtp[m_{i+1}]{\CtxD_{i+1}, \ep{t}{\roleq_{i+1}} : T_{i+1}}{Q_i}$ for $i = 1,\dots,k-1$
\end{itemize}
Using \refrule{t-par} we deduce $\wtp[1 + \sum_{i=1}^{h-1}{n_i} + m_1]{\Ctx_1,\dots,\Ctx_{h-1},\CtxD_1, \ep{t}{\roleq_1} : T_1}{\pres{s}{\procs{P} \ppar R}}$.  We conclude $\wtp[m]{\Ctx}{\pres{t}{\pres{s}{\procs{P} \ppar R} \ppar \procs{Q}}}$ with another application of \refrule{t-par} by taking $m \eqdef n$.

\proofrule{s-cast-comm} 
Then $P = \pcast{u}{\pcast{v}{R}} \pcong \pcast{v}{\pcast{u}{R}} = Q$. We can assume $u \ne v$ or else $P = Q$.
From rule \refrule{t-cast} we deduce that there exist $\Ctx_1, S, T, n_1, n_u$ such that
\begin{itemize}
\item $\Ctx = \Ctx_1, u : S$
\item $S \subt[n_u] T$
\item $n = n_u + n_1$
\item $\wtp[n_1]{\Ctx_1, u : T}{\pcast{v}{R}}$
\end{itemize} 
From rule \refrule{t-cast} we deduce that there exist $\Ctx_2, S', T', n_2, n_v$ such that
\begin{itemize}
\item $\Ctx_1 = \Ctx_2, v : S'$
\item $S' \subt[n_v] T'$
\item $n_1 = n_v + n_2$
\item $\wtp[n_2]{\Ctx_2, u : T, v : T'}{R}$
\end{itemize}
We derive $\wtp[n_u + n_2]{\Ctx_2, u : S, v : T'}{\pcast{u}{R}}$ with one application of \refrule{t-cast} and we conclude with another application of \refrule{t-cast} by taking $m \eqdef n$.

\proofrule{s-cast-new}
Then $P = \pres{s}{\pcast{\ep{s}{\role}}{R} \ppar \procs{P}} \pcong \pres{s}{R \ppar \procs{P}} = Q$.
From rule \refrule{t-par} we deduce that there exist $\CtxD, n'$ and $\Ctx_i, \roleq_i, n_i$ for $i = 1,\dots,h$ such that
\begin{itemize}
\item $\Ctx = \CtxD, \Ctx_1, \dots, \Ctx_h$ for $i = 1,\dots,h$
\item $n = 1 + n' + \sum_{i=1}^h n_i$
\item $\Map{\role}{S} \ppar \prod_{i = 1}^h \Map{\roleq_i}{S_i} \ft$
\item $\wtp[n']{\CtxD, \ep{s}{\role} : S}{\pcast{\ep{s}{\role}}{R}}$
\item $\wtp[n_i]{\Ctx_i, \ep{s}{\roleq_i} : S_i}{P_i}$ for $i = 1,\dots,h$
\end{itemize} 
From rule \refrule{t-cast} we deduce that there exist $T, m', m_s$ such that
\begin{itemize}
\item $S \subt[m_s] T$
\item $n' = m_s + m'$
\item $\wtp[m']{\CtxD, \ep{s}{\role} : T}{R}$
\end{itemize}
From $\Map{\role}{S} \ppar \prod_{i = 1}^h \Map{\roleq_i}{S_i} \ft$, $S \subt[m_s] T$ and \cref{def:ssubt} we deduce 
$\Map{\role}{T} \ppar \prod_{i = 1}^h \Map{\roleq_i}{S_i} \ft$.
We conclude with an application of \refrule{t-par} by taking $m \eqdef 1 + m' + \sum_{i=1}^h n_i \le n$.

\proofrule{s-cast-swap}
Then $P = \pres{s}{\pcast{\ep{t}{\role}}{R} \ppar \procs{P}} \pcong \pcast{\ep{t}{\role}}{\pres{s}{R \ppar \procs{P}}} = Q$ and $t \ne s$.
From rule \refrule{t-par} we deduce that there exist $\Ctx_i, \roleq_i, n_i$ for $i = 1,\dots,h$ such that
\begin{itemize}
\item $\Ctx = \Ctx_1, \dots, \Ctx_h$ for $i = 1,\dots,h$
\item $n = 1 + \sum_{i=1}^h n_i$
\item $\prod_{i = 1}^h \Map{\roleq_i}{S_i} \ft$
\item $\wtp[n_1]{\Ctx_1, \ep{s}{\roleq_1} : S_1}{\pcast{\ep{t}{\role}}{R}}$
\item $\wtp[n_i]{\Ctx_i, \ep{s}{\roleq_i} : S_i}{P_i}$ for $i = 2,\dots,h$
\end{itemize} 
From rule \refrule{t-cast} we deduce that there exist $\CtxD, T, n', m_t$ such that
\begin{itemize}
\item $\Ctx_1 = \CtxD, \ep{t}{\role} : S$
\item $S \subt[m_t] T$
\item $n_1 = m_t + n'$
\item $\wtp[n']{\CtxD, \ep{t}{\role} : T, \ep{s}{\roleq_1} : S_1}{R}$
\end{itemize}
We derive $\wtp[1 + n' + \sum_{i=2}^h n_i]{\CtxD, \ep{t}{\role} : T,\Ctx_2,\dots,\Ctx_h}{\pres{s}{R \ppar \procs{P}}}$ with an application of \refrule{t-par}. We conclude with an application of \refrule{t-cast} by taking $m \eqdef n$.

\proofrule{s-call} 
Then $P = \pinvk{A}{\seqof u} \pcong R\subst{\seqof u}{\seqof x} = Q$ and
$\Definition{A}{\seqof x}{R}$.
From \refrule{t-call} we conclude that there exist $\seqof S$ and $m$ such that
$\tass{A}{\seqof{S}}{m}$ and $\Ctx = \seqof{u : S}$ and $\wtp[m]{\seqof{u :
S}}{Q}$ and $m \leq n$.
\end{proof}

\begin{lemma}[Subject Reduction]
	\label{lem:subj_red}
	If\/ $\wtp[n] \Ctx {P}$ and $P \red Q$, then $\wtp[m] \Ctx {Q}$ for some $m$.
\end{lemma}
\begin{proof}
By induction on the derivation of $P \red Q$ and by cases on the last rule applied.

\proofrule{r-choice}
Then $P = P_1 \choice P_2 \red P_k = Q$ and $k \in \set{1,2}$.
From \refrule{t-choice} we deduce that $\wtp[m]{\Ctx}{Q}$ for some $m$.

\proofrule{r-signal}
Then $ P = \pres{s}{\pwait{\ep{s}{\role}}{Q} \parop \pclose{\ep{s}{\roleq_1}} \parop \cdots \parop \pclose{\ep{s}{\roleq_h}}} \red Q$.
From \refrule{t-par}, \refrule{t-wait} and \refrule{t-close} we deduce that there exist $m$ and $n_i$ for $i=1,\dots,h$ such that
\begin{itemize}
\item $n = 1 + m + \sum_{i=1}^h n_i$
\item $\wtp[m]{\Ctx, \ep{s}{\role} : \End[\In]}{\pwait{\ep{s}{\role}}{Q}}$
\item $\wtp[m]{\Ctx}{Q}$
\item $\wtp[n_i]{\ep{s}{\roleq_i} : \End[\Out]}{\pclose{\ep{s}{\roleq_i}}}$ for
$i=1,\dots,h$
\end{itemize}
There is nothing left to prove.

\proofrule{r-channel}
Then $P = \pres{s}{\poch{\ep{s}{\rolep}}{\roleq}{v}{P'} \parop
\pich{\ep{s}{\roleq}}{\rolep}{x}{Q'} \parop \procs{R}} \red \pres{s}{P' \parop
Q'\subst{v}{x} \parop \procs{R}} = Q$.
From \refrule{t-par} we deduce that there exist $\Ctx_i, S_i, \role_i, n_i$ for $i=1,\dots,h$ such that
\begin{itemize}
\item $\Ctx = \Ctx_1,\dots,\Ctx_h$
\item $n = 1 + \sum_{i=1}^h n_i$
\item $\prod_{i=1}^h \Map{\role_i}{S_i} \ft$
\item $\role = \role_1$ and $\roleq = \role_2$
\item $\wtp[n_1]{\Ctx_1, \ep{s}{\role} : S_1}{\poch{\ep{s}{\rolep}}{\roleq}{v}{P'}}$
\item $\wtp[n_2]{\Ctx_2, \ep{s}{\roleq} : S_2}{\pich{\ep{s}{\roleq}}{\rolep}{x}{Q'}}$
\item $\wtp[n_i]{\Ctx_i, \ep{s}{\role_i} : S_i}{R_i}$ for $i = 3,\dots,h$
\end{itemize}
From \refrule{t-channel-out} and \refrule{t-channel-in} we deduce that there exist $S_v, T_1, T_2, \CtxD_1$ such that
\begin{itemize}
\item $S_1 = \roleq\Out{S_v}.T_1$
\item $\Ctx_1 = \CtxD_1, v : S_v$
\item $\wtp[n_1]{\CtxD_1, \ep{s}{\rolep} : T_1}{P'}$
\item $S_2 = \rolep\In{S_v}.T_2$
\item $\wtp[n_2]{\Ctx_2, \ep{s}{\roleq} : T_2, x : S_v}{Q'}$
\end{itemize} 
Using \cref{lem:substitution} we deduce $\wtp[n_2]{\Ctx_2, \ep{s}{\roleq} : T_2,
v : S_v}{Q'\subst{v}{x}}$. Using \cref{def:coherence} we deduce $\Map{\rolep}{T_1}
\parop \Map{\roleq}{T_2} \parop \prod_{i=3}^h \Map{\role_i}{S_i} \ft$. We conclude with
one application of \refrule{t-par} taking $m \eqdef n$.

\proofrule{r-pick}
Then $P = \pres{s}{\pobranch[i\in I]{\ep{s}{\role}}{\roleq}{\Tag_i}{\PP_i} \parop \procs{Q}} \red \pres{s}{\pobranch{\ep{s}{\role}}{\roleq}{\Tag_k}{\PP_k}\parop \procs{Q}} = Q$ and $k \in I$. 
From \refrule{t-par} we deduce that there exist $\Ctx_i, \role_i, n_i, S_i$ for $i=1,\dots,h$ such that
\begin{itemize}
\item $\Ctx = \Ctx_1, \dots, \Ctx_h$
\item $n = 1 + \sum_{i=1}^h n_i$
\item $\prod_{i=1}^h \Map{\role_i}{S_i} \ft$
\item $\role = \role_1$ and $\roleq = \role_i$ for some $i \in \set{2,\dots,h}$
\item $\wtp[n_1]{\Ctx_1, \ep{s}{\role} : S_1}{\pobranch[i\in I]{\ep{s}{\role}}{\roleq}{\Tag_i}{\PP_i}}$
\item $\wtp[n_i]{\Ctx_i, \ep{s}{\role_i} : S_i}{Q_i}$ for $i=2,\dots,h$
\end{itemize}
From \refrule{t-tag} we deduce that there exist $T_i$ for all $i \in I$ such that
\begin{itemize}
\item $S_1 = \Tags\roleq\Out \Tag_i.T_i$
\item $\wtp[n_1]{\Ctx_1, \ep{s}{\role} : T_i}{P_i}~{}^{(i\in I)}$
\end{itemize}
From the hypothesis that $k \in I$ we deduce that $\wtp[n_1]{\Ctx_1, \ep{s}{\role} : T_k}{P_k}$ and from \refrule{t-tag} we deduce 
$\wtp[n_1]{\Ctx_1, \ep{s}{\role} : \roleq\Out\Tag_k.T_k}{\pobranch{\ep{s}{\role}}{\roleq}{\Tag_k}{\PP_k}}$. 
From \cref{def:coherence} we deduce that $\roleq\Out\Tag_k.T_k \parop \prod_{i=2}^h \Map{\role_i}{S_i} \ft$. We conclude with an application of \refrule{t-par} taking $m \eqdef n$.

\proofrule{r-tag}
Then $P = \pres{s}{\pobranch{\ep{s}{\role}}{\roleq}{\Tag_k}{\PP'} \parop \pibranch[i\in I]{\ep{s}{\roleq}}{\role}{\Tag_i}{Q_i} \parop \procs{R}} 
        \red 
        \pres{s}{P' \parop Q_k \parop \procs{R}} = Q$ and $k \in I$.
From \refrule{t-par} we deduce that there exist $\Ctx_i, S_i, \role_i, n_i$ for $i=1,\dots,h$ such that
\begin{itemize}
\item $\Ctx = \Ctx_1,\dots,\Ctx_h$
\item $n = 1 + \sum_{i=1}^h n_i$
\item $\prod_{i=1}^h \Map{\role_i}{S_i} \ft$
\item $\role = \role_1$ and $\roleq = \role_2$
\item $\wtp[n_1]{\Ctx_1, \ep{s}{\role} : S_1}{\pobranch{\ep{s}{\role}}{\roleq}{\Tag_k}{\PP'}}$
\item $\wtp[n_2]{\Ctx_2, \ep{s}{\roleq} : S_2}{\pibranch[i\in I]{\ep{s}{\roleq}}{\role}{\Tag_i}{Q_i}}$
\item $\wtp[n_i]{\Ctx_i, \ep{s}{\role_i} : S_i}{R_i}$ for $i = 3,\dots,h$
\end{itemize}
From \refrule{t-tag} we deduce that there exist $S'_1$ and $T_i$ for every $i \in I$ such that
\begin{itemize}
\item $S_1 = \roleq\Out \Tag_k.S'_1$
\item $\wtp[n_1]{\Ctx_1, \ep{s}{\role} : S'_1}{P'}$
\item $S_2 = \Tags\rolep\In \Tag_i.T_i$
\item $\wtp[n_2]{\Ctx_2, \ep{s}{\roleq} : T_i}{Q_i}~{}^{(i\in I)}$
\end{itemize}
From \cref{def:coherence} we deduce that $\Map{\role}{S'_1} \parop \Map{\roleq}{T_k} \parop \prod_{i=2}^h \Map{\role_i}{S_i} \ft$. We conclude with an application of \refrule{t-par} by taking $m \eqdef n$.

\proofrule{r-par}
Then $P = \pres{s}{P' \parop \procs{R}} \red \pres{s}{Q' \parop \procs{R}} = Q$ and $P' \red Q'$.
From \refrule{t-par} we deduce that there exist $\Ctx_i, \role_i, S_i, n_i$ for $i=1,\dots,h$ such that
\begin{itemize}
\item $\Ctx = \Ctx_1,\dots,\Ctx_h$
\item $n = 1 + \sum_{i=1}^h n_i$
\item $\prod_{i=1}^h \Map{\role_i}{S_i} \ft$
\item $\wtp[n_1]{\Ctx_1, \ep{s}{\role_1} : S_1}{P'}$
\item $\wtp[n_i]{\Ctx_i, \ep{s}{\role_i} : S_i}{R_i}$ for $i = 2,\dots, h$
\end{itemize}
Using the induction hypothesis on $\wtp[n_1]{\Ctx_1, \ep{s}{\role_1} : S_1}{P'}$ and $P' \red Q'$ we deduce $\wtp[n'_1]{\Ctx_1, \ep{s}{\role_1} : S_1}{Q'}$ for some $n'_1$. We conclude with an application of \refrule{t-par} taking $m \eqdef 1 + n'_1 + \sum_{i=2}^h n_i$.

\proofrule{r-cast}
Then $P = \pcast{u}{P'} \red \pcast{u}{Q' } = Q$ and $P' \red Q'$.
From \refrule{t-cast} we deduce that there exist $S, T, \Ctx', n', m_u$ such that
\begin{itemize}
\item $\Ctx = \Ctx', u : S$
\item $S \subt[m_u] T$
\item $n = m_u + n'$
\item $\wtp[n']{\Ctx', u : T}{P'}$
\end{itemize}
Using the induction hypothesis on $\wtp[n']{\Ctx', u : T}{P'}$ and $P' \red Q'$ we deduce $\wtp[m']{\Ctx', u : T}{Q'}$ for some $m'$. We conclude with an application of \refrule{t-cast} taking $m \eqdef m_u + m'$.

\proofrule{r-struct}
Then $P \pcong P' \red Q' \pcong Q$.
From \cref{lem:subj_cong} we deduce that $\wtp[n']{\Ctx}{P'}$ for some $n' \le n$. Using the induction hypothesis on $\wtp[n']{\Ctx}{P'}$ and $P' \red Q'$ we deduce $\wtp[m']{\Ctx}{Q'}$ for some $m'$. We conclude using \cref{lem:subj_cong} once more.
\end{proof}
\subsection{Measure}

\begin{definition}[rank]
  \label{def:rank}
  The \emph{rank} of a session map $M = \prod_{i=1}^h \Map{\role_i}{S_i}$, written $\rank{M}$, is the element of $\Nat
  \cup \set\infty$ defined as
  \begin{center}
  	\begin{math}
  		\rank{M} \eqdef \min | M \wlred{\In\terminated}|
  	\end{math}
  \end{center}
  where $|M \wlred{\action} N|$ denotes the length of the sequence $\tau,\dots,\tau,\action$ and we postulate that $\min\emptyset = \infty$.
\end{definition}

\begin{definition}[Measure]
The measure of a process is a lexicographically ordered pair of natural numbers
$(m , n)$ where:
\begin{itemize}
\item $m$ is an upper bound to the number of sessions that the process may open
and of weights of casts that the process may perform \emph{in the future} before
it terminates;
\item $n$ is the overall effort for terminating the sessions that have been
already opened \emph{in the past}, \ie the sum of their rank (\cref{def:rank}).
\end{itemize}
\end{definition}

We now introduce a refined set of typing rules for processes that allow us to
associate them with their measure, not just with their rank.
\begin{mathpar}
    \inferrule[mt-thread]{
        \mathstrut
    }{
        \wtpn{(n, 0)}\Ctx{P}
    }
    \wtp[n]\Ctx{P}
    \and
    \inferrule[mt-cast]{
        \wtpn\Measure{\Ctx, u : T}{P}
    }{
        \wtpn{\Measure + (n,0)}{\Ctx, u : S}{\pcast{u} P}
    }
    ~
    S \subt[n] T
    \and
    \inferrule[mt-par]{
        \wtpn{\Measure_i}{\Ctx_i, \ep{s}{\role_i} : S_i}{P_i}~{}^{(i=1,\dots,h)}
    }{
        \wtpn{\sum_{i=1}^h \Measure_i + (0, \rank{\set{\Map{\role_i}{S_i}}_{i=1,\dots,h}})}{
            \Ctx_1,\dots,\Ctx_h
        }{
            \pres{s}{P_1 \parop \dots \parop P_h}
        }
    }
    ~ \coherent{\set{\Map{\role_i}{S_i}}_{i=1..h}}
\end{mathpar}

The idea behind these rules is that they distinguish between \emph{past} and
\emph{future} of a process by looking at its structure. Indeed, unguarded
sessions have been created, casts have not been performed yet and sessions that
occur guarded have not been created yet.
\refrule{mt-thread} adopts the rank of the process inside the usual typing
judgment (\cref{tab:ts}) as first component of the measure. This rule has lower
priority with respect to the other rules so that it is applied to processes that
are not casts or restrictions.
In \refrule{mt-cast} the first component of the measure is increased by the
weight of the cast.
\refrule{mt-par} increases the second component of the measure by the rank of
the involved session.

\begin{lemma}
    \label{lem:measure_rank}
    The following properties hold:
    \begin{enumerate}
        \item $\wtp[n]\Ctx{P}$ implies $\wtpn\Measure\Ctx{P}$ for some
        $\Measure \leq (n, 0)$;
        \item $\wtpn\Measure\Ctx{P}$ implies $\wtp[n]\Ctx{P}$ for some $n$ such that $\Measure \leq (n, 0)$.
    \end{enumerate}
\end{lemma}
\begin{proof}
    We prove item 1 by induction on the structure of $P$. 
    The proof of item 2 is by a straightforward induction over $\wtpn\Measure\Ctx{P}$.
    
\proofcase{Case $P = \pres{s}{\procs{P}}$}
From \refrule{t-par} we deduce that there exist $\Ctx_i, \role_i, S_i, n_i$ for $i = 1,\dots,h$ such that
\begin{itemize}
\item $\Ctx = \Ctx_1,\dots,\Ctx_h$
\item $n = 1 + \sum_{i=1}^h n_i$
\item $\prod_{i=1}^h \Map{\role_i}{S_i} \ft$
\item $\wtp[n_i]{\Ctx_i, \ep{s}{\role_i} : S_i}{P_i}~{}^{(i=1,\dots,h)}$
\end{itemize} 
Using the induction hypothesis on $\wtp[n_i]{\Ctx_i, \ep{s}{\role_i} : S_i}{P_i}~{}^{(i=1,\dots,h)}$ we deduce that there exist $\Measure_i$ for $i=1,\dots,h$ such that 
\begin{itemize}
\item $\wtpn{\Measure_i}{\Ctx_i, \ep{s}{\role_i} : S_i}{P_i}~{}^{(i=1,\dots,h)}$
\item $\Measure_i \le (n_i,0)$ for $i=1,\dots,h$
\end{itemize}
We conclude with one application of \refrule{mt-par} by taking $\Measure \eqdef \sum_{i=1}^h \Measure_i + (0, \rank{\prod_{i=1}^h \Map{\role_i}{S_i}})$ and observing that $\Measure < (n_1,0) + (n_2,0) + \dots + (n_h,0) + (1,0) = (n,0)$.

\proofcase{Case $P = \pcast{u}{Q}$}
From \refrule{t-cast} we deduce that there exist $\CtxD, S, T, m, m_u$ such that
\begin{itemize}
\item $\Ctx = \CtxD, u : S$
\item $S \subt[m_u] T$
\item $n = m_u + m$
\item $\wtp[m]{\CtxD, u : T}{Q}$
\end{itemize}
Using the induction hypothesis on $\wtp[m]{\CtxD, u : T}{Q}$ we deduce $\wtpn{\MeasureN}{\CtxD, u : T}{Q}$ for some $\MeasureN \le (m,0)$. We conclude with an application of \refrule{mt-cast} by taking $\Measure \eqdef \MeasureN + (m_u,0)$ and observing that $\Measure \le (m,0) + (m_u,0) = (n,0)$.

\proofcase{In all the other cases} We conclude with an application of \refrule{mt-thread} by taking $\Measure \eqdef (n,0)$.
\end{proof}

\begin{lemma}
	\label{lem:measure_pcong}
	If $\wtpn\MeasureM\Ctx P$ and $P \pcong Q$, then there exists $\MeasureN \le \MeasureM$ such that $\wtpn\MeasureN\Ctx Q$.
\end{lemma}
\begin{proof}
By induction on the derivation of $P \pcong Q$ and by cases on the last rule applied. We only consider the base cases.

\proofrule{s-par-comm} 
Then $P = \pres{s}{\procs{P} \ppar P' \ppar Q' \ppar \procs{Q}} \pcong \pres{s}{\procs{P} \ppar Q' \ppar P' \ppar \procs{Q}} = Q$.
From rule \refrule{mt-par} we deduce that there exist $\Ctx_i, \role_i, S_i, \Measure_i$ for $i = 1,\dots,h$ such that
\begin{itemize}
\item $\Ctx = \Ctx_1,\dots,\Ctx_h$
\item $\Measure = \sum_{i=1}^h \Measure_i + (0 , \rank{\prod_{i=1}^h \Map{\role_i}{S_i}})$
\item $\prod_{i=1}^h \Map{\role_i}{S_i} \ft$
\item $\wtpn{\Measure_i}{\Ctx_i, \ep{s}{\role_i} : S_i}{P_i}$ for $i = 1,\dots,k$
\item $\wtpn{\Measure_{k+1}}{\Ctx_{k+1}, \ep{s}{\role_{k+1}} : S_{k+1}}{P'}$
\item $\wtpn{\Measure_{k+2}}{\Ctx_{k+2}, \ep{s}{\role_{k+2}} : S_{k+2}}{Q'}$
\item $\wtpn{\Measure_i}{\Ctx_i, \ep{s}{\role_i} : S_i}{Q_i}$ for $i = k+3,\dots,h$
\end{itemize}
We conclude $\wtpn{\MeasureN}\Ctx{Q}$ with one application of \refrule{mt-par} by taking $\MeasureN \eqdef \Measure$.

\proofrule{s-par-assoc}
Then $P = \pres{s}{\procs{P} \ppar \pres{t}{R \ppar \procs{Q}}} \pcong \pres{t}{\pres{s}{\procs{P} \ppar R} \ppar \procs{Q}} = Q$ and $s \in \fn{R}$.
From rule \refrule{mt-par} we deduce that there exist $\Ctx_i, \role_i, S_i, \Measure_i$ for $i = 1,\dots,h$ such that
\begin{itemize}
\item $\Ctx = \Ctx_1,\dots,\Ctx_h$
\item $\Measure = \sum_{i=1}^h \Measure_i + (0 , \rank{\prod_{i=1}^h \Map{\role_i}{S_i}})$
\item $\prod_{i=1}^h \Map{\role_i}{S_i} \ft$
\item $\wtpn{\Measure_i}{\Ctx_i, \ep{s}{\role_i} : S_i}{P_i}$ for $i = 1,\dots,h - 1$
\item $\wtpn{\Measure_h}{\Ctx_h, \ep{s}{\role_h} : S_h}{\pres{t}{R \ppar \procs{Q}}}$
\end{itemize}
From rule \refrule{mt-par} and the hypothesis that $s \in \fn{R}$ we deduce that there exist $\CtxD_i, \roleq_i, T_i, \MeasureN_i$ for $i = 1,\dots,k$ such that
\begin{itemize}
\item $\Ctx_h = \CtxD_1,\dots,\CtxD_k$
\item $\Measure_h = \sum_1^k \MeasureN_i + (0 , \rank{\prod_{i=1}^k \Map{\roleq_i}{T_i}})$
\item $\prod_{i=1}^k \Map{\roleq_i}{T_i} \ft$
\item $\wtpn{\MeasureN_1}{\CtxD_1, \ep{s}{\role_h} : S_h, \ep{t}{\roleq_1} : T_1}{R}$
\item $\wtpn{\MeasureN_{i+1}}{\CtxD_{i+1}, \ep{t}{\roleq_{i+1}} : T_{i+1}}{Q_i}$ for $i = 1,\dots,k-1$
\end{itemize}
Using \refrule{t-par} we deduce $\wtpn{\sum_{i=1}^{h-1}{\Measure_i} + \MeasureN_1 + \rank{\prod_{i=1}^h \Map{\role_i}{S_i}}}{\Ctx_1,\dots,\Ctx_{h-1},\CtxD_1, \ep{t}{\roleq_1} : T_1}{\pres{s}{\procs{P} \ppar R}}$.  We conclude $\wtpn{\MeasureN}{\Ctx}{\pres{t}{\pres{s}{\procs{P} \ppar R} \ppar \procs{Q}}}$ with another application of \refrule{mt-par} by taking $\MeasureN \eqdef \Measure$.

\proofrule{s-cast-comm} 
Then $P = \pcast{u}{\pcast{v}{R}} \pcong \pcast{v}{\pcast{u}{R}} = Q$. We can assume $u \ne v$ or else $P = Q$.
From rule \refrule{mt-cast} we deduce that there exist $\Ctx_1, S, T, \Measure_1, m_u$ such that
\begin{itemize}
\item $\Ctx = \Ctx_1, u : S$
\item $S \subt[m_u] T$
\item $\Measure = \Measure_1 + (m_u , 0)$
\item $\wtpn{\Measure_1}{\Ctx_1, u : T}{\pcast{v}{R}}$
\end{itemize} 
From rule \refrule{t-cast} we deduce that there exist $\Ctx_2, S', T', \Measure_2, m_v$ such that
\begin{itemize}
\item $\Ctx_1 = \Ctx_2, v : S'$
\item $S' \subt[m_v] T'$
\item $\Measure_1 = \Measure_2 + (m_v , 0)$
\item $\wtpn{\Measure_2}{\Ctx_2, u : T, v : T'}{R}$
\end{itemize}
We derive $\wtpn{\Measure_2 + (m_u,0)}{\Ctx_2, u : S, v : T'}{\pcast{u}{R}}$ with one application of \refrule{mt-cast} and we conclude with another application of \refrule{mt-cast} by taking $\MeasureN \eqdef \Measure$.

\proofrule{s-cast-new}
Then $P = \pres{s}{\pcast{\ep{s}{\role}}{R} \ppar \procs{P}} \pcong \pres{s}{R \ppar \procs{P}} = Q$.
From rule \refrule{mt-par} we deduce that there exist $\CtxD, \Measure', S$ and $\Ctx_i, \roleq_i, S_i, \Measure_i$ for $i = 1,\dots,h$ such that
\begin{itemize}
\item $\Ctx = \CtxD, \Ctx_1, \dots, \Ctx_h$
\item $\Measure = \Measure' + \sum_{i=1}^h \Measure_i + (0 , \rank{\Map{\role}{S} \ppar \prod_{i = 1}^h \Map{\roleq_i}{S_i}})$
\item $\Map{\role}{S} \parop \prod_{i = 1}^h \Map{\roleq_i}{S_i} \ft$
\item $\wtpn{\Measure'}{\CtxD, \ep{s}{\role} : S}{\pcast{\ep{s}{\role}}{R}}$
\item $\wtpn{\Measure_i}{\Ctx_i, \ep{s}{\roleq_i} : S_i}{P_i}$ for $i = 1,\dots,h$
\end{itemize} 
From rule \refrule{mt-cast} we deduce that there exist $T, \MeasureN', m_s$ such that
\begin{itemize}
\item $S \subt[m_s] T$
\item $\MeasureN' = \Measure' + (m_s , 0)$
\item $\wtpn{\MeasureN'}{\CtxD, \ep{s}{\role} : T}{R}$
\end{itemize}
From $\Map{\role}{S} \ppar \prod_{i = 1}^h \Map{\roleq_i}{S_i} \ft$, $S \subt[m_s] T$ and \cref{def:ssubt} we deduce $\Map{\role}{T} \ppar \prod_{i = 1}^h \Map{\roleq_i}{S_i} \ft$. We conclude with an application of \refrule{mt-par} by taking $\MeasureN = \MeasureN' + \sum_{i=1}^h \Measure_i + (0 , \rank{\Map{\role}{S} \ppar \prod_{i = 1}^h \Map{\roleq_i}{S_i}}) \le n$.

\proofrule{s-cast-swap}
Then $P = \pres{s}{\pcast{\ep{t}{\role}}{R} \ppar \procs{P}} \pcong \pcast{\ep{t}{\role}}{\pres{s}{R \ppar \procs{P}}} = Q$ and $t \ne s$.
From rule \refrule{mt-par} we deduce that there exist $\Ctx_i, \roleq_i, \Measure_i$ for $i = 1,\dots,h$ such that
\begin{itemize}
\item $\Ctx = \Ctx_1, \dots, \Ctx_h$
\item $\Measure = \sum_{i=1}^h \Measure_i + (0 , \rank{\prod_{i = 1}^h \Map{\roleq_i}{S_i}})$
\item $\prod_{i = 1}^h \Map{\roleq_i}{S_i} \ft$
\item $\wtpn{\Measure_1}{\Ctx_1, \ep{s}{\roleq_1} : S_1}{\pcast{\ep{t}{\role}}{R}}$
\item $\wtpn{\Measure_i}{\Ctx_i, \ep{s}{\roleq_i} : S_i}{P_i}$ for $i = 2,\dots,h$
\end{itemize} 
From rule \refrule{mt-cast} we deduce that there exist $\CtxD, T, \Measure', m_t$ such that
\begin{itemize}
\item $\Ctx_1 = \CtxD, \ep{t}{\role} : S$
\item $S \subt[m_t] T$
\item $\Measure_1 = \Measure' + (m_t , 0)$
\item $\wtpn{\Measure'}{\CtxD, \ep{t}{\role} : T, \ep{s}{\roleq_1} : S_1}{R}$
\end{itemize}
We derive $\wtpn{\Measure' + \sum_{i=2}^h \Measure_i + (0 , \rank{\prod_{i = 1}^h \Map{\roleq_i}{S_i}})}{\CtxD, \ep{t}{\role} : T,\Ctx_2,\dots,\Ctx_h}{\pres{s}{R \ppar \procs{P}}}$ with an application of \refrule{mt-par}. We conclude with an application of \refrule{mt-cast} by taking $m \eqdef n$.

\proofrule{s-call} 
Then $P = \pinvk{A}{\seqof u} \pcong R\subst{\seqof x}{\seqof u} = Q$ and $\Definition{A}{\seqof x}{R}$.
From \refrule{mt-thread} we deduce that $\wtpn{n}{\Ctx}{\pinvk{A}{\seqof u}}$ for some $n$ such that $\Measure = (n , 0)$. Using \cref{lem:subj_cong} we deduce $\wtp[m]{\Ctx}{Q}$ for some $m \le n$.
Using \cref{lem:measure_rank} we deduce that $\wtpn{\MeasureN}{\Ctx}{Q}$ for some $\MeasureN \le (m , 0)$. We conclude observing that $\MeasureN \le (m , 0) \le (n , 0) = \Measure$.
\end{proof}
\subsection{Normal Forms}


We introduce \emph{process contexts} to easily refer to unguarded sub-processes:
\[
	\textbf{Process context}
	\quad
	\PCtxC, \PCtxD ~~::=~~ \Hole \mid \pres{s}{\procs{P} \parop \PCtxC \parop \procs{Q}} \mid \pcast{u}\PCtxC
\]

\begin{definition}[Choice Normal Form]
	\label{def:choice_normal_form}
	We say that $P_1 \choice P_2$ is an \emph{unguarded choice} of $P$ if there
	exists $\PCtxC$ such that $P \pcong \PCtxC[P_1 \choice P_2]$. We say that
	$P$ is in \emph{choice normal form} if it has no unguarded choices.
\end{definition}

\begin{definition}[Thread Normal Form]
	\label{def:thread_normal_form}
	A process is in \emph{thread normal form} if it is generated by the grammar below:
	\[
		\begin{array}{@{}rcl@{}}
			\Pnf, \Qnf & ::= & \pcast{u}\Pnf \mid \Ppar
			\\
			\Ppar, \Qpar & ::= & \pres{s}{\procs{\Ppar}} \mid \Pth
			\\
			\Pth & ::= &
			\pdone \mid
			\pclose{u} \mid \pwait{u}{P} \mid 
			\pbranch[i\in I]{\chvar}{\rolep}\Pol{\la_i}{\PP_i} \mid 
			\poch{u}{\rolep}{v}{P} \mid
			\pich{u}{\rolep}{x}{P}
		\end{array}
	\] 
\end{definition}

Intuitively, a process is in \emph{thread normal form} if it consists of an initial prefix of casts followed by a parallel composition of threads, where a thread is either $\pdone$ or a process waiting to perform an input/output action on some channel $u = \ep{s}{\role}$ for some $\role$. In this latter case, we say that the thread is an $s$-thread.

\begin{definition}[Proximity Normal Form]
	\label{def:pnf}
	We say that $\Pnf$ is in \emph{proximity normal form} if $\Pnf = \PCtxC[\pres{s}{\procs{\Pth}}]$ for some $\PCtxC$, $s$, $\procs{\Pth}$ where
	each $\Pth_i$ for $i=1,\dots,h$ is a $s$-thread.
\end{definition}


\begin{lemma}
	\label{lem:cnf2}
	If $\wtp[n]\Ctx{P}$ and $\wtpi\Ctx{P}$, then there exists $Q$ in choice
	normal form such that $P \wred Q$ and $\wtp[m]\Ctx{Q}$ for some $m \le n$.
\end{lemma}
\begin{proof}
By induction on $\wtpi\Ctx{P}$ and by cases on the last rule applied.

\proofcase{Case $P$ is already in choice normal form} 
We conclude taking $Q \eqdef P$ and $m \eqdef n$.

\proofrule{t-call}
Then $P = \pinvk{A}{\seqof{u}}$ and $\Definition{A}{\seqof{x}}{R}$.
We deduce $\Ctx = \seqof{u : S}$, $\tass{A}{\seqof{S}}{n'}$ and $\wtpi\Ctx{R\subst{\seqof u}{\seqof x}}$. Moreover, it must be the case that
$\wtp[n']\Ctx{R\subst{\seqof u}{\seqof x}}$ and $n' \leq n$ since \refrule{t-call} is used in the coinductive judgment as well.
Using the induction hypothesis we deduce that there exist $Q$ in choice normal form and $m \le n'$ such that $R\subst{\seqof u}{\seqof x} \wred Q$ and $\wtp[m]\Ctx{Q}$.
We conclude by observing that $P \wred Q$ using \refrule{r-struct} and that $m \leq n' \leq n$.

\proofrule{co-choice}
Then $P = P_1 \choice P_2$.
We deduce $\wtpi\Ctx{P_k}$ with $k \in \set{1,2}$.
Moreover, it must be the case that $\wtp[n]\Ctx{P_k}$ since \refrule{t-choice} is used in the coinductive judgment.
Using the induction hypothesis we deduce that there exist $Q$ in choice normal form and $m \leq n$ such that $P_k \wred Q$ and $\wtp[m]\Ctx{Q}$.
We conclude by observing that $P \red P_k$ by \refrule{r-choice}.

\proofrule{t-choice}
Analogous to the previous case but we consider the premise in which the rank is the same of the conclusion to keep sure that it does not increase.

\proofrule{t-par}
Then $P = \pres{s}{P_1 \parop \dots \parop P_h}$. 
We deduce 
\begin{itemize}
\item $\Ctx = \Ctx_1, \dots, \Ctx_h$
\item $\wtpi{\Ctx_i, \ep{s}{\role_i} : S_i}{P_i}$ for $i=1,\dots,h$
\item $\prod_{i=1}^h \Map{\role_i}{S_i} \ft$
\end{itemize}
Furthermore, it must be the case that $\wtp[n_i]{\Ctx_i, \ep{s}{\role_i} : S_i}{P_i}$ for $i=1,\dots,h$ 
and $n = 1 + \sum_{i=1}^h n_i$ since \refrule{t-par} is used in the coinductive judgment as well.
Using the induction hypothesis we deduce that there exist $Q_i$ in choice normal form and $m_i \leq n_i$ such that $P_i \wred Q_i$ and $\wtp[m_i]{\Ctx_i, \ep{s}{\role_i} : S_i}{Q_i}$ for $i=1,\dots,h$.
We conclude by taking $m \eqdef 1 + \sum_{i=1}^h m_i$ and $Q \eqdef \pres{s}{Q_1 \parop \cdots \parop Q_h}$ with one application of \refrule{t-par}, observing that $m = 1 + \sum_{i=1}^h m_i \leq 1 + \sum_{i=1}^h n_i = n$ and that $P \wred Q$ by \refrule{r-par}.

\proofrule{t-cast}
Then $P = \pcast{u} P'$.
Analogous to the previous case, just simpler.
\end{proof}

\begin{lemma}
	\label{lem:cnf1}
	If $\wtp[n]\Ctx{P}$, then there exists $Q$ in choice normal form such that $P \wred Q$ and $\wtp[m]\Ctx{Q}$ for some $m \le n$.
\end{lemma}
\begin{proof}
	Consequence of \cref{lem:cnf2} noting that $\wtp[n]\Ctx{P}$ implies $\wtpi\Ctx{P}$.
\end{proof}

\begin{lemma}
	\label{lem:cnf}
	If $\wtpn\Measure\Ctx{P}$, then there exist $Q$ in choice normal form and $\MeasureN \leq \Measure$ such that $P \wred Q$ and $\wtpn\MeasureN\Ctx{Q}$.
\end{lemma}
\begin{proof}
By induction on $\wtpn\Measure\Ctx{P}$ and by cases on the last rule applied.

\proofrule{mt-thread}
Then $P$ is a thread. We deduce that
\begin{itemize}
\item $\Measure = (n , 0)$ for some $n$
\item $\wtp[n]\Ctx{P}$
\end{itemize}
From \cref{lem:cnf1} we deduce that there exist $Q$ and $m \le n$ such that $P \wred Q$ and $\wtp[m]\Ctx{Q}$. From \cref{lem:measure_rank} we deduce $\wtpn\MeasureN\Ctx{Q}$ for some $\MeasureN \le (m , 0)$. We conclude observing that $\MeasureN \le (m , 0) \le (n , 0) = \Measure$.

\proofrule{mt-cast}
Then $P = \pcast{u}{P'}$. We deduce that
\begin{itemize}
\item $\Ctx = \CtxD, u : S$
\item $S \subt[n] T$
\item $\Measure = \Measure' + (n,0)$
\item $\wtpn{\Measure'}{\Ctx', u : T}{P'}$
\end{itemize}
Using the induction hypothesis we deduce that there exist $Q'$ and $\MeasureN' \le \Measure'$ such that $P' \wred Q'$ and $\wtpn{\MeasureN'}{\Ctx', u : T}{Q'}$. We conclude with an application of \refrule{mt-cast} taking $Q \eqdef \pcast{u}{Q'}$, $\MeasureN \eqdef \MeasureN' + (n,0)$ and observing that $P \wred Q$ using \refrule{r-cast}. 

\proofrule{mt-par}
Then $P = \pres{s}{P_1 \parop \dots \parop P_h}$. 
We deduce 
\begin{itemize}
\item $\Ctx = \Ctx_1, \dots, \Ctx_h$
\item $\Measure = \sum_{i=1}^h \Measure_i + (0, \rank{\prod_{i=1}^h \Map{\role_i}{S_i}})$
\item $\wtpn{\Measure_i}{\Ctx_i, \ep{s}{\role_i} : S_i}{P_i}$ for $i=1,\dots,h$
\item $\prod_{i=1}^h \Map{\role_i}{S_i} \ft$
\end{itemize}
Using the induction hypothesis we deduce that there exist $Q_i$ in choice normal form and $\MeasureN_i \leq \Measure_i$ such that $P_i \wred Q_i$ and 
$\wtpn{\MeasureN_i}{\Ctx_i, \ep{s}{\role_i} : S_i}{Q_i}$ for $i=1,\dots,h$.
We conclude by taking $\MeasureN \eqdef \sum_{i=1}^h \MeasureN_i + (0, \rank{\prod_{i=1}^h \Map{\role_i}{S_i}})$ and $Q \eqdef \pres{s}{Q_1 \parop \cdots \parop Q_h}$ with one application of \refrule{mt-par}, observing that $\MeasureN = \sum_{i=1}^h \MeasureN_i + (0, \rank{\prod_{i=1}^h \Map{\role_i}{S_i}}) \leq \sum_{i=1}^h \Measure_i + (0, \rank{\prod_{i=1}^h \Map{\role_i}{S_i}}) = \Measure$ and that $P \wred Q$ by \refrule{r-par}.
\end{proof}

\begin{lemma}
	\label{lem:tnf}
	If $\wtpi\Ctx P$ and $P$ is in choice normal form, then there exists $\Pnf$ such that $P \pcong \Pnf$. 
\end{lemma}
\begin{proof}
By induction on $\wtpi\Ctx P$ and by cases on the last rule applied.

\proofcase{Cases \refrule{t-choice} and \refrule{co-choice}} These cases are impossible from the hypothesis that $P$ is in choice normal form.

\proofcase{Cases \refrule{t-done}, \refrule{t-wait}, \refrule{t-close}, \refrule{t-channel-in}, \refrule{t-channel-out}, \refrule{t-tag}, \refrule{co-tag}}
Then $P$ is a thread and is already in thread normal form and we conclude by reflexivity of $\pcong$.

\proofrule{t-call}
Then there exist $A$, $Q$, $\seqof u$ and $\seqof S$ such that
\begin{itemize}
\item $P = \pinvk{A}{\seqof u}$
\item $\Definition{A}{\seqof x}Q$
\item $\Ctx = \seqof{u : S}$
\item $\wtpi{\seqof{u : S}}{Q\subst{\seqof u}{\seqof x}}$
\end{itemize}
Using the induction hypothesis on $\wtpi{\seqof{u : S}}{Q\subst{\seqof u}{\seqof x}}$ we deduce that there exists $\Pnf$ such that $Q\subst{\seqof u}{\seqof x} \pcong \Pnf$.
We conclude $P \pcong \Pnf$ using \refrule{s-call} and the transitivity of $\pcong$.

\proofrule{t-par}
Then there exist $s$ and $P_i, \Ctx_i, S_i, \role_i$ for $i=1,\dots,h$ such that
\begin{itemize}
\item $P = \pres{s}{P_1 \parop \cdots \parop P_h}$
\item $\Ctx = \Ctx_1, \dots,\Ctx_h$
\item $\wtpi{\Ctx_i, \ep{s}{\role_i} : S_i}{P_i}$ for $i=1,\dots,h$
\end{itemize}
Using the induction hypothesis on $\wtpi{\Ctx_i, \ep{s}{\role_i} : S_i}{P_i}$ we deduce that there exist $\Pnf_i$ such that $P_i \pcong \Pnf_i$ for $i=1,\dots,h$.
By definition of thread normal form, it must be the case that $\Pnf_i = \pcast{\seqof{u_i}} \Ppar_i$ for some $\seqof{u_i}$ and $\Ppar_i$. Let $\seqof{v_i}$ be the same sequence as $\seqof{u_i}$ except that occurrences of $\ep{s}{\role_i}$ have been removed.
We conclude by taking $\Pnf \eqdef \pcast{\seqof{v_1}\dots\seqof{v_h}}\pres{s}{\Ppar_1 \parop \dots \parop \Ppar_h}$ and using the fact that $\pcong$ is a pre-congruence and observing that
	\[
		\begin{array}{rcll}
			P & = & \pres{s}{P_1 \parop \cdots \parop P_h} & \text{by definition of $P$}
			\\
			& \pcong & \pres{s}{\Pnf_1 \parop \cdots \parop \Pnf_h} & \text{using the induction hypothesis}
			\\
			& = & \pres{s}{\pcast{\seqof{u_1}}\Ppar_1 \parop \cdots \parop \pcast{\seqof{u_h}}\Ppar_h}
			& \text{by definition of thread normal form}
			\\
			& \pcong & \pcast{\seqof{v_1} \dots \seqof{v_h}}\pres{s}{\Ppar_1 \parop \cdots \parop \Ppar_h}
			& \text{by \refrule{s-cast-new}, \refrule{s-cast-swap}, \refrule{s-par-comm}}
			\\
			& = & \Pnf & \text{by definition of $\Pnf$}
		\end{array}
	\]

\proofrule{t-cast}
Then there exist $u$, $Q$, $\Ctx'$, $S$ and $T$ such that
\begin{itemize}
\item $P = \pcast{u} Q$
\item $\Ctx = \Ctx', u : S$
\item $\wtpi{\Ctx', u : T}{Q}$
\item $S \subt T$
\end{itemize}
Using the induction hypothesis on $\wtpi{\Ctx', u : T}{Q}$ we deduce that there exists $\Qnf$ such that $Q \pcong \Qnf$.
We conclude by taking $\Pnf \eqdef \pcast{u}\Qnf$ using the fact that $\pcong$ is a pre-congruence.
\end{proof}

\begin{lemma}[Proximity]
  \label{lem:proximity}
  If $s\in\fn{P} \setminus \bn\PCtxC$, then $\pres{s}{\PCtxC[P] \parop \procs{Q}} \pcong \PCtxD[\pres{s}{P \parop \procs{Q}}]$ for some $\PCtxD$.
\end{lemma}
\begin{proof}
By induction on the structure of $\PCtxC$ and by cases on its shape.

\proofcase{Case $\PCtxC = \Hole$}
We conclude by taking $\PCtxD \eqdef \Hole$ using the reflexivity of $\pcong$. 

\proofcase{Case $\PCtxC = \pres{t}{\procs{P'} \parop \PCtxC' \parop \procs{Q'}}$}
From the hypothesis $s \in \fn{P} \setminus \bn\PCtxC$ we deduce $s \ne t$ and $s \in \fn{P} \setminus \bn{\PCtxC'}$.
Using the induction hypothesis and \refrule{s-par-comm} we deduce that there exists $\PCtxD'$ such that 
$\pres{s}{\PCtxC'[P] \parop \procs{Q}} \pcong \PCtxD'[\pres{s}{P \parop \procs{Q}}]$.
Take $\PCtxD \eqdef \pres{t}{\PCtxD' \parop \procs{P'} \parop \procs{Q'}}$. We conclude
  \[
    \begin{array}{rcll}
      \pres{s}{\PCtxC[P] \parop \procs{Q}}
      & = & \pres{s}{\pres{t}{\procs{P'} \parop \PCtxC'[P] \parop \procs{Q'}} \parop \procs{Q}}
      & \text{by definition of $\PCtxC$}
      \\
      & \pcong & \pres{s}{\procs{Q} \parop \pres{t}{\PCtxC'[P] \parop \procs{P'} \parop \procs{Q'}}}
      & \text{by \refrule{s-par-comm}}
      \\
      & \pcong & \pres{t}{\pres{s}{\procs{Q} \parop \PCtxC'[P]} \parop \procs{P'} \parop \procs{Q'}}
      & \text{by \refrule{s-par-assoc} and $s \in \fn{\PCtxC'[P]}$}
      \\
      & \pcong & \pres{t}{\pres{s}{\PCtxC'[P] \parop \procs{Q}} \parop \procs{P'} \parop \procs{Q'}}
      & \text{by \refrule{s-par-comm}}
      \\
      & \pcong & \pres{t}{\PCtxD'[\pres{s}{P \parop \procs{Q}}] \parop \procs{P'} \parop \procs{Q'}}
      & \text{using the induction hypothesis}
      \\
      & = & \PCtxD[\pres{s}{P \parop \procs{Q}}]
      & \text{by definition of $\PCtxD$}
    \end{array}
  \]
  
\proofcase{Case $\PCtxC = \pcast{\ep{t}{\role}}\PCtxC'$ and $s \ne t$}
Using the induction hypothesis we deduce that there exists $\PCtxD'$ such that $\pres{s}{\PCtxC'[P] \parop \procs{Q}} \pcong \PCtxD'[\pres{s}{P \parop \procs{Q}}]$.
Take $\PCtxD \eqdef \pcast{\ep{t}{\role}}\PCtxD'$. We conclude
  \[
    \begin{array}{rcll}
      \pres{s}{\PCtxC[P] \parop \procs{Q}}
      & = & \pres{s}{\pcast{\ep{t}{\role}}\PCtxC'[P] \parop \procs{Q}}
      & \text{by definition of $\PCtxC$}
      \\
      & \pcong & \pcast{\ep{t}{\role}}\pres{s}{\PCtxC'[P] \parop \procs{Q}}
      & \text{by \refrule{s-cast-swap} and $t \ne s$}
      \\
      & \pcong & \pcast{\ep{t}{\role}}\PCtxD'[\pres{s}{P \parop \procs{Q}}]
      & \text{using the induction hypothesis}
      \\
      & = & \PCtxD[\pres{s}{P \parop \procs{Q}}]
      & \text{by definition of $\PCtxD$}
    \end{array}
  \]

\proofcase{Case $\PCtxC = \pcast{\ep{s}{\role}}\PCtxC'$}
Using the induction hypothesis we deduce that there exists $\PCtxD$ such that $\pres{s}{\PCtxC'[P] \parop \procs{Q}} \pcong \PCtxD[\pres{s}{P \parop \procs{Q}}]$.
We conclude
  \[
    \begin{array}[b]{rcll}
      \pres{s}{\PCtxC[P] \parop \procs{Q}}
      & = & \pres{s}{\pcast{\ep{s}{\role}}\PCtxC'[P] \parop \procs{Q}}
      & \text{by definition of $\PCtxC$}
      \\
      & \pcong & \pres{s}{\PCtxC'[P] \parop \procs{Q}}
      & \text{by \refrule{s-cast-new}}
      \\
      & \pcong & \PCtxD[\pres{s}{P \parop \procs{Q}}]
      & \text{using the induction hypothesis}
    \end{array}
    \qedhere
  \]
\end{proof}

\begin{lemma}[Quasi - Deadlock Freedom]
	\label{lem:dl_freedom}
	If $\wtpn\Measure\EmptyCtx\Pnf$, then $\Pnf = \pdone$ or $\Pnf \pcong \Qnf$ for some $\Qnf$ in proximity normal form.
\end{lemma}
\begin{proof}
By induction on the derivation of $\wtpn\MeasureM\EmptyCtx\Pnf$ we deduce that $\Pnf$ consists of $s_1,\dots,s_h$ sessions and $\sum_{i=1}^h k_i - h + 1$ threads where $k_i$ is the number of roles in $s_i$. The scenarios in which no communication is possible are those in which for each session $s_i$ there are less than $k_i$ $s_i$-threads. If we assume that for each $s_i$ there are $k_i - 1$ threads, then we obtain 
\[
\sum_{i=1}^h k_i - h + 1 - \sum_{i=1}^h(k_i - 1) = \sum_{i=1}^h k_i - h + 1 - \sum_{i=1}^hk_i + h = 1
\]
$s_i$-thread for some $s_i$; hence, there exist $k_i$ $s_i$-threads. In other words, there exist $\PCtxD,\PCtxC_1,\dots,\PCtxC_{k_i}$ and $\Pth_1,\dots,\Pth_{k_i}$ $s_i$-threads such that $\Pnf = \PCtxD[\pres{s_i}{\PCtxC_1[\Pth_1] \parop \cdots \parop \PCtxC_{k_i}[\Pth_{k_i}]}]$. We conclude
	\[
		\begin{array}{rcll}
			\Pnf & = & \PCtxD[\pres{s_i}{\PCtxC_1[\Pth_1] \parop \cdots \parop \PCtxC_{k_i}[\Pth_{k_i}]}]
			& \text{by definition of $\Pnf$}
			\\
			& \pcong & \PCtxD[\PCtxD_1[\pres{s_i}{\Pth_1 \parop \PCtxC_2[\Pth_2] \parop \cdots \parop \PCtxC_{k_i}[\Pth_{k_i}]}]]
			& \text{by \cref{lem:proximity}}
			\\
			& \pcong & \PCtxD[\PCtxD_1[\pres{s_i}{\PCtxC_2[\Pth_2] \parop \Pth_1 \parop \cdots \parop \PCtxC_{k_i}[\Pth_{k_i}]}]]
			& \text{by \refrule{s-par-comm}}
			\\
			& \dots \\
			& \pcong & \PCtxD[\PCtxD_1[\PCtxD_2[ \dots \PCtxD_{k_i}[\pres{s_i}{\Pth_{k_i} \parop \cdots \parop \Pth_2 \parop \Pth_1}] \dots ]]]
			& \text{for some $\PCtxD_2,\dots,\PCtxD_{k_i}$} 
			\\
			& \eqdef & \Qnf
		\end{array}
	\]
	The fact that $\Qnf$ is in thread normal form follows from the observation
	that $\Pnf$ does not have unguarded casts (it is a closed process in thread
	normal form) so the pre-congruence rules applied here and in
	\cref{lem:proximity} do not move casts around. We conclude that $\Qnf$ is in
	proximity normal form by its shape.
\end{proof}

\cref{lem:dl_freedom} is dubbed ``quasi-deadlock freedom'' because it does not
say that $\Qnf$ reduces. Indeed, a process in proximity normal form is only
\emph{ready to communicate} thanks to its shape (see reduction rules). We can
prove that a well typed process of such kind actually reduces by observing that
\refrule{t-par} requires that the involved session is coherent. This result is
the key ingredient for proving \cref{lem:pnf_helpful_direction}.

\subsection{Soundness}

\begin{lemma}
	\label{lem:pnf_helpful_direction}
	If $\wtpn\Measure\Ctx\Pnf$ where $\Pnf$ is in proximity normal form, then there exist $Q$ and $\MeasureN < \Measure$ such that $\Pnf \wred^+ Q$ and $					\wtpn\MeasureN\Ctx{Q}$.
\end{lemma}
\begin{proof}
From the hypothesis that $\Pnf$ is in proximity normal form we know that $\Pnf = \PCtxC[\pres{s}{\Pth_1 \parop \cdots \parop \Pth_h}]$ for some $\PCtxC$, $s$ and $\Pth_1, \dots, \Pth_h$ $s$-threads.
We reason by induction on $\PCtxC$ and by cases on its shape.
	
\proofcase{Case $\PCtxC = \Hole$}
From \refrule{mt-thread} and \refrule{mt-par} we deduce that there exist $\Ctx_i, S_i, \role_i, n_i$ for $i=1,\dots,h$ such that
\begin{itemize}
\item $\Ctx = \Ctx_1,\dots,\Ctx_h$
\item $\prod_{i=1}^h \Map{\role_i}{S_i} \ft$
\item $\Measure = (\sum_{i=1}^h n_i , \rank{\prod_{i=1}^h \Map{\role_i}{S_i}})$
\item $\wtp[n_i]{\Ctx_i, \ep{s}{\role_i} : S_i}{\Pth_i}$ for $i=1,\dots,h$
\end{itemize}
From the hypothesis that $\prod_{i=1}^h \Map{\role_i}{S_i} \ft$ we deduce $\prod_{i=1}^h \Map{\role_i}{S_i} \wlred{\In\terminated}$. 
We now reason on the rank of the session and on the shape of $S_i$. For the sake of simplicity, we implicitly apply \refrule{s-par-comm} at process level.\\\\
If $\rank{\prod_{i=1}^h \Map{\role_i}{S_i}} = 1$, then $\prod_{i=1}^h \Map{\role_i}{S_i} \lred{\In\terminated}$ using \refrule{l-terminate}.
\begin{itemize}
\item \proofcase{Case $S_1 = \End[\In]$ and $S_j = \End[\Out]$ for $j=2,\dots,h$}
Then 
	\begin{itemize}
	\item $\Ctx_j = \EmptyCtx$ and $\Pth_j = \pclose{\ep{s}{\role_j}}$ for $j=2,\dots,h$
	\item $\Pth_1 = \pwait{\ep{s}{\role_1}}{Q}$
	\item $\wtp[n_1]{\Ctx_1}{Q}$
	\end{itemize}
From \cref{lem:measure_rank} we deduce that $\wtpn{\MeasureN}{\Ctx_1}{Q}$ for some $\MeasureN \le (n_1 , 0)$. We conclude observing that $\Pnf \red Q$ by \refrule{r-signal} and that $\MeasureN \le (n_1 , 0) < (\sum_{i=1}^h n_i , \rank{\prod_{i=1}^h \Map{\role_i}{S_i}}) = \Measure$.
\end{itemize}
If $\rank{\prod_{i=1}^h \Map{\role_i}{S_i}} > 1$, then $\prod_{i=1}^h \Map{\role_i}{S_i} \lred{\tau} \dots \lred{\In\terminated}$ first using \refrule{l-tau} and \refrule{l-sync}. Observe that \refrule{l-pick} is never used since we are considering the minimum reduction sequence; a synchronization through \refrule{l-pick} and \refrule{l-sync} would lead to a longer reduction. Then $S_1 \xlred{\Map{\role_1}{\role_2\Out\Tag_k}}$ and $S_2 \xlred{\Map{\role_2}{\role_1\In\Tag_k}}$ for some $\Tag_k$ or $S_1 \xlred{\Map{\role_1}{\role_2}\Out S}$ and $S_2 \xlred{\Map{\role_1}{\role_2}\In S}$.
\begin{itemize}
\item \proofcase{Case $S_1 = \Tags\role_2\Out \Tag_i.S'_i$ and $S_2 = \JTags\role_1\In \Tag_j.T_j$ with $k \in I$}
From the hypothesis that $\prod_{i=1}^h \Map{\role_i}{S_i} \ft$ we deduce $I \subseteq J$. From \cref{def:coherence} we deduce $\prod_{i = 3}^h \Map{\role_i}{S_i} \parop \Map{\role_1}{S'_{k}} \parop \Map{\role_2}{T_{k}} \ft$ and from \refrule{t-tag} we deduce that 
	\begin{itemize}
	\item $\Pth_1 = \pbranch[i \in I]{\ep{s}{\role_1}}{\role_2}{\Out}{\Tag_i}{P'_i}$
	\item $\Pth_2 = \pbranch[j \in J]{\ep{s}{\role_2}}{\role_1}{\In}{\Tag_j}{Q_j}$
	\item $\wtp[n_1]{\Ctx_1, \ep{s}{\role_1} : S'_i}{P'_i}$ for all $i \in I$
	\item $\wtp[n_2]{\Ctx_2, \ep{s}{\role_2} : T_j}{Q_j}$ for all $j \in J$
	\end{itemize}
Let $Q \eqdef \pres{s}{P'_k \parop Q_k \parop P_3 \parop \dots \parop P_h}$ and observe that $\Pnf \wred^+ Q$ by \refrule{r-pick} and \refrule{r-tag}. From \cref{lem:measure_rank} we deduce that there exist $\Measure_1 \le (n_1 , 0), \Measure_2 \le (n_2 , 0)$ such that
	\begin{itemize}
	\item $\wtpn{\Measure_1}{\Ctx_1, \ep{s}{\role_1} : S'_k}{P'_k}$
	\item $\wtpn{\Measure_2}{\Ctx_2, \ep{s}{\role_2} : T_k}{Q_k}$
	\end{itemize}
Let $\MeasureN \eqdef \Measure_1 + \Measure_2 + (\sum_{i=3}^h n_i, \rank{\Map{\role_1}{S'_k} \parop \Map{\role_2}{T_k} \parop \prod_{i=3}^h \Map{\role_i}{S_i}})$. We conclude with one application of \refrule{mt-par} observing that
\[
\begin{array}{rcll}
	\MeasureN & = & \Measure_1 + \Measure_2 + (\sum_{i=3}^h n_i, \rank{\Map{\role_1}{S'_k} \parop \Map{\role_2}{T_k} \parop \prod_{i=3}^h \Map{\role_i}{S_i}})
	\\
	& \le & (\sum_{i=1}^h n_i, \rank{\Map{\role_1}{S'_k} \parop \Map{\role_2}{T_k} \parop \prod_{i=3}^h \Map{\role_i}{S_i}})
	& \text{by \cref{lem:measure_rank}}
	\\
	& < & (\sum_{i=1}^h n_i, \rank{\prod_{i=1}^h \Map{\role_i}{S_i}}) 
	& \text{before reductions}
	\\
	& = & \Measure
\end{array}
\]
\end{itemize}
\begin{itemize}
\item \proofcase{Case $S_1 = \role_2\Out{S}.T_1$ and $S_2 = \role_1\In{S}.T_2$}
\end{itemize}
From the hypothesis that $\prod_{i=1}^h \Map{\role_i}{S_i} \ft$ and \cref{def:coherence} we deduce $\Map{\role_1}{T_1} \parop \Map{\role_2}{T_2} \parop \prod_{i=3}^h \Map{\role_i}{S_i} \ft$ and from \refrule{t-channel-out} and \refrule{t-channel-in} we deduce that
	\begin{itemize}
	\item $\Pth_1 = \poch{\ep{s}{\role_1}}{\role_2}{u}{P'_1}$
	\item $\Pth_2 = \pich{\ep{s}{\role_2}}{\role_1}{x}{P'_2}$
	\item $\wtp[n_1]{\Ctx_1, \ep{s}{\role_1} : T_1}{P'_1}$
	\item $\wtp[n_2]{\Ctx_2, \ep{s}{\role_2} : T_2, x : S}{P'_2}$
	\end{itemize}
Let $Q \eqdef \pres{s}{P'_1 \parop P'_2\subst{u}{x} \parop \Pth_3 \parop \cdots \parop \Pth_h}$ and observe that $\Pnf \red Q$ by \refrule{r-channel}. Using \cref{lem:substitution} we deduce $\wtp[n_2]{\Ctx_2, \ep{s}{\role_2} : T_2, u : S}{P'_2\subst{u}{x}}$ and from \cref{lem:measure_rank} we deduce that there exist $\Measure_1 \le (n_1 , 0), \Measure_2 \le (n_2 , 0)$ such that
	\begin{itemize}
	\item $\wtpn{\Measure_1}{\Ctx_1, \ep{s}{\role_1} : T_1}{P'_1}$
	\item $\wtpn{\Measure_2}{\Ctx_2, \ep{s}{\role_2} : T_2, u : S}{P'_2\subst{u}{x}}$
	\end{itemize}
Let $\MeasureN \eqdef \Measure_1 + \Measure_2 + (\sum_{i=3}^h n_i, \rank{\Map{\role_1}{T_1} \parop \Map{\role_2}{T_2} \parop \prod_{i=3}^h \Map{\role_i}{S_i}})$. We conclude with one application of \refrule{mt-par} observing that
\[
\begin{array}{rcll}
	\MeasureN & = & \Measure_1 + \Measure_2 + (\sum_{i=3}^h n_i, \rank{\Map{\role_1}{T_1} \parop \Map{\role_2}{T_2} \parop \prod_{i=3}^h \Map{\role_i}{S_i}})
	\\
	& \le & (\sum_{i=1}^h n_i, \rank{\Map{\role_1}{T_1} \parop \Map{\role_2}{T_2} \parop \prod_{i=3}^h \Map{\role_i}{S_i}})
	& \text{by \cref{lem:measure_rank}}
	\\
	& < & (\sum_{i=1}^h n_i, \rank{\prod_{i=1}^h \Map{\role_i}{S_i}}) 
	& \text{before reductions}
	\\
	& = & \Measure
\end{array}
\]

\proofcase{Case $\PCtxC = \pres{t}{\procs{\Ppar} \parop \PCtxD \parop \procs{\Qpar}}$}
Let $\Rnf \eqdef \PCtxD[\pres{s}{\Pth_1 \parop \dots \parop \Pth_h}]$ and observe that $\Rnf$ is in proximity normal form. From \refrule{mt-par} we deduce that there exist $\Ctx_i, S_i, \Measure_i, \role_i$ for $i=1,\dots,h$ and $k \le h$ such that
\begin{itemize}
\item $\Ctx = \Ctx_1,\dots,\Ctx_h$
\item $\wtpn{\Measure_1}{\Ctx_1, \ep{t}{\role_1} : S_1}{\Rnf}$
\item $\wtpn{\Measure_i}{\Ctx_i, \ep{t}{\role_i} : S_i}{\Ppar_i}$ for $i=1,\dots,k$
\item $\wtpn{\Measure_i}{\Ctx_i, \ep{t}{\role_i} : S_i}{\Qpar_i}$ for $i=k+1,\dots,h$
\item $\Measure = \sum_{i=1}^h \Measure_i + (0 , \rank{\prod_{i=1}^h \Map{\role_i}{S_i}})$
\end{itemize}
Using the induction hypothesis on $\wtpn{\Measure_1}{\Ctx_1, \ep{t}{\role_1} : S_1}{\Rnf}$ we deduce that there exists $Q'$ and $\MeasureN' < \Measure_1$ such that
\begin{itemize}
\item $\Rnf \wred^+ Q'$
\item $\wtpn{\MeasureN'}{\Ctx_1, \ep{t}{\role_1} : S_1}{Q'}$
\end{itemize}
We conclude taking $Q \eqdef \pres{t}{Q' \parop \procs\Ppar \parop \procs\Qpar}$ and $\MeasureN \eqdef \MeasureN' + \sum_{i=2}^h \Measure_i + (0 , \rank{\prod_{i=1}^h \Map{\role_i}{S_i}})$ and observing that $\MeasureN < \Measure$ and $\Pnf \wred^+ Q$ by \refrule{r-par}.

\proofcase{Case $\PCtxC = \pcast{\ep{t}{\roleq}}{\PCtxD}$}
Observe that $t \ne s$. Let $\Rnf \eqdef \PCtxD[\pres{s}{\Pth_1 \parop \cdots \parop \Pth_h}]$ and note that $\Rnf$ is in proximity normal form. From \refrule{mt-cast} we deduce that there exists $\CtxD, \Measure', S, T, m_t$ such that
\begin{itemize}
\item $\Ctx = \CtxD, \ep{t}{\roleq} : S$
\item $S \subt[m_t] T$
\item $\Measure = \Measure' + (m_t,0)$
\item $\wtpn{\Measure'}{\CtxD, \ep{t}{\roleq} : T}{\Rnf}$
\end{itemize}
Using the induction hypothesis on $\wtpn{\Measure'}{\CtxD, \ep{t}{\roleq} : T}{\Rnf}$ we deduce that there exist $Q'$ and $\MeasureN' < \Measure'$ such that $\Rnf \wred^+ Q'$ and $\wtpn{\MeasureN'}{\CtxD, \ep{t}{\roleq} : T}{Q'}$. We conclude taking $Q \eqdef \pcast{\ep{t}{\roleq}}{Q'}$ and $\MeasureN \eqdef \MeasureN' + (m_t , 0)$ and observing that $\MeasureN < \Measure$ and $\Pnf \wred^+ Q$ by \refrule{r-cast}.
\end{proof}

\begin{lemma}
	\label{lem:helpful_direction}
	If $\wtpn{\Measure}{\EmptyCtx}{P}$, then either $P \pcong \pdone$ or $P \wred^+ Q$ and $\wtpn{\MeasureN}{\EmptyCtx}{Q}$ for some $Q$ and $\MeasureN < \Measure$.
\end{lemma}
\begin{proof}
Using \cref{lem:cnf} we deduce that there exist $P'$ in choice normal form such that $P \wred P'$ and $\wtpn{\MeasureM'}\EmptyCtx{P'}$ and $\MeasureM' \leq \MeasureM$.
By \cref{lem:measure_rank} we deduce $\wtp\EmptyCtx{P'}$. Using \cref{lem:tnf} we deduce that there exist $\Pnf$ such that $P' \pcong \Pnf$.

If $\Pnf = \pdone$ there is nothing left to prove.

If $\Pnf \neq \pdone$, by \cref{lem:dl_freedom} we deduce $\Pnf \pcong \Qnf$ for some $\Qnf$ in proximity normal form.
From \cref{lem:measure_pcong} we deduce $\wtpn{\MeasureM''}\EmptyCtx\Qnf$ for some $\MeasureM'' \leq \MeasureM'$.
Using \cref{lem:pnf_helpful_direction} we conclude that $\Qnf \wred^+ Q$ and $\wtpn\MeasureN\EmptyCtx Q$ for some $Q$ and $\MeasureN < \MeasureM'' \leq \MeasureM' \leq \MeasureM$.
\end{proof}

\begin{lemma}
	\label{lem:weak_termination}
	If $\wtp[n]\EmptyCtx{P}$, then either $P \pcong \pdone$ or $P \wred^+ \pdone$.
\end{lemma}
\begin{proof}
From \cref{lem:measure_rank} we deduce that there exists $\MeasureM \le (n,0)$ such that $\wtpn\MeasureM\EmptyCtx P$.
We proceed doing an induction on the lexicographically ordered pair $\MeasureM$.
From \cref{lem:helpful_direction} we deduce either $P \pcong \pdone$ or $P \wred^+ Q$ and $\wtpn\MeasureN\EmptyCtx{Q}$ for some $\MeasureN < \MeasureM$.
In the first case there is nothing left to prove.
In the second case we use the induction hypothesis to deduce that either $Q \pcong \pdone$ or $Q \wred^+ \pdone$.
We conclude using either \refrule{r-struct} or the transitivity of $\wred^+$, respectively.
\end{proof}

\begin{proof}[Proof of \cref{thm:soundness}]
Immediate consequence of \cref{lem:subj_red,lem:weak_termination}.
\end{proof}

\end{document}